\documentclass[12pt,a4paper,english]{article}
\usepackage{amssymb,amsmath,graphicx,multicol}
\usepackage{marginnote}
\pdfoutput=1
\usepackage{amsmath}   
\usepackage{blkarray}
 
\usepackage{xcolor} 
\usepackage{verbatim}
\usepackage{tikz}
\usepackage{amsmath, amssymb, amsfonts, mathrsfs}
\usepackage{blkarray}
\usepackage[T1]{fontenc}
\usepackage{graphicx}
\usepackage{amssymb}
\usepackage{esint}
\usepackage{a4wide}
\usepackage{psfrag}
\usepackage{babel}
\usepackage{epstopdf}
\usepackage{color}
\usepackage[
colorlinks=true, 
linkcolor=black,  
citecolor=black,   
urlcolor=teal    
]{hyperref}
\usepackage{braket}
\usepackage{ytableau}
\usepackage{caption}
\usepackage{subcaption}
\usepackage{bm}
\usepackage{array}
\usepackage[
    style=phys,
]{biblatex}
\DeclareFieldFormat[article]{title}{\mkbibemph{#1}}
\newcommand{\EQ}{\begin{equation}}
\newcommand{\EN}{\end{equation}}
\newcommand{\be}{\begin{equation}}
\newcommand{\ee}{\end{equation}}
\newcommand{\bea}{\begin{eqnarray}}
\newcommand{\eea}{\end{eqnarray}}
\newcommand{\bi}{\begin{itemize}}
\newcommand{\aaaa}{\mathfrak{a}}
\newcommand{\ei}{\end{itemize}} 
\newcommand{\bc}{\begin{column}{0.50\textwidth}}
\newcommand{\ec}{\end{column}}
\newcommand{\bcs}{\begin{columns}} 
\newcommand{\ecs}{\end{columns}}
\makeatletter

\newcommand{\half}{\frac{1}{2}}
\newcommand{\inv}[1]{\frac{1}{#1}}
\newcommand{\br}[1]{\left( #1 \right)}
\newcommand{\sbr}[1]{\left[ #1 \right]}
\newcommand{\cbr}[1]{\left\{ #1 \right\}}
\newcommand{\dd}{\mathrm{d}}

\usepackage{etoolbox} 

\newif\ifappendix
\appendixfalse

\pretocmd{\appendix}{\appendixtrue}{}{}

\makeatletter
\renewcommand{\@seccntformat}[1]{%
  \ifappendix
    Appendix \csname the#1\endcsname:\;
  \else
    \csname the#1\endcsname\quad
  \fi}
\makeatother

\begin{document}

\vspace*{-1.5cm}
\begin{center}
\LARGE

Statistical Signatures of Integrable and Non-Integrable \\ 
Quantum Hamiltonians

\end{center}
\vspace{0.5cm}
\begin{center}
{\large Feng He\footnote{fhe@sissa.it}, Arthur Hutsalyuk\footnote{hutsalyuk@gmail.com}, Giuseppe Mussardo\footnote{mussardo@sissa.it}, 
Andrea Stampiggi\footnote{astampig@sissa.it}
\vspace{0.9cm}}

{\sl International School for Advanced Studies (SISSA),\\
Via Bonomea 265, 34136 Trieste, Italy\\[2mm]
INFN, Sezione di Trieste\\[2mm]
}

\end{center}

\vspace{0.5cm}
\begin{center}
{\bf Abstract}\\[5mm]
\end{center}
Integrability has long served as a cornerstone concept in classical mechanics, where it possesses a precise and unambiguous definition. Extending this notion to the quantum domain, however, remains a far more subtle and elusive problem. In particular, deciding whether a given quantum Hamiltonian -- viewed simply as a matrix of its elements -- does or does not define an integrable system is far from obvious. Yet this question is crucial: it bears directly on non-equilibrium dynamics, spectral correlations, the behaviour of correlation functions, and other fundamental properties of many-body quantum systems.
In this work, we develop a statistical framework for addressing quantum integrability from a purely probabilistic standpoint. Our approach begins with the observation that a necessary signature of integrability is the finite probability of encountering vanishing energy gaps in the spectrum. On this basis, we formulate a twofold protocol capable of distinguishing between integrable and non-integrable Hamiltonians. The first step consists of a systematic Monte Carlo decimation of the spectrum, designed to reveal the emergence (or absence) of Poissonian level spacing statistics. The iterative decimation compresses the Hilbert space exponentially, and its termination point determines whether the spectrum is governed by a mixed distribution or approaches the Poisson limit. In the second step, the distinction is instead obtained by analysing the distributions of $k$-step gaps which can help in discriminating between Poisson and mixed statistics.
This procedure applies to Hamiltonians of arbitrary finite size, regardless of whether their structure involves a finite number of blocks or an exponentially fragmented Hilbert space. As a concrete benchmark, we implement the protocol on a class of quantum Hamiltonians constructed from the permutation group $\mathcal{S}_N$, thereby demonstrating both its effectiveness and its broad applicability.

\newpage

\section{Introduction}

Integrability has long been a fundamental concept in the characterization of systems in classical mechanics, where it admits a clear and well-defined meaning \cite{Gutzwiller}.  However, as well known, extending this notion to quantum mechanics is far from straightforward \cite{SutherlandBM,Weigert,Caux}.

The standard criteria for quantum integrability are typically grounded in the quantum Yang–Baxter equation \cite{Baxter,Baxter-8V,Baxter-8V-stat}, which underpins the Bethe ansatz approach to solving quantum Hamiltonians. Since Bethe’s seminal work \cite{Bethe} and its subsequent generalizations to broader classes of systems \cite{Yang-YB,Sutherland68,Sutherland,SFT}, there have been extensive efforts to develop systematic methods for identifying and classifying integrable models \cite{AFZ1983,Baz,BPZ,Zamo1989,Zamsigma,ZZ,Ogievetsky,JimboMiwa,Jimbo}. Among these, Reshetikhin’s criterion (see \cite{Grabowski95,deLeeuw_2020,deLeeuwRetorePalettaPribytok,deLeeuw_2019}) has been widely applied, providing a constructive framework to establish integrability and generate sets of commuting conserved charges \cite{deLeeuwPaletta,deLeeuwRetoreRyan,deLeeuwPalettaPozsgay,deLeeuwCorcoran,GomborPozsgay,RibeiroProsen,Buca_2021,Korepin24,deLeeuw2506,deLeeuw2410,Korepin1809,Korepin2401,Korepin2508}.
In practice, however, implementing these criteria is often a highly nontrivial task since it typically requires conjecturing the form of the first nontrivial conserved charge, a step that severely limits their applicability as a general diagnostic tool. This limitation underscores the need for simple and effective algorithms capable of detecting integrability in situations where only the spectrum—often obtained numerically—is available.

As discussed below, in identifying quantum integrable models one may face various degrees of difficulty -- particularly in absence of a classical counterpart, where one essentially deals with sets of numerical data, such as sequences of eigenvalues of various observables or matrix elements expressed in a chosen basis of the Hilbert space. There is an extensive body of literature on this subject (see, for instance, \cite{Wigner1955,Wigner1967,DysonI,DysonII,DysonIII,Porter,Mehta1,Mehta2,Brody,Bohigas,BGS,OxfordHandbook,Berry,BerryLesHouches1981,BerryLesHouches1989,Porter2,BerryRobnik,Vernier,BKM,Takacs,SS,Bhosale1} and references therein), mostly devoted to the analysis of models and Hamiltonians of chaotic or non-integrable dynamics (hereafter, the two terms will be used interchangeably). In this context, the prospect of identifying integrable dynamics through statistical analysis—what Porter aptly termed ``statistical spectroscopy'' \cite{Porter}—is both intriguing and conceptually compelling. Guiding tools are considerations from probability theory, together with well-defined questions related, for instance, to the level spacings of energy levels, the nature of their fluctuations or the outcome of specific tests. 

This paper is organised as follows: after the discussion in Section \ref{S2-integrability} on the main obstacles in properly defining quantum integrability, we present in Section \ref{S2-quant-integrable} a probabilistic argument that allows us to support the hypothesis, at least at a preliminary level, regarding whether a given model is integrable or not. As we will see, this argument relies solely on the existence of a finite probability of encountering a zero energy gap in the energy spectrum. 
For this reason, it is necessary to further substantiate this preliminary result through additional statistical analyses, primarily based on symmetry considerations, in order to distinguish between genuinely integrable dynamics and those exhibiting mixed characteristics. We discuss the role played by symmetries in Section \ref{S4-symmetries}, while in Section \ref{S5-Gap-level} we will examine the Poisson and Wigner-Dyson gap distributions -- corresponding, respectively, to the energy-level spacings of integrable and chaotic random matrices -- as well as the higher order spacing distributions of these two classes of models.  
In Section \ref{S6-superposition} we discuss the various gap distributions that arise from the superposition of the spectra of different types.  
As we will show, the emergence of such mixed distributions warrants particular attention and careful analysis, as their identification can be rather subtle. In fact, the superposition of a sufficiently large number of spectra of different types can closely approximate a Poisson distribution, making it indistinguishable from it within statistical uncertainty and, therefore, potentially leading to the erroneous conclusion that the underlying Hamiltonian is integrable when, in fact, it is not. Our discussion is based on the early analysis of this problem presented in \cite{Mehta1,Porter2,BerryRobnik} and, more recently, \cite{Vernier}. In Section \ref{S7-protocol}, we address the question of whether the observed spectrum follows a genuine Poisson distribution or a mixed distribution. As we will see, 
such a question can be resolved through a twofold protocol. The first step involves a systematic Monte Carlo-inspired decimation of the energy levels designed to test for the presence of Poisson statistics in the level spacings. Iterating this decimation rapidly reduces the dimension of the Hilbert space associated with the spectrum, scaling it down exponentially. If the process terminates before reaching its cutoff, i.e. a prescribed minimum number of levels, the spectrum can be identified as following a mixed distribution. If, however, the procedure reaches its final stage — yielding a number of levels of the assigned minimal size— the distinction between Poisson and mixed statistics can then be drawn by computing the original $k$-step energy gaps and comparing their distributions with those predicted for Poisson and mixed statistics, which differ in a clearly discernible manner. For any finite Hamiltonian matrix —regardless of its size— this protocol (especially the Monte Carlo decimation) remains equally effective, whether the system comprises a finite number of independent blocks or a fragmented Hilbert space corresponding to an exponential number of blocks.

To provide concrete examples of our analysis, in this paper we employ quantum Hamiltonians defined in terms of the finite permutation group. The use of this group offers several advantages, as will become evident in the discussion presented in Section \ref{s_permutation}. Finally, Section \ref{examples} presents the main results of our investigation, while the concluding remarks are provided in Section \ref{conclusions}. As a general remark, far from being rigorous, this paper combines analytic results together with heuristic arguments, statistical reasoning together with exact numerical diagonalization in order to point out some important features of integrable and non-integrable quantum Hamiltonians.

\section{\label{S2-integrability} Integrability in Classical and Quantum Worlds}
In this section, we begin by reviewing the main features of integrability in classical systems, both with finite and infinite degrees of freedom. We then argue that, in the quantum context, the only systems with an effective approach to check the presence of integrability are relativistic quantum field theories. The question of how to approach integrability in other quantum settings is addressed at the end of the section.

\subsection{Classical systems with a finite number of degrees of freedom}
Integrability is readily defined in classical mechanical systems, particularly those with a finite number 
$N$ of degrees of freedom, where the phase space is $2N$-dimensional and parametrized by generalized coordinates 
$q_i$ and their conjugate momenta $p_i$ ($i=1,2,\ldots, N$). For these systems, beyond the conceptual clarity of their definition, classical integrability is also characterized by a well-established procedure for the explicit solution of their equations of motion—arguably the primary reason for the enduring interest in integrable systems. Let's briefly summarise these two key points: 
\begin{itemize}
\item In classical mechanics, a system is said to be integrable if it possesses $N$ independent integrals of motion $Q_i(q, p)$ ($i=1,2,\ldots, N$)
that are in involution—that is, their mutual Poisson brackets vanish
\be
\{ Q_i, Q_j\} \,=\,0\,\,\,.
\ee
When the Hamiltonian $H(q,p)$ is time-independent, it is typically one of these conserved quantities.
\item
The existence of these integrals of motion confines the dynamics to a submanifold ${\mathcal S}$ of the phase space with dimension 
$D = 2 N - N = N$, which, under quite general conditions \cite{Arnold}, has the topology of an $N$-dimensional torus. 
This allows one to introduce a new set of canonical variables, known as action-angle variables, denoted $(I_i, \varphi_i)$ ($i=1,2,\ldots, N$). 
The action variables $I_i$, defined as 
\be 
I_i \,=\, \frac{1}{2 \pi} \oint_{\gamma_i} {\mathbf p} \cdot d{\mathbf q}\,\,\, ,
\ee
with $\gamma_i$ the $i$-th irreducible cycle of the torus, are specific functions of the integrals of motion $Q_i$, such that 
the Hamiltonian becomes a function of the actions alone, $H(I_1,I_2,\ldots I_N)$. In terms of these variables, the Hamiltonian equations of motion 
take a particularly simple form, and their solutions are straightforward to obtain
\bea
&& \dot I_i \,=\, -\left(\frac{\partial H}{\partial \varphi_i}\right) \,=\, 0 \hspace{5mm}
\longrightarrow 
\hspace{5mm} I_i \,=\, {\rm constant}\,\,\,, \\
&&\dot{\varphi_i} \,=\, \left(\frac{\partial H}{\partial I_i}\right) \, \equiv \omega_i
\hspace{5mm}
\longrightarrow 
\hspace{5mm}
\varphi_i \,=\, \omega_i \, t + \varphi_0 \,\,\, .
\eea
\end{itemize}
Given that the motion is restricted to the $N$-dimensional torus, in integrable models ergodicity is absent, in contrast to generic non-integrable systems, which densely span the phase space over the course of time. Despite this feature, it is, however, worth stressing that for large $N$,  it may be difficult to visually distinguish the motion of an integrable model from that of a non-integrable one. The reason is that all single-value functions $G({\bold q},{\bold p})$ of the system (in particular, the original coordinates and momenta), expressed in terms of the action-angle variables, are periodic functions of $\varphi_i$ and, therefore, are given in terms of a multiple Fourier series. Substituting for $\varphi_i$ their function of time, the time dependence of $G$ takes the form
\be 
G(t) \,=\, \sum_{n_1 = - \infty}^{\infty} \cdots \sum_{n_N = - \infty}^{\infty} {\mathcal G}_{n_1,\ldots,n_N} \, 
e^{\left[i t \left(n_1 \,\omega_1 + \cdots n_N \,\omega_N\right)\right]}  \,\,\, .
\ee
Since the frequencies $\omega_i$ are not, in general, commensurable, this function is not strictly periodic but only conditionally periodic. 
Therefore, the corresponding visual intricacy of motion may hide the integrability structure of the model. This feature is obviously more pronounced 
when the number of degrees of freedom tends to infinity, $N \rightarrow \infty$. Examples of dynamical systems with infinitely many degrees of freedom are 
field theories, both in the relativistic and non-relativistic versions.

\subsection{Classical systems with infinite degrees of freedom}
As classical systems with an infinite number of degrees of freedom, we consider here only field theories, namely those systems associated with local variables $\phi_a(x,t)$ that 
depend on the space and time coordinates $(x, t)$ and, eventually, on an additional internal index $a$ that distinguishes their type. For reasons that will soon become clear, we restrict our attention only to relativistically invariant and ($1 + 1$) dimensional field theories. The dynamics of these systems are encoded in the local Lagrangian density ${\mathcal L}(\phi_a,\partial_t\phi_a,\partial_x\phi_a)$, a function of the field $\phi_a$ and its time and space derivatives. Using ${\mathcal L}$, we can also define the conjugate momentum $\pi_a(x,t)$ of the field $\phi_a(x,t)$
\be
\pi_a(x,t)\,=\, \frac{\partial {\mathcal L}}{\partial(\partial_t \phi_a)} \,\,\,,
\ee
which satisfies the equal-time functional Poisson bracket with the field $\phi$ 
\be
\{\phi_a(x,t),\pi_b(y,t)\} \,=\, \delta_{ab}\, \delta(x-y)\,.
\ee
We can define the local Hamiltonian density $H(x,t)$, which depends on $x$ and $t$ through $\phi(x,t)$ and $\pi(x,t)$, via a Legendre transform of ${\mathcal L}$
\be
{\mathcal H}(x,t) \,=\, \sum_a \phi(x, t) \partial_t\phi_a(x,t) - {\mathcal L}(x,t)\,\,\,.
\ee
Integrating over space gives the total Hamiltonian $H$, which is the generator of time evolution
\be
H \,=\,\int dx \, {\mathcal H}(x,t) \,\,\,.
\ee      
As for classical systems with a finite number of degrees of freedom, we declare that a classical field theory is integrable if it is supported by an infinite number of functionally independent  conserved charges ${Q}_i(\phi_a,\pi_a)$ (where ${Q}_1= H$), which satisfy 
\be
\{Q_i, Q_j\} \,=\, 0\,\,\,.
\ee
In particular, the vanishing of the Poisson bracket with $H$ implies that they are constants of motion, a condition that is guaranteed if, associated with each of these charges, 
there are two local densities $(\rho_i(x,t),j_i(x,t))$ that satisfy the conservation law 
\be
\partial_t \rho_i(x,t) \,=\, \partial_x j_i(x,t) \,\,\,
\label{conslaws}
\ee
since $Q_i$ can be expressed in this case as 
\be
Q_i \,=\, \int dx \, \rho(x,t) \,\,\,.
\label{loccharges}
\ee
For models which are known to be classically integrable, such as the Sin(h)-Gordon model, their action-angle variables and their explicit solutions of the equation of motion are 
provided by the Inverse Scattering Methods \cite{Faddev,Novikov}. Explicit expressions of the conserved densities $\rho_i(x,t)$ for the Sinh-Gordon model can be found in \cite{DeLucaMussardo} and, hereafter, in Appendix~\ref{AppendixA}.  
 When properly generalized, these classical integrable systems yield non-trivial examples of quantum integrable models \cite{GMbook}. But how can one identify and possibly classify {\em all} quantum integrable systems with infinite degrees of freedom? To the best of our knowledge, a clear and well-defined answer exists only in the context of relativistic quantum field theories.

\subsection{Integrable Quantum Field Theories}
Integrable quantum field theories are supported by an infinite number of local conserved charges $Q_a$ ($a = 1, 2, \ldots)$ which commute with each other
\be
[ Q_a, Q_b ] \,=\, 0 
\,\,\,.
\label{zerocommutators}
\ee
While we will return later to the nature of these charges and the vanishing of their commutators, it is worth emphasizing for now that their existence implies that all scattering processes—regardless of the number of particles involved—are purely elastic and factorizable \cite{ZZ}.  This holds despite the fact that, in general, relativistic field theories permit particle production whenever sufficient energy is available in the center-of-mass frame. For kinematical reasons\footnote{The eigenvalues of the local charges on multi-particle states, which are their common eigenvectors, are generically given by the sum of higher powers of their momenta. Using the different action of these charges on states with different momenta, one can arbitrarily shift the point of space-time where the interactions take place \cite{Witten-Shankar}. In three or higher dimensions, for theories relative to localized particle excitations this means that we are dealing with free theories.}, non-trivial examples of integrable relativistic quantum field theories exist only in $(1+1)$ dimensions. The list includes purely bosonic models, such as the Sinh-Gordon and, more generally, Toda models \cite{GMbook,Zamo1989,AFZ1983,BCDS1990I,CM1989I,CM1989II,Ch1990,M1992,D1997,OotaI,FKS2000,DeliusII}, fermionic systems, such as the Gross-Neveu model \cite{ZZ}, as well as supersymmetric versions of all these examples (see, for instance, \cite{Witten-Shankar,Sc1990,A1994,Hollo}). This list increases even further if one also adds to it theories with soliton excitations, such as the Sine Gordon model or various sigma models based on group manifolds \cite{ZZ,Zamsigma,Ogievetsky}. 

The elasticity property of scattering amplitudes has an immediate consequence for relativistic QFT: if the scattering is to be elastic, all production and decay processes must be forbidden as a consequence of the peculiar values of the parameters of the model. Vice versa, if for a given model we are able to show the existence of production processes, this fact alone automatically provides explicit proof that such a model is {\em not} integrable. This criterion has the advantage of being checkable for any process involving a finite number of particles in a {\em finite} number of steps, as originally shown in a seminal paper by P.~Dorey \cite{D1997}. 

For relativistic quantum field theories which are uniquely defined by Feynman diagrams (as in {\em Diagrammar} by 't Hooft and Veltman \cite{tHooft:1973wag}), we have then the following 
{\em integrable sieve}: chosen a class of models -- specified by the number of fields present in the Lagrangian and its symmetry -- the systematic analysis of all production processes acts,  as a matter of fact, as a {\em sieve}, in the sense that at the end of this procedure, we are left with very special Lagrangians which -- by construction -- 
are integrable. On the basis of this algorithmic procedure, it has been possible, for instance, to exclude from the list of integrable models all Lagrangians with purely polynomial interactions, alias Landau-Ginzburg (LG) theories. On the other hand, it has led to the definition of new quantum integrable field theories using models which are generalisation of the Sinh-Gordon model, namely the affine simply-laced Toda field theories \cite{GMbook,Zamo1989,AFZ1983,BCDS1990I,CM1989I,CM1989II,Ch1990,M1992,D1997,OotaI,FKS2000,DeliusII,Khastgir,MussardoTateo}: these theories are built up in terms of the simple roots of the ADE Lie algebras, and the Sinh-Gordon model (corresponding to the algebra $A_1$) is indeed their simplest representative. For the identification of non-relativistic local and continuous field theories based on the considerations made above, see \cite{GMB,GMB2}.  

\subsection{Generic Quantum Hamiltonians}

In what follows, we focus on generic quantum systems that do not necessarily have a (semi)-classical limit and we are mostly interested in the Hamiltonians $H$ of the systems  explicitly represented as hermitian matrices, i.e., $N \times N$ arrays of complex numbers, where we will eventually consider the $N\rightarrow \infty$ limit. In the absence of any alternative formulations of these models—such as representations in terms of coordinates and momenta—or a lucky insight, perhaps regarding a Bethe Ansatz solution, the question arises: what constitutes an appropriate criterion for determining whether the system under consideration is quantum integrable? In the present work, we adopt as a starting point the criterion similar to the one proposed by Caux and Mossel (CM) \cite{Caux}. Namely, we assume that our Hamiltonian $H_{N}$ is a member of a sequence of Hamiltonians of increasing dimensions $(H_{N_1}, H_{N_2}, \ldots , H_{N_k}\ldots)$, with $N_1 < N_2 < \ldots N_k$, for which it is possible to define a sequence of sets of operators $(\{Q_a^{(N_1)}\}, \{Q_a^{(N_2)}\}, \ldots , \{Q_a^{(N_k)}\})$ such that 
\begin{enumerate}
\item All operators $\{Q_a^{(N_j)}$\} commute with each other and with their Hamiltonian $H_{N_j}$;
\item the operators in $\{Q_a^{(N_j)}\}$ are algebraically independent; 
\item being all these charges $\{Q_a^{(N_j)}\}$ on the same footing (i.e., any of them can play the role of quantum Hamiltonian), they must share 
the same statistical features for any $N$ and $j$, as, for instance, the statistical distribution of their unfolded spectra;
\item the cardinality ${\mathcal C}^{(N_j)}$ of the set $\{Q_a^{(N_j)}\}$ becomes unbounded in the infinite
size limit $N_j \rightarrow \infty$.
\end{enumerate}
An open question regards, of course, the number $n$ of conserved charges $\{Q_a^{(N_j)}\}$ in any given $N_j$-dimensional Hilbert space: is, for instance, the dimension $N_j$ of the Hilbert space itself? Or, if the Hilbert space is constructed as the tensor product of $M_j$ $q$-dimensional linear spaces, i.e., $N_j = q^{M_j}$, is instead the number $M_j$ of these vector spaces? In the next Section we will see that this last option seems to be the one selected by our statistical argument. At any rate, the number $n$ of conserved charges must be {\em maximal}, meaning with that we should be able to differentiate the set of Hamiltonian eigenvectors in terms of the quantum numbers of the conserved charges.

Concerning algebraic independence, it is easy to see that if any Hamiltonian $H$ in the sequence of Hamiltonians $(H_{N_1}, H_{N_2}, \ldots , H_{N_k}\ldots)$ has a non-degenerate set of eigenvalues $E_i $ ($ i =1, 2, \ldots N$), then any operator $Q$ which commutes with $H$ is functionally dependent on it. Indeed, if $[Q, H] = 0$, then $Q$ also commutes with all powers of $H$, i.e., $[Q, H^k] = 0$ ($k = 1, 2, \ldots N$). Hence $Q$ can be expressed as $Q = {\mathcal P}(H)$, where ${\mathcal P}$ is a polynomial of order $(N-1)$. To see this, the operator $H$ admits the spectral decomposition 
\be
H \,=\,\sum_{i=1}^N E_i \, \mid E_i \rangle \, \langle E_i \mid \,=\, 
\sum_{i=1}^N E_i\, P_i \,\,,
\ee
where the projectors $P_i$ on the one-dimensional eigenstate $\mid E_i \rangle $ can be written as 
\be 
P_i \,=\, \prod_{j\neq i} \frac{H - E_j}{E_i - E_j}\,\,.
\label{projectors}
\ee 
Since $Q$ commutes with $H$ and its eigenstates $|\: E_i \rangle$ are not degenerate, they are also eigenstates of $Q$ with eigenvalues $q_i$
\be 
Q \mid E_i \rangle \,=\, q_i \, \mid E_i \rangle \,\,.
\ee
Thus $Q$ is diagonal in the basis of $H$ and therefore it can be expressed in terms of $H$ and its powers $H^k$ as 
\be 
Q \,=\, \sum_{i=1}^N q_i \, P_i \,=\, \sum_{i=1}^N q_i \, \prod_{j\neq i} \frac{H - E_j}{E_i - E_j}\,\,.
\label{Qproduct}
\ee
Equivalently, since all eigenvalues $E_i$ are non degenerate, the Vandermonde matrix made of the eigenvalues of ${\bf I}$ and $H^k$ ($k=1,\cdots,N-1$) 
 \be 
{V}_N= 
\left(
\begin{array}{ccccc}
1 & E_1 & E_1^2 &  ...  & E_1^{N-1} \\
1 & E_2 & E_2^2 & ... & E_2^{N-1} \\
1 & E_3 & E_3^2 & ... & E_3^{N-1} \\
. & . & . & . &  . \\   
. & . & . & \,\,. &  . \\   
. & . & . & \,\,\,. & . \\   
1 & E_N & E_N^2 & ... & E_N^{N-1} \\   
\end{array}
\right)
\ee    
has the non-zero determinant $||V_N|| = \prod_{i < j} (E_i - E_j)$. Hence ${\bf 1}, H, H^2,\ldots H^{N-1}$ is a complete basis for the diagonal matrices and therefore any operator $Q$ simultaneously diagonalizable with $H$ can be written as a linear combination of the identity operator $I$ and higher powers of $H$, i.e. $Q$ is functionally dependent on $H$, as stated by a well-known result by von Neumann \cite{vonNeumann} 
\be
[Q,H]\,=\,0 
\,\,\,\,\,\,
\Longrightarrow     
Q\,=\, \sum_{k=0}^{N-1} p_k H^k \,=\, {\mathcal P}(H)\,\,\,,
\,\,\,
\ee
where the coefficients $p_k$ can be explicitly computed by comparing this expression with the one in Eq.~\eqref{Qproduct}. 

The argument does not obviously apply if the spectrum of $H$ is instead generally degenerate. In the coming Section, we see in detail how, in this case, one can show that there exists a maximal set of functionally independent operators $Q_a$ which commute with $H$ and themselves and whose number scales as $\log N$.

\section{\label{S2-quant-integrable} A Class of Quantum Integrable Hamiltonians\protect\footnote{The argument presented in this Section is originally due to Giuseppe Mussardo and Andrea De Luca, unpublished.}}\label{classQIM}
Consider a quantum Hamiltonian associated with a hermitian operator ${H}$. Once its real eigenvalues are sorted in an increasing order, one can study the statistics of the level spacings, i.e. the distribution of the variables 
\be
\hat s_i \,=\, E_{i+1} - E_i \,\,\,.
\ee 
\begin{figure}[t]
\begin{center}
\includegraphics[width=0.6\textwidth]{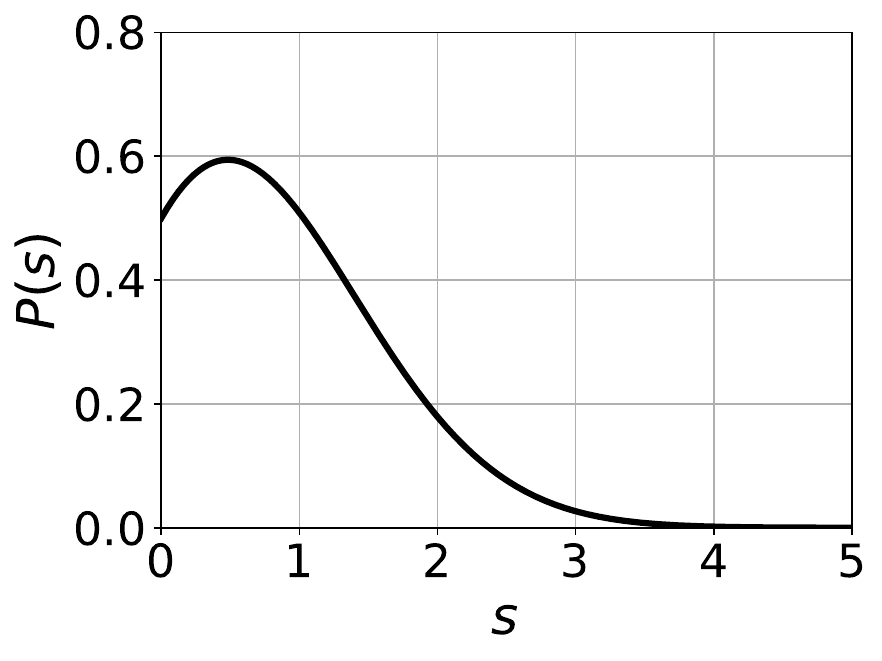}
\end{center}
\caption{An example of level spacing distribution $P(s)$ with a non-zero probability to have zero gaps in the spectrum.} 
\label{nonzeroprobability}
\end{figure}
It is customary to rescale such spacings\footnote{We will be more precise later on this procedure of rescaling the gaps, see Section~\ref{probdistr} and \ref{examples}.} by their mean $\langle \hat s_i \rangle $, i.e. $\hat s_i \rightarrow s_i = \hat s_i/\langle \hat s_i \rangle$, so that, considering the normalized spacings $s_i$, one can reasonably study and compare the spectra of Hamiltonians of different origins, extracting their universal features. In the Section \ref{probdistr}, we shall examine the various probability distributions $P(s)$ describing the fluctuations of the normalized level spacings, distinguishing the cases in which the system is integrable, chaotic, or exhibits a mixed nature. 

\subsection{Playing with block-form matrices}
\begin{figure}[t]
\centering
\includegraphics[width=0.5\textwidth]{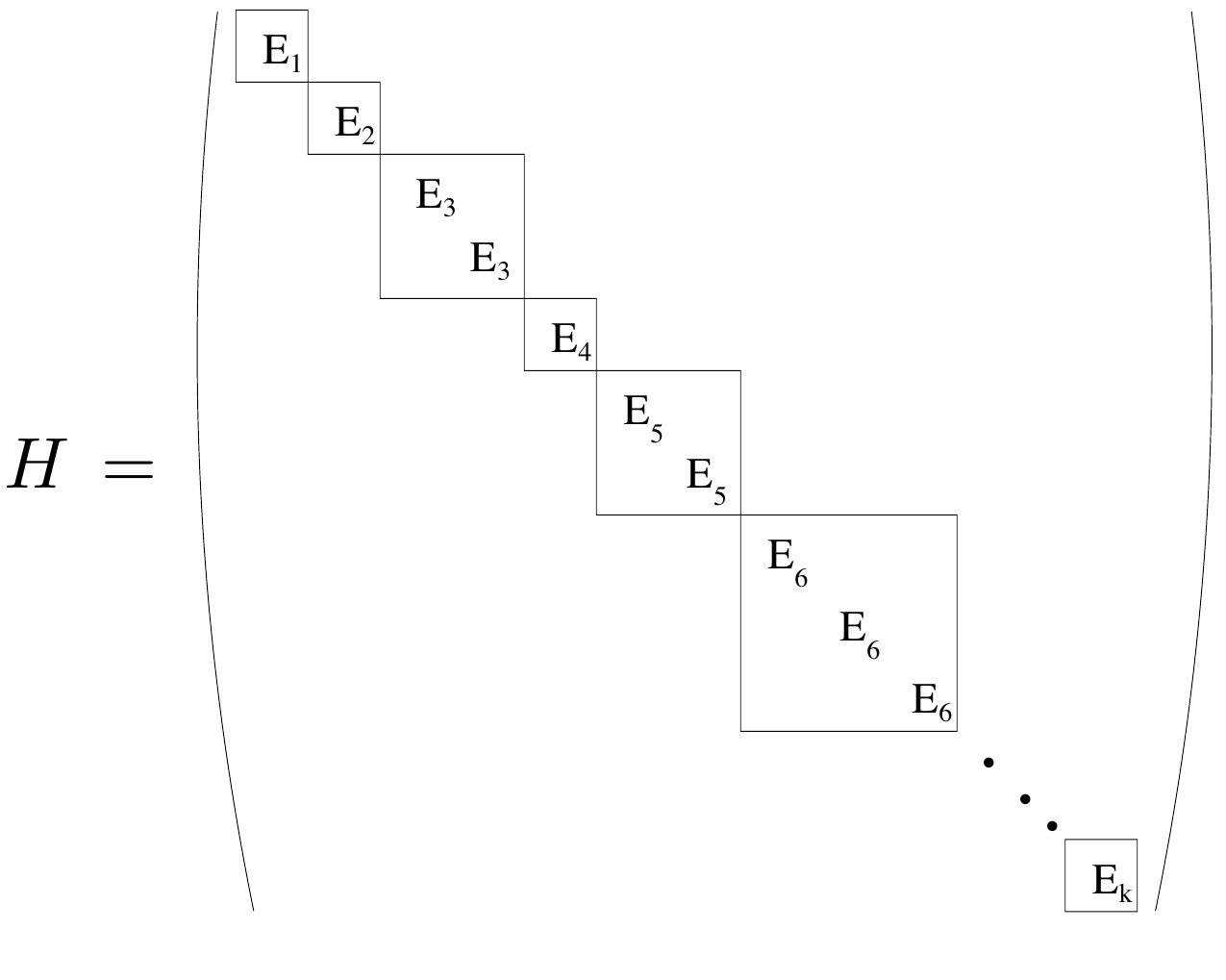}
\caption{Block form of an Hamiltonian with a degenerate spectrum according to the probability distribution $P(s)$ of its spacing levels and an energy resolution $\delta E$.}
\label{bbb}
\end{figure}

To establish a probabilistic argument ensuring the existence of a set of operators that commute with $H$ but are not functionally dependent on it, it suffices, for the present purpose, to assume the existence of a nonzero probability $\hat p_0$  of having vanishing spacings of $H$, i.e., $P(s = 0) \,= \,\hat p_0 \,\neq \, 0$, as exemplified by the distribution shown in Figure 1. 
As we argue in more detail in the next Section, the symmetries present in integrable models may induce degeneracies in the spectrum and, therefore, a non-zero probability of vanishing gaps. Let's see how we can define a maximal set of conserved charges which commute with $H$ and themselves. To this aim, we regard $H$ as part of an ensemble of Hamiltonians, with their level spacing described by the same probability distribution $P(s)$. Hence, we can generate a string of numbers $s_i$ according to the probability distribution $P(s)$. Once we choose an arbitrary value $E_1$ as a starting point, we can proceed by defining all successive energy levels $E_i$ by the recursive relation 
\be 
E_{i+1}\,=\,E_{i} + s_i 
\,\,\,\,\,\,\,\,\,\, 
, 
\,\,\,\,\,\,\,\,\,\, 
E_2 \,=\,E_1 +s_1 
\,\,\,.
\label{sequencepoisson}
\ee
We will collect the sequence of all gaps $s_i$ in a string ${\mathcal S}$ of real numbers. Imagining that we have an energy resolution $\delta E$ (so that energy levels which differ by $\delta E$ can be considered as effectively degenerate), the string ${\mathcal S}$ of the gaps takes the following form   
\be
{\mathcal S}\, = \, (x_1,\,x_2, \,0,\,x_3,\,0,\,x_4,\,0,\,x_5,\,0,\,0,\,x_6,\,x_7,\,\cdots) 
\label{poissond}
\ee
and contains a certain number of 0 (with probability $p_0 = P(\delta E)$), together with a sequence of real positive numbers $x > 0$ (with total probability $p = 1- p_0$). 
The presence of the $0$'s in the sequence ${\mathcal S}$ is the one which induces degeneracies in the energy levels $E_i$ defined by the recursive relation \eqref{sequencepoisson}: the corresponding diagonal matrix form of the Hamiltonian will be made of diagonal blocks of equal eigenvalues as in Figure~\ref{bbb}.  

In an $N \times N$ truncated representation of the Hamiltonian, it is quite simple to compute how many blocks $B_S$ of size $S$ are present, on average, in the matrix $H_N$ at leading order in $N$. We have to use the laws of probability: 
\begin{enumerate}
\item $B_1$ is the expected number of $1 \times 1$ blocks, alias the number of non-degenerate eigenvalues of $H_N$. Since there is a non-degenerate eigenvalue each time the two consecutive values of $s_i$ are both different from $0$, and each non-zero value $x_i$ is independently generated with probability $p$,  we have 
$B_1 \simeq N p^2$. 
\item $B_2$ is the expected number of $2 \times 2$ blocks in $H_N$. We have two degenerate eigenvalues each time we encounter the string $(\cdots,\,x_1,\,0,\,x_2,\,\cdots)$ in the sequence of the $s_i$'s, with $x_1 \neq 0$ and $x_2 \neq 0$. Hence, employing the independence of the corresponding probabilities, we have $B_2 \simeq N p^2 p_0$. 
\item Generalizing the previous argument, to obtain the expected  number $B_k$ of $k \times k$ blocks in $H_N$, one has to consider the probability of getting a string made of $(k-1)$ $0$'s, sandwiched between two non-zero numbers $x_l$ and $x_{l+1}$, i.e.
\begin{equation*}
    (\cdots,x_l,\,\overbrace{0,0,0,\cdots,0,0,0}^{k-times},x_{l+1},\cdots)\,\,.
\end{equation*}
Hence, $B_k \simeq N p^2 p_0^{k-1}$. 
\end{enumerate}
The above considerations can be refined by taking into account the edge effects of the blocks. One can write down recursive equations for their probabilistic values and the final exact expressions for the expected number of the various blocks are given by 
\begin{eqnarray}
&B_1 \,\,\,= &2p + (N-2)p^2;\nonumber \\
&B_2 \,\,\,= &2(1-p)p + (N-3)p^2(1-p);\cr\cr
&B_k \,\,\,= &2(1-p)^{k-1}p + (N-k-1)p^2(1-p)^{k-1}; \label{theoreticalblock}\\
& .. &  ...\cr\cr 
&B_{N-1} =& 2(1-p)^{N-2}p;\cr\cr
&\,\,\,B_{N}\, \,\,=& (1-p)^{N-1}.\nonumber
\end{eqnarray}

It is obvious that, probabilistically speaking, an $N\times N$ matrix $H_N$ cannot have  blocks of arbitrary size: in fact, there will be, on average, no block of size $B_l$ when $(N-l-1) p^2 p_0^{l-1} < 1$, a fact which is reasonably confirmed by the simulations (see Table \ref{tableblocks}). 
\begin{table}[t]
\begin{tabular}[b]{|| l | l | l | l | l | l | l ||} \hline 
 & $B_1^{ex} = 24290$ & $B_2^{ex} = 2469 $ & $B_3^{ex} = 220 $ & $B_4^{ex} = 19 $ 
 & $ B_5^{ex} = 6$ & $B_6^{ex} = 1$ \\ 
 $N=30.000$ & & & & & &  \\ 
& $B_1^{th} = 24300$ & $B_2^{th} = 2429 $ & $B_3^{th} = 242 $ & $B_4^{th} = 24 $ 
 & $ B_5^{th} = 2$ & $B_6^{th} = 0.2$ \\ \hline\hline
 & $B_1^{ex} = 40417$ & $B_2^{ex} = 4071 $ & $B_3^{ex} = 421 $ & $B_4^{ex} = 38 $ 
 & $ B_5^{ex} = 4$ & $B_6^{ex} = 1$ \\ 
 $N=50.000$ & & & & & &  \\ 
& $B_1^{th} = 40500$ & $B_2^{th} = 4050 $ & $B_3^{th} = 405 $ & $B_4^{th} = 40 $ 
 & $ B_5^{th} = 4$ & $B_6^{th} = 0.4$ \\ \hline\hline
 & $B_1^{ex} = 80798$ & $B_2^{ex} = 8102 $ & $B_3^{ex} = 866 $ & $B_4^{ex} = 87 $ 
 & $ B_5^{ex} = 8$ & $B_6^{ex} = 2$ \\ 
 $N=100.000$ & & & & & &  \\ 
& $B_1^{th} = 81000$ & $B_2^{th} = 8100 $ & $B_3^{th} = 810 $ & $B_4^{th} = 81 $ 
 & $ B_5^{ex} = 8$ & $B_6^{ex} = 0.8$ \\ \hline\hline
\end{tabular} 
\caption{Comparison of the number of blocks $B^{ex}_k$ obtained by a randomly generated $N$ gaps with $p_0 = 0.1$ and numbers $B_k^{th}$ given by the theoretical formula \eqref{theoreticalblock}.}
\label{tableblocks}
\end{table}

Once the Hamiltonian is degenerate, we are obviously unable to express an arbitrary operator $Q_N$ which commutes with $H_N$ in terms only of $H_N$ and its powers $H_N^k$ ($k=1,2,\cdots N-N_B$, where $N_B$ are the numbers of blocks): the only operators which are expressible as $Q_N = \sum_{k=0}^{N-N_B} \alpha_k H_N^k$ are those which have precisely the same block form of $H_N$. In order to find a maximal set of independent commuting operators $Q_k$ ($k=1,2,\ldots,n$)  where it holds
\be
[Q_i, Q_k] =0 \,\,\,,
\ee
and in which $Q_1$ identifies the original $H$, we can argue as follows. In integrable models, all conserved charges are essentially on an equal footing — i.e. each of them can, in principle, serve as the Hamiltonian. It follows that the level spacings of every conserved charge must exhibit the same level spacing statistics $P(s)$ as those of the original Hamiltonian; otherwise, the symmetry underlying their equivalence would be violated. Indeed, if one such charge, say $Q_r$, were to display, for instance, a spacing distribution with zero probability of vanishing gaps, then, upon adopting  $Q_r$ as the Hamiltonian and invoking the result of the previous section, we would be forced to conclude that all other conserved charges $Q_i$ are functions of $Q_r$ , and hence not independent—a contradiction.

Once it is established that all $Q_i$ share the same level‑spacing statistics, determining their total number $n$ and constructing them explicitly becomes straightforward. The essential role of each $Q_i$ is to lift the degeneracies of the others while preserving their mutual symmetry: since they possess the same number of degenerate‑eigenvalue blocks, it is the relative arrangement of these blocks that must differ. This requirement leads to a simple condition on the corresponding sequences ${\mathcal S}_1, {\mathcal S}_2, \ldots {\mathcal S}_n$ of level spacings associated with the $n$ conserved charges $Q_i$ generated according to the probability distribution $P(s)$ with cut‑off $\delta E$
\be
\begin{array}{lll} 
Q_1 & \Longrightarrow & {\mathcal S}_1 = (
\begin{array}{llllllllll}  x_1, & x_2, & 0, & x_3, & 0, & x_4, & 0, & x_5, & x_6, &\cdots
\end{array}) \\
Q_2 & \Longrightarrow & {\mathcal S}_2 = ( 
\begin{array}{llllllllll} y_1, & 0, & \,y_2, & y_3, & y_4, & 0, & \,y_5, & y_6, & 0, & \,\cdots
\end{array} ) \\
Q_3 & \Longrightarrow & {\mathcal S}_3 = ( 
\begin{array}{llllllllll}
0, & \,z_1, & z_2, & 0, & \,z_3, & \,z_4, & 0, & \,z_5, & z_6, & \cdots
\end{array} )\\
\cdots & \cdots & \cdots 
\end{array}
\ee
Once we organize these sequences in a matrix, putting one on top of the other,
\be 
V =
\left(\begin{array}{cccccccccc}
x_1  & x_2  & 0  & x_3  & 0  & x_4  & 0  & x_5  & x_6  &\cdots\\
y_1  & 0  & y_2  & y_3  & y_4  & 0  & y_5  & y_6  & 0  & \cdots\\
0  & z_1  & z_2  & 0  & z_3  & z_4  & 0  & z_5  & z_6  & \cdots \\
\cdots & \cdots & \cdots & \cdots & \cdots & \cdots & \cdots & \cdots & \cdots & 
\end{array}
\right)
\ee
it is easy to see that they will all remain degenerate if there will be an entire column made of 
$0$'s. The probability that this event will not happen is given by 
\be 
\hat P = (1-p_0^n)^N \,\,\,.
\ee  
This is the probability that in a column, not all elements are $0$, multiplied for the $N$ columns. If we want to be sure of this event, we can impose that 
\be 
\hat P > 1- \epsilon \,
\ee 
with $\epsilon \rightarrow 0$, and in this way we find that the number of independent conserved charges $Q_i$ needed to resolve all relative degeneracies scales only as 
$\log N$ 
\be 
n \simeq \frac{1}{\log p_0} (\log\epsilon - \log N)  
\ee 
($\log p_0 < 0$). So, for instance, taking $p_0 = 10^{-3}$, $\epsilon=10^{-6}$ and $N = 10^9$, one would need only 4 extra conserved charges in addition to the original Hamiltonian. These extra charges  completely resolve the degeneracy of $Q_1$ and simultaneously their own degeneracies. From the above construction, it is, moreover, obvious that all these operators are linearly and functionally independent of each other. Computationally speaking, it is very easy to find $n$ of such operators: once a sequence of random numbers is generated with the probability distribution $P(s)$, it is sufficient to generate a few of its permutations and to check that the corresponding operators associated with the new sequences do not have an overlapping column of all zeros.

The logarithmic dependence of the number of independent conserved charges is a welcome feature of the above construction: in fact, as anticipated, associating the above Hamiltonian with a quantum system of size $L$, with $q$ states per site, the system will have a total number of states equal to $N = q^L$. Therefore, the number of conserved charges found with the above construction scales only with $L$ and {\em not} with the dimension of the Hilbert space. Let us, however, mention that  a drawback of this stochastic construction is that it is quite difficult to disentangle the local nature of both the Hamiltonian and the associated conserved charges. 

\subsection{Locality of the Hamiltonians}
Since in the subsequent sections of this paper we will consider both local and non‑local Hamiltonians, it is worthwhile to comment on this important feature of quantum models and its implications for the structure of their matrix representation. By local models, we mean those that are local either in the spatial coordinate $x$ (for continuous systems) or in the lattice site $i$ (for discrete systems). In a generic basis of the Hilbert space, the matrix representation of a local Hamiltonian is typically dense, meaning that almost all of its entries are nonzero (as will be illustrated by an explicit example below). However, for any local theory, there exists a particular basis—referred to here as the local basis—in which the Hamiltonian is represented by a sparse matrix, i.e., one in which the vast majority of entries vanish. This sparsity is the defining structural feature of local Hamiltonians.

To clarify the concept of the local basis, let's consider the paradigmatic example of the one-dimensional quantum Ising model. Its quantum Hamiltonian for a lattice of $L$ sites is given by 
\begin{equation}
\label{Ising}
 H \equiv \sum_{i=1}^L \left(\sigma_z^i \sigma_z^{i+1} + h \sigma_x^i\right) = \sum_{i=1}^L
H_i\,\,,
\end{equation}
where $\sigma_a$ are the usual Pauli matrices. The last equality makes evident the local nature of this model: the Hamiltonian
has been written as a sum of operators involving only two lattice sites. The matrix representation of this Hamiltonian can be obtained 
once we fix a basis, with a typical choice being the set of common eigenstates of the $\sigma_z^i$ operators. They can be written as:
\be 
\ket{\uparrow \uparrow \ldots \uparrow}, \quad \ket{\uparrow \uparrow \ldots
\downarrow},  \ldots \quad \Rightarrow \quad \ket{m_1 m_2 \ldots m_L }\,\,,
\label{vectorIsing}
\ee
where each $m_i \in \{\uparrow,\downarrow\}$ corresponds to the two possible eigenstates of $\sigma_z^i$:
$$ \sigma_z^i \ket\uparrow = \ket\uparrow, \quad \sigma_z^i
\ket\downarrow = - \ket\downarrow \quad\Rightarrow \quad \sigma_z^i
\ket{m_i} = m_i \ket{m_i}\,\,\,. $$
Therefore, the Hilbert space is made up of $N = 2^L $ elements, with a possible basis given by the one above. We call this the real-space basis, characterized by the fact that its elements are simply tensor products of the states of each site. In this basis, the matrix elements of the Hamiltonian density $ H_i $ are given by
\be
 \mathcal{H}_{s,s'} = \bra{m_1 \ldots m_N} H_i \ket {m_1' \ldots m_N'} =
(m_i m_{i+1}\delta_{m_i, m_i'} + h\delta_{m_i, -m_i'})\prod_{k\neq i}
\delta_{m_k, m_k'}\,\,,
\label{nonzeroIsing}
\ee
where $s,s'$ are labelling the full set of indices $m_1,\ldots, m_N$.
From this expression, it is easy to deduce that on each row of the matrix there are 
$L+1$ non-zero entries and, therefore, the total number of non-zero elements of the $N\times N$ matrix $H_N$  is ${\mathcal N} =  (L+1) N$. Since the total number of matrix elements is $N^2$, the density of non-zero elements is given by 
\be 
\rho \,=\,\frac{{\mathcal N} }{N^2} \,=\,\left(\frac{\log N}{\log 2} +1\right)\frac{1}{N} \,\,\,,
\ee
while the density of the zeros of the matrix $H_N$ is 
\be
\rho_0 \,=\,1-\rho\,=\,1- \left(\frac{\log N}{\log 2} +1\right) \frac{1}{N}  \,\,\,. 
\label{densityzeros}
\ee
Therefore, for large values of $N$, the Hamiltonian matrix $H_N$ is a {\em sparse} matrix, i.e. a matrix with a very large number of zeros and very few non-zero entries. 

As a matter of fact, this statement is quite general for any local quantum Hamiltonian. Consider, for instance, the one coming from quantum field theory: the fraction of non-zero elements of the Hamiltonian is $O(e^{-L})$, i.e exponentially small in the thermodynamic limit $L \rightarrow \infty$. 

Take for simplicity a $(1+1)$ bosonic field theory, whose Hamiltonian can be written as 
\be 
H \,=\,\int {\mathcal H}(x) \, dx \,=\,\int  \left[\frac{1}{2} \Pi^2(x,t) + \frac{1}{2} (\partial_x \hat\varphi(x,t))^2
+V(\hat\varphi(x,t)) \right] \,dx \,\,\,,
\label{fieldtheory}
\ee
where $\Pi(x,t) = \frac{\partial\hat\varphi}{\partial t}$ is the canonical conjugate of the operator $\hat\varphi(x,t)$ and they satisfy the equal-time commutation relation 
\be 
[\hat\varphi(x,t),\Pi(y,t)]\,=\,i \delta(x-y) \,\,\,.
\label{commutationrelation}
\ee
The action of $\Pi(x)$ on the basis in which the operator $\hat\varphi(x)$ is diagonal  is given by 
\be
\Pi(x) \rightarrow -i \frac{\delta}{\delta \varphi(x)}\,\,\,.
\label{actionPi}
\ee
For building up the Hilbert space of such a theory, we can choose the local basis given by the coordinate representation: in this representation, at any given time $t$, the states $\mid \varphi \rangle$ are given by the values of the field $\hat\varphi(x,t)$ at the position $x$
\be
\begin{array}{c}
\hat\varphi(x) \mid \varphi \rangle \,=\,\varphi(x) \mid \varphi \rangle \,\,\,,\\
\mid \varphi \rangle \,=\,\prod_x \mid \varphi(x) \rangle \,\,\,,\\
\langle \varphi' \mid \varphi \rangle \,=\,\prod_x \delta(\varphi'(x) - \varphi(x)) \,\,\,,\\
\prod_x \int_{-\infty}^{\infty} d\varphi(x) \mid \varphi \rangle \langle \varphi \mid \,=\,1 \,\,\,.
\end{array}
\label{hilbertQFT}
\ee 
In order to set up a matrix representation of the Hamiltonian \eqref{fieldtheory} with indices which are discrete rather than continuous, it is convenient to discretize both the space (in terms of a lattice with a step $a$) and the values of the field $\varphi(x)$ (in terms of a set of $q$ values spaced by $\epsilon$). This means that $x$ will be restricted to the lattice points 
\be
x \,=\,m a \,\,\,,
\,\,\,\, m=1,2,\ldots L \,\,\,,
\ee
while, at these points, the field $\varphi$ will take the $q$ values  
\be 
\varphi_m \,=\,\left(-\frac{q-1}{2},-\frac{q+1}{2},\ldots, 0,\ldots \frac{q-3}{2},\frac{q-1}{2} \right) \,\epsilon\,\,\,.
\label{discretevaluefield}
\ee
In such a scheme, the local basis is spanned by the $N = q^L$ vectors associated with the values of the field at the $L$ lattice points 
\be
\mid \varphi \rangle \,=\,\mid \varphi_1,\varphi_2,\ldots ,\varphi_L \rangle \,\,\,,
\,\,\,\,\,\,\,\,
\varphi_i \in \left[-\frac{(q-1)}{2},\frac{(q-1)}{2}\right] \,\epsilon\,\,\,,
\label{Hilbertdis}
\ee
while the Hamiltonian \eqref{fieldtheory} becomes 
\begin{eqnarray}
H_N &\,=\, & a \sum_{m=1}^L \left[\frac{1}{2} \Pi^2_m  + \frac{1}{2} \left(\frac{\varphi_{m+1} - \varphi_{m-1}}{2a}\right)^2 + V(\varphi_m)\right] \nonumber \\
& = &  a \sum_{m=1}^L \left[\frac{1}{2} \Pi^2_m + \frac{1}{2} \left(\frac{\varphi^2_{m+1} + \varphi^2_{m-1} - 2 \varphi_{m-1} \varphi_{m+1}}{4 a^2}\right) + V(\varphi_m)\right]\,\,. \nonumber 
\end{eqnarray}
The last two terms in this Hamiltonian act locally on the states \eqref{Hilbertdis} whereas the operator $\Pi^2$ induces hopping between the states. To show this, observe that for any function $f(\{\varphi_i\})$ (where the $\varphi_i$'s are measured in units of $\epsilon$) it holds  
\be
\Pi_m f(\{\varphi_i\})\,=\, - i \frac{\partial f[\{\varphi_i\}]}{\partial \varphi_m} \,=\,-i \,\frac{f[\varphi_m + 1,\{\varphi_k\}] - f[\varphi_m - 
1,\{\varphi_k\}]}{2 \epsilon} \,\,\,,
\label{actionPim}
\ee
where, in the last expression, $\{\varphi_k\}$  denotes all the other values of the field that are different from $\varphi_m$. Iterating this definition, for the action of $\Pi_m^2$ we have 
\be 
\Pi^2_m f[\{\varphi_i\}] \,=\,-\frac{\partial^2 f[\{\varphi_i\}]}{\partial\varphi_m^2} \,=\,-
\frac{f[\varphi_m + 2,\{\varphi_k\}] + f[\varphi_m -2,\{\varphi_k\}] - 2 f[\varphi_m,\{\varphi_k\}]}{4 \epsilon^2}
\,\,\,.
\ee
Therefore, for the matrix elements of this operator on the local basis, we have 
\begin{eqnarray}
&& \langle \varphi_L' ,\ldots \varphi_m',\ldots \varphi_1' \mid 
\Pi^2_m \mid \varphi_1,\ldots,\varphi_m,\ldots , \varphi_L \rangle \,\,\nonumber \\
&& =C\,\delta_{\varphi_1,\varphi_1'} \,\ldots \delta_{\varphi_{m-1},\varphi_{m-1}'} 
 \delta_{\varphi_{m+1},\varphi_{m+1}'} \ldots \delta_{\varphi_{L},\varphi_{L}'} \,
\left[\delta_{\varphi_m+2,\varphi_m'} +  
\delta_{\varphi_m-2,\varphi_m'} - 2 \delta_{\varphi_m,\varphi_m'} \right] \nonumber
\end{eqnarray}
where $C$ is a normalization factor. Hence, this operator induces a hopping term among the states \eqref{Hilbertdis}. In conclusion, the Hamiltonian of any scalar field theory can be considered as a suitable generalization of the Hamiltonian of the Ising model, where $\Pi^2(i)$ plays the role of $\sigma^x_i$ and induces a spin-flip among the field values.

Let's now count the number of non-zero terms present in the $N \times N$ Hamiltonian $H_N$ of the bosonic quantum field theory: in each row, there are $3 L$ non-zero terms ($L$ of them are the diagonal terms, where all values of $\varphi$ are the same for the bra and ket states, while $2 L$ are those where the bra and ket states differ from each other for the value of the field $\varphi_m$ by $\pm 2$). Since there are $N$ rows, the total number of non-zero values of such a matrix closely follows the previous computation of the Ising case and it is equal to ${\mathcal N} = 3 L N$, i.e. the density of non-zero values is given by 
\be
\rho\,=\, \frac{{\mathcal N}}{N^2} = \frac{3}{\log q} \frac{1}{N} \log N \,\,\,.
\ee
while the density of zeros is 
\be
\rho_0 \,=\,1-\rho\,=\,1- \frac{3}{\log q} \frac{1}{N} \log N \,\,\,. 
\label{densityzerosQFT}
\ee
Therefore, for large values of $N$, as in the Ising case, the Hamiltonian matrix $H_N$ has a very large number of zero entries, i.e. it is a {\em sparse} matrix. 
 
Let's finally discuss the nature of the matrices associated with local conserved charges in an integrable quantum field theory, where local conserved charges $Q_s$ are derived from the conservation laws of local densities, as in Eq.\,\eqref{conslaws}, and 
since $Q_i$ can be expressed in this case as 
\be
Q_i \,=\, \int dx \, \rho_i(x,t) \,\,\,.
\ee
The densities $\rho_i$ are, in general, functions of the following operators: the canonical conjugate field $\Pi$, the field $\hat\varphi(x)$ and higher powers and higher space-derivatives of both of them. In particular, there is a maximum power $\Pi^k(x)$ ($k\leq i$) present in $\rho_i$.  While all expressions in $\rho_i$ involving the 
 field $\hat\varphi(x)$ and its derivatives act locally on the basis \eqref{Hilbertdis}, the higher powers $\Pi^k(x)$ of the conjugate field will induce a hopping term of maximum $k$-values in the local basis. Therefore, repeating the analysis done above, one arrives at the conclusion that in the local basis, the matrices of the local conserved charges are also sparse matrices, with a density of non-zero and zero entries given by 
 \be 
 \rho^{(s)} \,=\, \frac{k+1}{N \log q} \, \log N \,\,\,, 
 \,\,\,\,\,
 \rho_0^{(s)} \,=\,1-\rho^{(s)}\,=\,1- \frac{k+1}{N \log q} \, \log N \,\,\,. 
 \label{densityconservedcharges}
 \ee
 We conclude this subsection with a remark on the non‑local character of the projectors onto energy subspaces. As is evident from their explicit form in Eq.~\eqref{projectors}, these operators involve progressively higher powers of $H$ (up to the highest one). Since a generic $k$th power of $H$ is an operator with support extending over $k$ sites, it follows directly that these projectors are intrinsically non‑local.
 
\subsection{The basis of momenta}
The conclusions of the previous section are quite reasonable, but by changing the basis in terms of a unitary matrix $U$, the resulting matrix in the new basis 
of the Hamiltonian \eqref{Ising} can be obtained in terms of a unitary matrix $U$ 
\begin{equation}
 \label{changeofbasis}
\mathcal{H}' \,=\, U^\dag \,\mathcal H \,U
\end{equation}
and will no longer be generally sparse; i.e., for an arbitrary unitary matrix $U$, any notion of locality will be lost. 

However, let's focus our attention on a much more common case: will the Hamiltonian still be  sparse in the basis of momenta? We will show that there is a
subtle issue related to the definition of this change of basis. To be more specific, let us place the Ising Hamiltonian on a circle by introducing periodic
boundary conditions. It means the site $L+1$ is identified with the first one. In this case, if the system is homogeneous, it becomes translation invariant (in units of lattice spacing). 
Namely, there is the one-site shift operator $T$, defined by its action on the basis vectors 
\be 
T\ket{m_1,\ldots, m_L} = \ket{m_L, m_1, \ldots, m_{L-1}} 
\ee
which commutes with the Hamiltonian 
\be [H, T] \,= \,0\,. 
\ee
Clearly, we have:
\begin{equation}
 \label{nilpotent}
T^L = \mathbf{1}
\end{equation}
and therefore, we can define the total momentum $P$ as:
\be 
T \,=\, e^{ i P} 
\ee
with 
\be 
[H,P] \,=\, 0\, .
\ee
Being $T$ a unitary operator and using \eqref{nilpotent}, we easily deduce the
usual structure of the spectrum of the momentum operator $P$, i.e. $\{\frac{2\pi n}{L} , \; n = 0,\ldots, L-1\} $. The basis of momenta can be considered as the basis of the eigenstates of the total
momentum. However, these eigenvalues are degenerate and so many definitions are
possible. Two significant examples will explain the ambiguity. Let's
construct a complete set of eigenstates of $P$ in the following way. Let's pick
up a state $\ket v = \ket{m_1,\ldots m_L}$ in the real-space basis. We obtain
an invariant subspace for $T$ by considering the set of states:
$$ I_v = \operatorname{Span}\{\ket v, T \ket v, T^2 \ket v, \ldots , T^{L_v-1}
\ket v\} $$
where $L_v$ has been defined as:\footnote{The minimum exists because the set contains at least $L$. Moreover, $L_v$ is a divisor of $L$.}
$$L_v \equiv \min\{n \; : \,\,\, \; T^n \ket v = \ket v \}\,\, . $$
The eigenstates of $P$ inside $I_v$ can be written as:
\begin{equation}
 \label{fourier1}
\ket{\tilde v_n} = \sum_{k = 0}^{L_v-1} e^{\frac{2 \pi i n k}{L_v}} T^k\ket v
\quad \Rightarrow \quad P \ket{\tilde v_n} = \frac{2 \pi n}{L_v} \ket{\tilde v_n
}\,\,. 
\end{equation}
A full basis of eigenstates for $P$ can be obtained by repeating this procedure
for different states $\ket v$: we call this basis the \textit{rigid-translation
Fourier basis} (RTFB). It is easy to understand that the resulting
matrix in this new basis is still sparse. In fact, each state is a superposition
of at most $L$ states and it follows the corresponding transformation $U$
contains at most $L$ non-zero entries in each row and column. Using
\eqref{changeofbasis}, we conclude that the Hamiltonian matrix in this basis is
still sparse having at most $L^3$ non-zero entries in each row ($L^3 \ll 2^L$).

Let's now change the point of view and consider what happens in the second-quantization framework, where 
there is a set of operators which satisfy (where the $\pm$ stands for the fermionic and bosonic case)
\be 
[a_i, a^\dag_j ]_{\pm} = \delta_{ij} \,\,\,.
\ee  
They create and destroy a (free) excitation at position $i$. The real-space basis can be written in this formalism as
\begin{equation}
 \label{secondquantizationrealspace}
 (a^\dag_{1})^{n_1} (a^\dag_{2})^{n_2}
\ldots \ket{\Omega} 
\end{equation}
where $\ket{\Omega}$ is the reference vacuum state of the theory. 
The same formalism can be adopted in the Ising case by setting\footnote{This
transformation can be made more rigorous using the
Jordan-Wigner transformation.} $a^\dag_i \simeq
S^+_i $ and taking $\ket\Omega \equiv
\ket{\downarrow\ldots\downarrow}$. Here we can define an excitation with
defined momentum, by setting 
\be 
\aaaa^\dag_k = \sum_{j = 0}^{L-1} e^{i k j} a_j^\dag \,\,\,.
\ee

Inserting this expression in \eqref{secondquantizationrealspace}, 
we get a new basis of eigenstates of
the total momentum $P$
\begin{equation}
 \label{secondquantizationfourier}
 P \aaaa^\dag_{k_1}\aaaa^\dag_{k_2}
\ldots \ket{\Omega} = \left(\sum_i k_i\right)
\aaaa^\dag_{k_1}\aaaa^\dag_{k_2}
\ldots \ket{\Omega}\,.
\end{equation}
We call this the \textit{single-particle Fourier basis} (SPFB).
However, once we restrict ourselves to a subspace where $P$ is defined (appearing as a
block for the Hamiltonian matrix), there is a strong difference between this
case and the one defined in \eqref{secondquantizationrealspace}: in fact, here
not only is the full state, but even each excitation will have a defined momentum. To
better understand this difference, let's consider a two particle case (it will
be clear that there is no difference for the one particle case). Let's consider
a state $\ket v = S^\dag_{x_1} S^\dag_{x_2} \ket{\Omega} = \ket{x_1,x_2}$, which
is the state with only two up spins in positions $x_1,x_2$. In the RTFB, this
state will appear in $L$ states $\ket{\tilde v_n}$, with $n = 0,\ldots, L-1$,
obtained as superpositions of the rigid translations $T^k \ket{v} $, where the
distance $|x_2 - x_1|$ always remains the same. Instead, in the SPFB, each
two-particle state has a non-zero matrix element with $\ket{x_1,x_2}$:
$$ \bra{x_1,x_2}\aaaa_{k_1}^\dag \aaaa_{k_2}^\dag
\ket\Omega = \sum_{j_1,j_2} e^{i (k_1 j_1 + k_2 j_2)} \bra{x_1,x_2}
j_1,j_2\rangle = e^{i (k_1 x_1 + k_1 x_2)} + e^{i (k_1 x_2 + k_1 x_1)}\,\,\,. $$
Increasing the number of particles up to $M$, in the RTFB case, we always have one
summation with $\simeq L$ terms. Instead, in the SPFB, we will have $M$
summations, corresponding to $\simeq L^M$ terms.

The conclusions we can draw from these considerations are as follows:
\begin{itemize}
 \item A local Hamiltonian will appear as a sparse matrix in the real-space
basis.
 \item If we consider the Fourier basis obtained as a superposition of the rigid
translation of the real-space basis, the Hamiltonian will appear again as a
sparse matrix, albeit with a larger density of non-zero entries. This is true,
just because the change of basis we are considering is sparse.
 \item If we consider the Fourier basis obtained by taking the Fourier transform
of the free single particles, the change of basis in each block of defined
total momentum will not be sparse at all. So, in the general case, the
Hamiltonian matrix will be characterized by dense blocks of fixed total
momentum. So, except for this trivial symmetry, it will not be sparse at all.
\end{itemize}

\section{\label{S4-symmetries} Symmetries}\label{s_symmetries}
In the statistical analysis of quantum Hamiltonians, symmetries play a central role and must be properly accounted for to distinguish between integrable and non-integrable systems. It is essential to differentiate between what we term {\em global symmetries} and {\em dynamical symmetries}. 

By {\em global symmetries} we mean explicit transformations under which the Hamiltonian is invariant. These symmetries are often non-Abelian, may induce spectral degeneracies, but their consequences can be fully analysed within the framework of familiar group theory \cite{Hamermesh1962,Zee2016}. In the following, we shall also regard translational invariance—leading to the conservation of the momentum operator $P$ as part of the global symmetries (its group on a one-dimensional lattice of $L$ sites being $\mathbb{Z}_L$). 
In summary, by global symmetry, we mean any property that is manifest and cannot escape our direct observation.

{\em Dynamical symmetries}, by contrast, are typically Abelian and associated with hidden conserved quantities, which are characteristic of integrable models. The corresponding operators are typically extracted by the expansion of the transfer matrix with respect to the spectral parameter (see, for instance, \cite{Faddev,Novikov,DeLucaMussardo,Korepin}). 
In general, the explicit forms of these charges quickly become intricate: in continuous systems, they are expressed in terms of the dynamical variables and their derivatives with respect to space and time, whereas in lattice models, they are written in terms of multi-site operators. Several examples will serve to clarify the nature of these two classes of symmetries and their role in organising the structure of the Hilbert space.

\begin{itemize}

\item {\bf ${\mathbb Z}_2$ Field Theories}. 
Consider a real scalar bosonic field $\varphi(x,t)$ in $(1+1)$ dimensions, subjected to two different dynamics associated to the Hamiltonian densities, 
\begin{eqnarray}
&& {\mathcal H}_{LG} \,=\,\frac{1}{2} \left[(\pi)^2 + (\partial_x\varphi)^2 + m^2 \varphi^2 \right] + \frac{g}{4!} \varphi^4\, , \\
&&
\label{sh-G}
{\mathcal H}_{ShG} \,=\,\frac{1}{2} \left[(\pi)^2 + (\partial_x\varphi)^2\right] + \frac{m^2}{2\gamma} \left(\cosh(\gamma \varphi) -1\right)
\end{eqnarray} 
the first refers to the Landau-Ginzburg theory while the second to the Sinh-Gordon theory\footnote{The various quantities presented in these expression must be regarded as defined by a proper normal order.}. 
Both systems are invariant under the ${\mathbb Z}_2$ symmetry $\varphi \rightarrow - \varphi$ which therefore splits their Hilbert space into even and odd sectors, so that the matrix of their Hamiltonian takes the block form 
\be
H\,=\, 
\left[
\begin{array}{c|c}
\colorbox{blue}{\rule{0pt}{35pt}\rule{35pt}{0pt}} &
\colorbox{white}{\rule{0pt}{35pt}\rule{35pt}{0pt}} \\
\hline
\colorbox{white}{\rule{0pt}{35pt}\rule{35pt}{0pt}} &
\colorbox{red}{\rule{0pt}{35pt}\rule{35pt}{0pt}} 
\end{array}
\right]\,\,\, .
\ee

This ${\mathbb Z}_2$ symmetry is of course the global symmetry of both models. 

However, the Hamiltonian of the Sinh-Gordon theory possesses an infinite set of hidden, conserved local charges $Q_a$ (explicitly given in Appendix A), which are absent in the Landau-Ginzburg theory. These charges define a dynamical symmetry unique to the Sinh-Gordon model and not present in the Landau-Ginzburg case. As a consequence, while the Hilbert space of the Landau-Ginzburg theory decomposes only into two sectors—corresponding to even and odd parity—the Hilbert space of the Sinh-Gordon model splits into an infinite number of subspaces, each characterized by the quantum numbers associated with these conserved charges. Ultimately, this implies that the Hamiltonian of the Sinh-Gordon model has a block-diagonal structure composed of infinitely many blocks of varying dimension, both in the even and in the odd sector of the global ${\mathcal Z}_2$ symmetry 
\be
H_{ShG}\,=\, 
\begin{bmatrix}
H_0 & 0 & 0 & \cdots \\
0 & H_1 & 0 & \cdots \\
0 & 0 & H_2 & \cdots \\
\vdots & \vdots & \vdots & \ddots
\end{bmatrix}
\quad \text{with } H_n \in \mathbb{C}^{d_n \times d_n}
\,\,\, .
\ee
Moreover, in this case the matrix elements of the Hamiltonian density and other local operators can be computed exactly \cite{FringMussardo,KoubekMussardo}.  
\item {\bf ${\mathbb Z}_3$ Field Theory}. 
A less trivial example is provided by a vector of two scalar fields, defined in $(1+1)$ dimensions,  $\phi(x,t) = (\phi_1(x,t), \phi_2(x,t))$, which can also be combined into the complex field $\Phi = \phi_1 + i \phi_2$, subjected to two different dynamics associated to the Hamiltonian densities\footnote{In these expressions $\pi_i = \partial_t \phi_i$, while 
$\Pi = \partial_t \Phi$.} 
\be
{\mathcal H}_{LG} \,=\,\frac{1}{2} \left[(\Pi \, \Pi^{\dagger}) + (\partial_x\Phi \, \partial_x \Phi^{\dagger}) + m^2 \Phi \,\Phi^{\dagger} \right] + 
\frac{g_1}{3!} (\Phi^3 + \Phi^{\dagger 3}) + \frac{g_2}{4!} (\Phi \Phi^{\dagger})^2 \,,
\ee
\be
\hspace{-5cm}
{\mathcal H}_{Toda} \,=\,\frac{1}{2}\sum_{i=1}^2  \left[(\pi_i)^2 + (\partial_x\phi_i)^2\right] + \frac{m^2}{\beta} \sum_{k=0}^2 e^{\beta\,\alpha_k \cdot \phi} 
\ee
the first refers to a complex Landau-Ginzburg theory while the second to an Affine Toda Field Theory associated to the root system $A_2$ where, in the last expression, $\alpha_1 = (1,-1/\sqrt{3})$ and $\alpha_2 = (0,2/\sqrt{3})$ are the simple roots of the Lie Algebra $A_2$ while $\alpha_0 = -(\alpha_1 + \alpha_2) =
(-1, -1/\sqrt{3})$ is the (minus) maximal root of this Lie algebra. 

Both theories are invariant under a ${\mathbb Z}_3$ symmetry: this is explicitly manifest in $H_{LG}$ of the Landau-Ginzburg theory since it corresponds to the field transformations 
\be
\Phi \rightarrow e^{2 \pi i/3} \, \Phi 
\,\,\,\,\,\,\,\,
,
\,\,\,\,\,\,\,\,
\Phi^{\dagger} \rightarrow e^{-2 \pi i/3}\, \Phi^{\dagger}
\label{z3}
\ee
while, in the Affine Toda Field Theory, the three roots of the $A_2$ algebra are invariant under their ${\mathbb Z}_3$ cyclic rotation 
\be
\alpha_0 \rightarrow \alpha_1 
\,\,\,\,\,\,\,\,
,
\,\,\,\,\,\,\,\,
\alpha_1 \rightarrow \alpha_2 
\,\,\,\,\,\,\,\,
,
\,\,\,\,\,\,\,\,
\alpha_2 \rightarrow \alpha_0\,.
\ee
Hence, the potential 
\be
V(\phi) \,=\, \sum_{k=0}^2  e^{\beta\,\alpha_k \cdot \phi}
\ee
is invariant under these cyclic rotations of the exponentials or equivalently under the transformations of the fields \eqref{z3} given above. This ${\mathbb Z}_3$ symmetry is the global  symmetry of both models and therefore their Hamiltonian splits in three different blocks, corresponding to the sectors with ${\mathbb Z}_3$ charges ${\mathcal C} = (0, e^{2\pi i/3}, e^{- 2\pi i/3})$
\be
H\,=\, \left[
\begin{array}{c|c|c}
\colorbox{gray}{\rule{0pt}{35pt}\rule{35pt}{0pt}} &
\colorbox{white}{\rule{0pt}{35pt}\rule{35pt}{0pt}} &
\colorbox{white}{\rule{0pt}{35pt}\rule{35pt}{0pt}} \\ \hline
\colorbox{white}{\rule{0pt}{35pt}\rule{35pt}{0pt}} &
\colorbox{red}{\rule{0pt}{35pt}\rule{35pt}{0pt}} &
\colorbox{white}{\rule{0pt}{35pt}\rule{35pt}{0pt}} \\ \hline
\colorbox{white}{\rule{0pt}{35pt}\rule{35pt}{0pt}} &
\colorbox{white}{\rule{0pt}{35pt}\rule{35pt}{0pt}} &
\colorbox{green}{\rule{0pt}{35pt}\rule{35pt}{0pt}} 
\end{array}
\right]\,.
\ee
However, the Hamiltonian of the Affine Toda Field theory possesses an infinite set of hidden, conserved local charges $Q_a$ \cite{Toda1,Toda2}, which are instead absent in the Landau-Ginzburg theory. These charges define a dynamical symmetry unique to the Affine Toda Field model and are not present in the Landau-Ginzburg case. As a consequence, while the Hilbert space of the Landau-Ginzburg theory decomposes only into three sectors mentioned above, the Hilbert space of the Affine Toda Field theory splits into an infinite number of subspaces, each characterized by the quantum numbers associated with these conserved charges. Hence, the Hamiltonian of the Affine Toda Field theory model has ultimately a block-diagonal structure composed of infinitely many blocks of varying dimension in the three charge sectors of the global ${\mathcal Z}_3$ symmetry.
\be
H_{Toda}\,=\, 
\begin{bmatrix}
h_0 & 0 & 0 & 0 & \cdots \\
0 & h_1 & 0 & 0 & \cdots \\
0 & 0 & h_2 & 0 & \cdots \\
0 & 0 & 0 & h_3 & \cdots \\
\vdots & \vdots & \vdots & \ddots
\end{bmatrix}
\quad \text{with } h_n \in \mathbb{C}^{d_n \times d_n}\,\,.
\ee
\item {\bf $SU(2)$ Heisenberg lattice model}. 
Let's now consider the one-dimensional spin chain Hamiltonian (Heisenberg model) defined on a lattice of $L$ sites with periodic boundary conditions 
\be
H \,=\, J \,\sum_{k=1}^L \vec{S}_k \cdot \vec{S}_{k+1}\,\,\,,
\label{Heisenberg}
\ee
where the angular momentum operators $\vec{S}_i$ satisfy the $SU(2)$ commutation relations 
\be
[(S_a)_k , (S_b)_l] \,=\, i \, \delta_{k,l} \, \epsilon_{a b c} \, (S_c)_k
\,\,\,\,\,\,\,
,
\,\,\,\,\,\,\,
a, b, c = 1,2,3\,.
\ee
Such a Hamiltonian commutes with the total angular momentum 
\be
{\vec S}_{tot} \,=\, \sum_{k=1}^N \vec{S}_k 
\ee
and therefore the total Hilbert space can be decomposed into the irreducible representations of $SU(2)$ identified by the value $S(S+1)$ of $\vec{S}^2$ and the $(2 S +1)$ states of the irreducible representation labelled by the eigenvalues of $S_z$. Correspondingly the matrix of the Hamiltonian \eqref{Heisenberg} splits in blocks of different size. With periodic boundary conditions, an additional global symmetry of the model is 
the translation by one site, which leads to the conservation of the one-site shift operator ${T}$ which satisfies 
\be
T \, \vec{S}_i \, T^{-1} \,=\, \vec{S}_{i+1} 
\,\,\,\,\,\,\,
,
\,\,\,\,\,\,\,\,
(\vec{S}_{i+L} \,=\, \vec{S}_i)\,\,.
\ee
Expressing $T$ as 
\be
T \,=\, e^{i P}\,\,\,,
\ee
the Hilbert space splits into $L$ eigenspaces of the momentum operator $P$ corresponding to the quantum numbers 
\be
p_n \,=\, \frac{2 \pi \, n}{L}  
\,\,\,\, 
,
\,\,\,\,
n\,=\, 0,1,\ldots L-1
\,\,\,.
\ee
Hence, $SU(2)$ and translation are in general the global symmetries of the Heisenberg model \eqref{Heisenberg}. 

If the operators ${\vec S}_i$ transform according to the spin-$1/2$ irreducible representation of $SU(2)$, the Heisenberg model also admits a dynamical symmetry generated by the infinite number of conserved quantities obtained by the derivatives of the logarithm of the transfer matrix $\tau(\lambda)$ with respect the spectral parameter $\lambda$ \cite{Korepin}
\be
Q_n \,=\, \frac{d^n}{d\lambda^n} \, \log \tau(\lambda)\,\,.
\ee
The presence of this dynamical symmetry (absent for all other values of the spin of the operators $\vec{S}_i$) further splits the Hilbert space and reduces the Hamiltonian to blocks of smaller size, ultimately one-dimensional, corresponding to the states which are solutions of the Bethe Ansatz equations, alias the common eigenvectors of all the conserved charges $Q_n$.   

\item {\bf $S_L$ Sutherland Permutation Hamiltonian}. As a further example of a lattice model, consider a one-dimensional system, to be analysed in greater detail later, consisting of a lattice of $L$ sites, each occupied by a single particle of a distinct colour, and governed by the Hamiltonian
\be
H\,=\, J\, \sum_{k=1}^L P_{k,k+1} \,\,\,,
\label{Sutherland}
\ee
where $P_{k,k+1}$ is the permutation operator of two neighbour sites. The global symmetry of this model is the permutation group $S_L$ of $L$ objects.  Accordingly, the Hilbert space splits in terms of the irreducible representations of this discrete group, given by the Young diagrams $Y_a$ ($a=1,2,\ldots)$ (see, for instance, Figure~\ref{perm44}). The number of the  irreducible representations is equal to the number ${\mathcal P}(L)$ of distinct integer partitions of the natural number $L$, a quantity which grows exponentially with $L$ \cite{Abrowitz}
\be\label{eq_integer_partitions_asymptotics}
{\mathcal P}(L) \simeq \frac{1}{4 L \sqrt{3}}\, \exp\left(\pi \sqrt{\frac{2 L}{3}}\right)
\,\,\,.
\ee
Furthermore, as reviewed in Appendix C, it is noteworthy that the dimensions of some irreducible representations grow exponentially with respect to $L$. 

With periodic boundary conditions, the discrete momentum operator $P$ is also among the global symmetries, and the Hilbert space correspondingly decomposes into 
$L$ distinct sectors, each labelled by a specific momentum quantum number.

In addition to this global symmetry, as show originally by Sutherland \cite{Sutherland}, the Hamiltonian \eqref{Sutherland} is also supported by a dynamical symmetry, namely it has an infinite number of conserved charges which permits its (nested) Bethe-Ansatz exact diagonalization.  In light of this dynamical symmetry the Hamiltonian splits further in smaller size blocks.

\begin{figure}[t]
\centering
\includegraphics[width=0.8\textwidth]{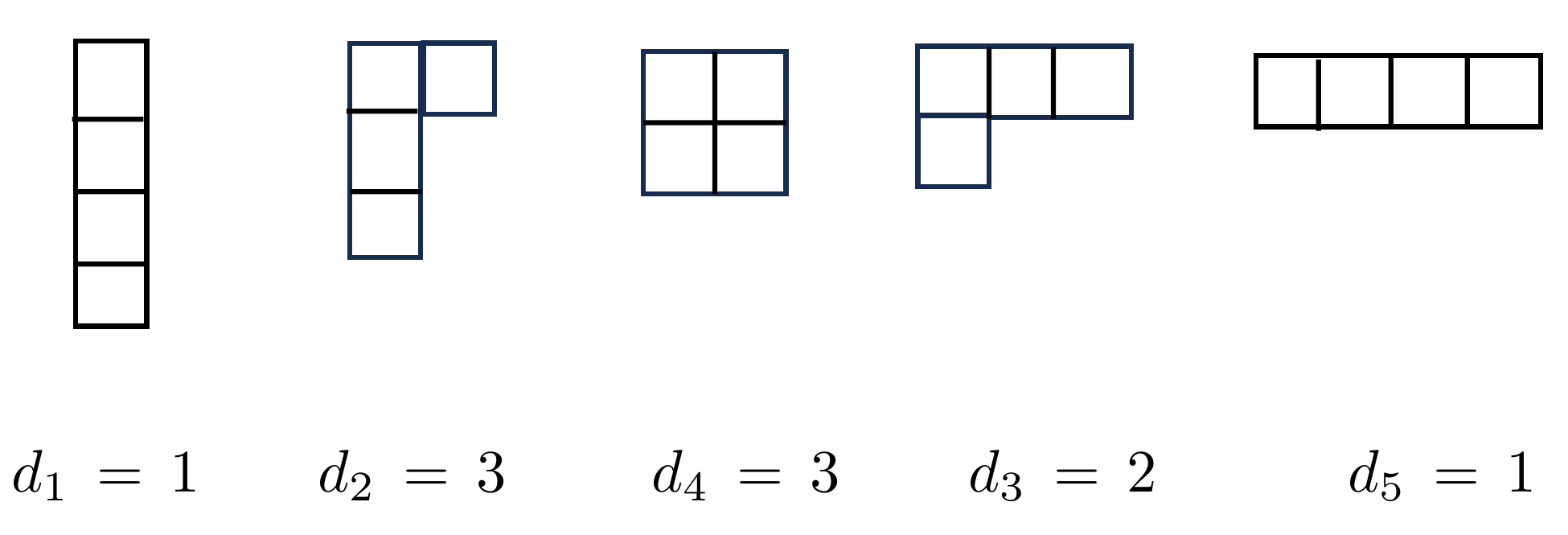}
\caption{Young diagrams associated to all irreducible representations of the permutation group $S_4$ of 4 objects. $d_i$'s are the dimensions of the irreducible representations.} \label{perm44}  
\end{figure} 

\item {\bf Hilbert space fragmentation}. In addition to the block structures of the Hamiltonian coming either from static or dynamical symmetries, we have also to consider the possibility of Hilbert space fragmentation due to some kind of kinetic or other types of constraints (see, for instance \cite{frag1,frag2,frag3} and references therein), more formally defined in terms of the 
commutant algebra \cite{frag3}. As a definition of Hilbert space fragmentation we mean that the $N$-dimensional Hilbert space ${\mathcal H}$ of the system splits into sectors ${\mathcal H} \,=\, \oplus_{k=1}^{N_H} {\mathcal H}_k$ such that 
\begin{itemize}
\item there no matrix elements in the Hamiltonian between different sectors;
\item the number $N_H$ of the sectors grows exponentially in the thermodynamic limit $N \rightarrow \infty$;
\item the block structure is not necessarily due to either global or dynamical symmetries.
\end{itemize}
To these properties, we can add a further specification. Namely, a system has strong Hilbert space fragmentation if the ratio of the dimension of the largest sector and the one of the full Hilbert space vanishes in the limit $N\rightarrow \infty$, while a system has a weak Hilbert space fragmentation if the dimensions of the largest sector remains comparable with the full Hilbert space. 
Notice that the Young diagram decomposition of the Sutherland Hamiltonian may be regarded as a phenomenon of Hilbert space fragmentation, given that there are an exponential number of sectors.

It is well known that fragment sectors are often encountered using a Krylov basis \cite{frag1,NandiPrat,Pozsgay,Krylov}: starting from some initial state $|\Psi_0\rangle$, one construct a set of vectors by applying iteratively increasing powers of the Hamiltonian 
\be
{\mathcal V} \,=\, \{|\Psi_k\rangle \,=\, H^k\, |\Psi_0\rangle
\,\,\,\,\,
,
\,\,\,\,\,
k=0,1,\ldots, n
\}
\ee
where $n$ is the smallest integer such that $H^{n+1} |\Psi_0\rangle$ is linearly dependent on the previous vectors of the set. If the system is ergodic, it is expected that the set 
${\mathcal V}$ spans the entire Hilbert space. On the other hand, if there are some global symmetries and $|\Psi_0\rangle$ is an eigenvector of the corresponding operators, it is expected that the space ${\mathcal V}$ spans the full sector with a given set of quantum numbers. Finally, if the Hilbert space is fragmented, the Krylov spaces as ${\mathcal V}$ are expected to be exponentially small with respect to the dimension of the Hilbert space. 

We shall not elaborate further on the various fragmentation phenomena of the Hilbert space, referring the reader to the specialised literature on the subject. The point to emphasise, however, is that our protocol, which will be presented in Section \ref{S7-protocol}, is capable of determining whether the Hilbert space under consideration is fragmented.

\end{itemize}

\section{\label{S5-Gap-level} Level Gap Distributions}\label{probdistr}
As the reader may have noticed, the statistical argument outlined in Section \ref{classQIM} for assessing whether a Hamiltonian is integrable hinges on the presence of a non-zero probability for vanishing level spacings. However, as we will show, the situation is considerably more nuanced and demands careful scrutiny. 

In this section, we assume that all global symmetries of the system have been fully accounted for, so that we are working within a single irreducible representation of these symmetries. Accordingly, we consider Hamiltonians acting on states belonging to the same global symmetry class. On this basis, in the following, we will analyse the level spacing distributions characteristic of genuinely chaotic and integrable systems. The entirety of our analysis is grounded in statistical arguments, as we are primarily concerned with the statistical properties of different Hamiltonians. Specifically, we address questions such as: How many energy levels occur per unit energy interval? What is the probability distribution of nearest-neighbour spacings? What is the likelihood of observing two consecutive gaps with specified values? And so on. To enable meaningful comparisons across different spectra, the first essential step is to normalise them so that the mean nearest-neighbour spacing is the same—conventionally set to unity. This is achieved by rectifying the spectrum through a procedure known as unfolding, which we describe in what follows.

\subsection{Rectifying the spectrum, alias unfolding}

In the statistical analysis of quantum spectra, the unfolding procedure is a crucial step for removing system-specific global trends in the energy level distribution and isolating the universal local fluctuations, such as level repulsion or clustering. This allows for   meaningful comparison across different Hamiltonians, and between theoretical predictions (e.g., Poisson or Wigner-Dyson statistics) and actual data. Universal statistical features (like those that indicate integrability or chaos) appear only after removing these large-scale variations. 

Thus, unfolding transforms a spectrum $\{E_n\}$ into a new sequence $\{e_n\}$ with a uniform mean level spacing, usually normalized to 1. It is known that this rectification of the spectrum is not unique and can be a delicate operation, since the choice of the smoothing method adopted can influence later analyses (see, for instance \cite{gomezMisleadingSignaturesQuantum2002}). 

There are two aspects to consider: the first consists of taking into account the local curvature of the density of the $N$ energy levels, here denoted as $\rho_0(E)$ (and normalized to 1). This suggests defining the unfolded energy levels $e_i$ in terms of the cumulative distribution of the level density as 
\be \label{eq_cumulative_doe}
e_i \,=\, N \, \int_{-\infty}^{E_i} \rho_0(E') \, dE' \,\,\,.
\ee
The second aspect concerns, in most cases, the ignorance of the analytical expression of the energy level density $\rho_0(E)$ which, therefore, must be estimated numerically.  This is done using the histogram of the spectrum, constructed with bins whose size should be larger than the mean level spacing but significantly smaller than the energy scale over which $\rho_0(E)$ varies appreciably. This numerical approach can introduce additional sources of error: overfitting $\rho_0(E)$  (i.e., using bins that are too large) may smooth out meaningful spectral fluctuations, while underfitting it (i.e. using excessively small bins) may imprint non-universal features onto the unfolded spectrum. The reader may find more details about this issue in Section~\ref{examples}.

From now on, we will use the notation $\{E_i\}$ for the original energies and $\{e_i\}$ for the unfolded energies. 

\subsection{Hamiltonian lines}\label{HamiltonianLines}
Consider a quantum Hamiltonian $H(\lambda)$ that depends on a real parameter $\lambda$. The corresponding energy levels $E_i(\lambda)$ 
can be represented as curves in the $(E, \lambda)$ plane. In general, two arbitrary curves in a plane—obtained by varying a single parameter—are expected to intersect. However, this is typically not the case when the curves represent eigenvalues of a quantum Hamiltonian. In fact, degeneracy between two energy levels generally requires the variation of two independent parameters, not just one (see, for instance \cite{BerryLesHouches1981,BerryLesHouches1989}). Let's present the familiar argument of this statement together with a simple derivation of the behaviour of the probability density of the gap. 

Let's imagine there exists a value $\lambda_*$ where two energy levels $E_1(\lambda_*)$ and $E(\lambda_*)$ get very close. Making a proper rotation in the Hilbert space, 
in the vicinity of this value of $\lambda$, we can restrict our attention to the $2\times 2$ block matrix, which involves these two levels, ignoring the rest of the spectrum 
\be
M \,=\, 
\left(
\begin{array}{ll}
E_1 & V \\
V^* & E_2 
\end{array}
\right) \,\,\,.
\ee
Diagonalizing it, we have the true energy levels of this two-level system 
\be
\hat E_{1,2} \,=\,\frac{E_1 + E_2}{2} \pm \frac{1}{2} \sqrt{(E_1 - E_2)^2 + |V|^2} 
\,\,\,. 
\ee
\begin{figure}[t]
\begin{center}
$\begin{array}{ccc}
\includegraphics[width=0.55\textwidth]{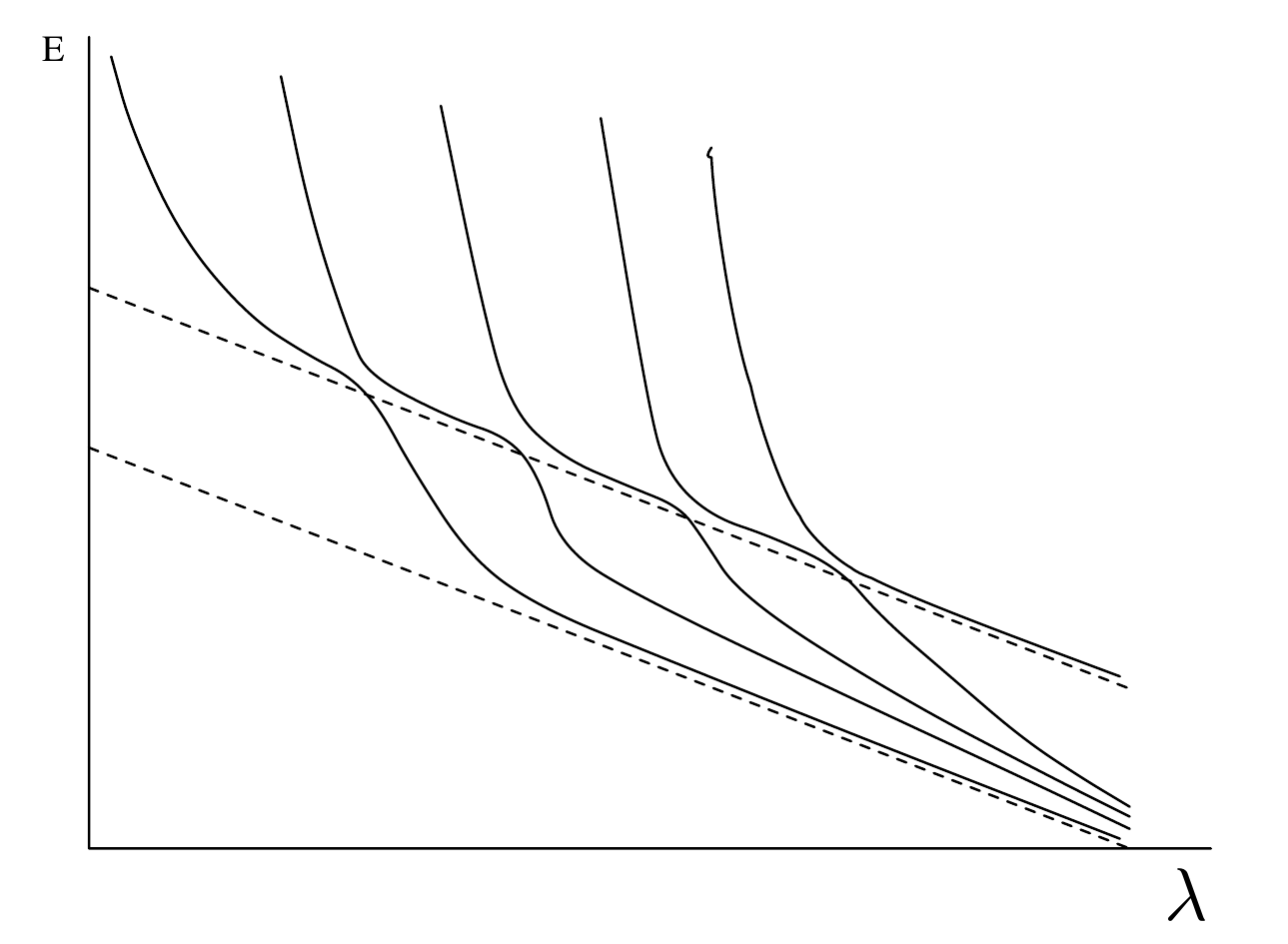}  
\end{array}$
\end{center}
\label{levelcrossing}
\caption{Typical phenomenon of level repulsion in a quantum non-integrable Hamiltonian which depends on a parameter $\lambda$.}
\end{figure} 
Therefore, in order to have a crossing of the two levels, we need to impose two conditions: (i) $E_1 = E_2$ and (ii) $V = 0$. Given for granted the first of them, i.e. $E_1 = E_2$, 
whether or not the second condition is satisfied generally delineates two distinct classes of Hamiltonians: those describing integrable systems (when $V = 0$ for a symmetry reason, e.g. the two states transform according to two different irreducible representations of the group of symmetry) and those characterizing chaotic ones ($V \neq 0$). In the latter case, 
nearby eigenstates "hybridize" due to the absence of conservation laws, which creates avoided crossings and leads to level repulsion. Indeed, with $V \neq 0$, let's estimate the probability density of finding a level at a distance $\Delta$ from a given one 
\be
\hat P(\Delta) \,\propto \, \int dE_-\,\, {\mathcal D V} \, 
\delta\left(\Delta - \sqrt{(E_1 - E_2)^2 + |V|^2}\right) \,\,
\ee
where $E_- = E_1 - E_2$ while for the integration on the variable $V$ we have two cases, whether $V$ is real or complex: 
\be
{\mathcal D} V \,=\,  \left\{
\begin{array}{cll}
d V &,& V \, \, {\rm real} \\
d V_1 \, dV_2 & , & V \,\, {\rm complex} \,\,\, ,\,\, V= V_1 + i V_2 
\end{array}
\right.
\\ .
\ee
Integrating over $E_-$, we are left with 
\be 
\hat P(\Delta) \propto \int {\mathcal D}V \, 
\sqrt{1 - \frac{|V|^2}{\Delta^2}} \, \theta(\Delta - |V|) 
\ee
which ends up in  
\be 
\hat P(\Delta) \,\propto \,
\left\{
\begin{array}{cll}
\Delta &,& V \,\, {\rm real}\\
\Delta^2 &,& V\,\, {\rm complex}
\end{array}
\right.
\\ \\ .
\ee

\subsection{Wigner-Dyson distributions for chaotic systems}
Random matrix theory has proven to be a highly effective framework for analysing chaotic quantum systems \cite{Wigner1955,Wigner1967,DysonI,DysonII,DysonIII,Porter,Mehta1,Mehta2,Brody,Bohigas,BGS,OxfordHandbook}. 
Rephrasing Wigner’s perspective \cite{Wigner1955,Wigner1967}, the idea is to conceptualize a complex system as a ``black box'' comprising a large number of interacting degrees of freedom governed by unknown dynamics. Consequently, one develops a statistical framework that relinquishes exact knowledge of the system’s microscopic details, aiming instead to construct a mathematical formalism in which all possible interaction laws are treated as equally probable. 
A further motivation to look at random matrix theory comes from the Bohigas-Giannoni-Schmit conjecture \cite{Bohigas} that says that quantum chaos manifests itself in the same statistical fingerprints as random matrices.
We are then interested in the statistics of the energy levels and, in particular, hereafter in the statistics of the separations between adjacent energy levels. The random matrix approach relies on the assumption that we can take the Hamiltonian $H$ to be a random matrix with stochastic matrix elements restricted only by the symmetry of the problem. Suppose we take a generic Hamiltonian matrix $H$ of size $N \times N$, $H = H^{\dagger} = U \Lambda U^{\dagger}$, where $U$ is the diagonalizing unitary matrix $U \in U(N)$ and $\Lambda = {\rm diag}\,(\lambda_1,\ldots,\lambda_N)$ is the diagonal matrix containing the real eigenvalues $\lambda_i$. The probability assigned to such a matrix is given by\footnote{The measure $d H$ is equal to $\prod_{1\leq i < j\leq N} dH^{R}_{ij} dH^{I}_{ij} \, 
\prod_{k=1}^N dH_{kk}$ if  $H$ is a complex hermitian matrix ($H_{k<l} = H_{k<l}^R + i H_{k < l}$) while it is equal to $\prod_{1\leq i < j\leq N} dH_{ij}  \, 
\prod_{k=1}^N dH_{kk}$ if the matrix is real.}
\be
dP(H) \,=\, d H\, \exp[- {\rm Tr}\,H^2]\,\,\,.
\ee
The partition function is then given by 
\be
Z \, =\, \int d H \, \exp[- {\rm Tr}\, H^2]\,\,\,,
\ee
and, given its diagonalization, can be written in terms of the eigenvalues 
\be
Z\,=\,c_N\, \prod_{j=1}^N \,\int_{-\infty}^{\infty} d\lambda_j \, e^{-\frac{\beta}{2}\lambda^2_j} \, \prod_{1 \leq k < l \leq N} |\lambda_l - \lambda_k|^\beta 
\label{randompartition}
\ee
where $c_N \,=\, \int dU$ is the volume of the unitary group $U(N)$ while 
\be
\beta \,=\, \left\{
\begin{array}{ccl}
1 & , & {\rm if} \, H \, \,{\rm is \,\,real,} \\
2 & , & {\rm if}\, H \, \,{\rm is \, \, complex.}
\end{array}
\right.
\ee 
$\beta = 1$ corresponds to the Gaussian Orthogonal Ensemble (GOE), while $\beta = 2$ to the Gaussian Unitary Ensemble (GUE). Hence, the joint probability density function of the eigenvalues of a real or hermitian random matrix from the Gaussian ensembles can be written as 
\be
P_{N}^{( \beta)}(\lambda_1,\ldots,\lambda_N)\,=\, {\rm const}\, \exp\left(-\frac{\beta}{2}\sum_{j=1}^N \lambda_j^2\right) \, 
\prod_{1\leq j < k \leq N} |\lambda_j - \lambda_k|^{\beta}\,\,\,.
\label{probabilityeigenvalues}
\ee 
From this expression, it is clear that the eigenvalues are no longer independent variables but are coupled and repel each other. Expressing $Z$ as 
\be
Z\,=\,c_N \, \prod_{j=1}^N \int_{-\infty}^{\infty} d\lambda_j \, \exp[- S(\{\lambda_j\})] \,\,\,,
\ee
where 
\be
S(\{\lambda_j\}) \,=\, \frac{\beta}{2}\sum_{j=1}^N \lambda_j^2 - \beta\sum_{1 \leq k \neq l \leq N} \log |\lambda_l - \lambda_k| 
\,\,\,,
\ee
they behave like charged Coulomb particles in two dimensions at a temperature $\beta$, confined to the real line and subject to a confining quadratic potential. All questions related to, for instance, the probability density of the levels, their correlation functions, and so on, and so forth, can be derived from the expression \eqref{probabilityeigenvalues} of the joint probability. Let's discuss, in particular, the level spacing probability distribution and the higher order spacing distributions. 

\subsubsection{Level spacing probability distribution}
Let $s_n$ be the spacing between two consecutive unfolded energies 
\be
s_n \,=\, e_{n+1} - e_n\,\,\,.
\ee
Using Eq.\,\eqref{probabilityeigenvalues}, in the infinite limit $N\rightarrow \infty$  the Wigner-Dyson probability $P^{(\beta)}_1(s)$ to find two next-neighbour unfolded energy levels at a distance $s$ is given by \cite{Mehta1,Mehta2}
\be
P^{(\beta)}_1(s) \,=\, \left\{
\begin{array}{lll}
\displaystyle \frac{\pi}{2} \,s\, e^{-\frac{\pi s^2}{4}} &, & \displaystyle \beta = 1,\,\,\,\,\,\,\,\,({\rm GOE}), \\
& & \\
\displaystyle \frac{32}{\pi^2} \, s^2 \, e^{-\frac{4 s^2}{\pi}}&,& \displaystyle\beta =2 ,\,\,\,\,\,\,\,(\rm{GUE}).
\end{array}
\right.
\label{WD111}
\ee
 Notice that these distributions capture the level repulsion at small spacings, i.e., for $s \rightarrow 0$ they vanish as $s$ and $s^2$ near the origin, respectively, as predicted by the argument discussed in Section \ref{HamiltonianLines}. They have a maximum around $s\simeq 1$ (at $s_* = \sqrt{2/\pi} = 0.7978..$ for $\beta =1$ and at $s_* \,=\,\sqrt{\pi/4} = 0.8862..$ for $\beta =2$) and decrease exponentially at infinity. 

\subsubsection{ Wigner's surmise}
There is a heuristic derivation by Wigner \cite{Wigner1955,Wigner1967} of the probability distribution of the energy gaps, which goes as follows. For a random sequence, the probability that a level will be in the small interval ($E+s, E+ s+ds$) is, of course, proportional to $ds$ and will be independent of whether or not there is a level at $E$. However, if there is level repulsion, the argument must be modified, since the probability we are looking for is the one concerning the occurrence of the two events $A$ and $B$:
\begin{enumerate}
\item[A:] no-level in ($E,E + s$);
\item[B:] one level in ($E + s,E + s + ds$). 
\end{enumerate}
Therefore, we are concerned with the joint probability of the events $A \cup B$, given by  $P(A \cup B) = P(A/B)P(B)$, where $P(A/B)$ is the conditional probability of $A$ given $B$. Hence, 
we have $p(s)ds = P(A \cup B)$ and the level spacing probability density $p(s)$ is such that
\be 
p(s) ds \,= \, {\rm Prob}({\rm one \,\,level\,\, in\,\, dI/\,no\,\,level\,\, in\,\, I}) \,\,{\rm Prob}({\rm no\,\, level\,\,in\,\, I})\,\,\,.
\ee
In terms of $p(s)$, the probability of having no level in $I$ is given by 
\be
\label{no-level-dI}
{\rm Prob}({\rm no\,\, level\,\,in\,\, I})\,=\, \int_s^{\infty} ds' \, p(s') \,\,\,,
\ee
namely, we need to have all level spacings larger than $s$ and not to have a level in the interval $(E, E + s)$. Hence, with the notation 
\be
\mu(s) ds\,=\, {\rm Prob}({\rm one \,\,level\,\, in\,\, dI/\,no\,\,level\,\, in\,\, I})\,\,\,,
\ee
we have 
\be
p(s)  \,=\, \mu(s)  \, \int_s^{\infty} ds' \, p(s') \,\,\,.
\ee
The differential equation that comes from this relation can be easily solved 
\be
p(s) \,=\, {\mathcal N} \mu(s) \, e^{-\int_0^s \mu(t) dt}\,\,\,.
\label{Wignersurmise}
\ee
This probability distribution must satisfy two conditions: 
\begin{enumerate} 
\item normalization
\begin{equation}
\int p(s) \,ds \,=\, 1;
\end{equation}
\item Unit mean gap 
\begin{equation}
\int p(s) \,s\, ds \,=\, 1.
\end{equation}
\end{enumerate}
As evident from this equation, the probability distribution $p(s)$ is fully determined once it is specified the quantity $\mu(s)$. 
However, it should be noted that the resulting $p(s)$ is expected to accurately capture the behaviour only in the limit of small $s$. 
With this remark in mind, if we consider the case where $\mu = const$,
imposing the two conditions above yields the Poisson distribution   
\begin{equation}
    p(s) \,=\, e^{-s}
    \,\,\,,
    \end{equation}
a distribution which we will meet in Section~\ref{Poissonsubsection} in relation to the energy level distribution of integrable models.  
On the other hand, taking for $\mu(s)$ a linear repulsive law for the levels, $\mu(s) \simeq \alpha s$ and imposing the two conditions above, the unit mean gap condition fixes $\alpha$ to have the value $\alpha = \pi/2$, so that we recover the GOE gap distribution. 
However, it is important to note that choosing
$\mu(s) \simeq \alpha s^2$ does {\em not} reproduce the full GUE distribution. Instead, it yields a different distribution that agrees with the GUE only in the small-$s$ limit.

Finally, assuming that $\mu(s)$ has a small $s$ behaviour $\mu(s) 
\simeq \alpha (1+\nu) s^{\nu}$, one gets the Brody set of distributions \cite{Brody} 
\begin{equation}
    p_B(s) \,=\, (1+\nu) \, a s^{\nu} \, \exp(-a s^{1+\nu})
\,\,\,\,\,\,\,\,
,
\,\,\,\,\,\,\,\,
a \,=\, \left[\Gamma\left(\frac{2+\nu}{1+\nu}\right)\right]^{1+\nu},
\end{equation}
where $\nu$ is the so-called ``Brody parameter''. The interesting feature of this family of distributions is that it interpolates between the Poisson distribution ($\nu =0$) and the GOE one ($\nu  = 1$). 

\subsubsection{Higher order spacing distributions}
\label{GUE-determinant}
\begin{figure}[t]
\begin{center}
\includegraphics[width=0.6\textwidth]{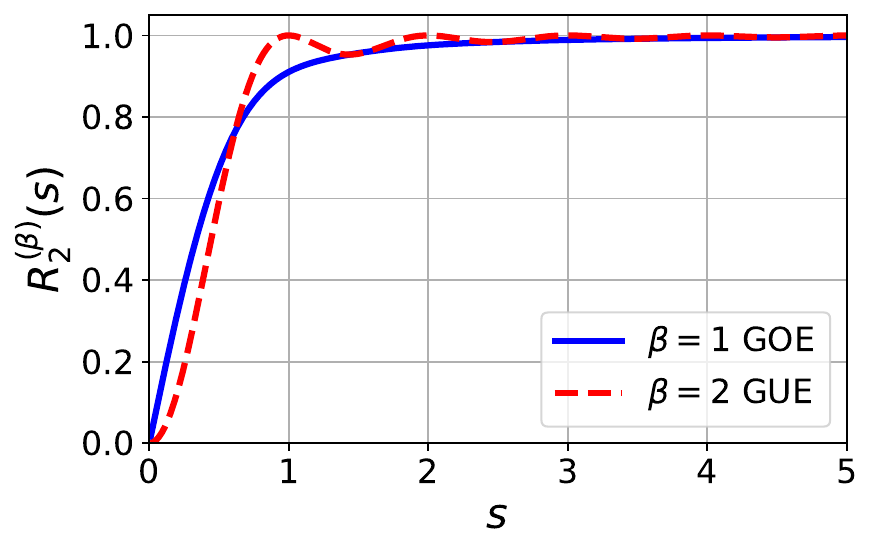} 
\end{center}
\caption{2-point correlation function $R_{2}^{(\beta)}(s)$ for the GUE and GOE. 
Note that both curves vanish at the origin, reflecting level repulsion, and rapidly saturate to the asymptotic value 
1, which characterizes the spectral rigidity.}
\label{2ptcorrelation}
\end{figure}

Important information about the global distribution of energy levels is encoded   in their correlation functions. In terms of the $P_{N}^{(\beta)}$ of Eq.\,\eqref{probabilityeigenvalues}, the $n$-point correlation function of a $N \times N$ random matrix is given by  \cite{DysonI,DysonII,DysonIII}
\be
R_{n}^{(\beta)}(x_1,\ldots,x_n) \,=\, \frac{N!}{(N-n)!} \,\int_{-\infty}^{\infty} \cdots \int_{-\infty}^{\infty} \, P_{N}^{(\beta)}(x_1,\ldots, x_N) \, dx_{n+1} \cdots dx_N\,\,\,.
\ee
This expression expresses the probability density of finding a level (regardless of its labelling) around each of the points $x_1, x_2, \ldots x_n$, once we integrate out the remaining levels. Exact evaluations of these distributions can be found in the classical works of Mehta \cite{Mehta1,Mehta2}. Unfolding the energy spectrum and going to the $N \rightarrow \infty$ limit, $R_1$ becomes a constant (equal to 1, since it corresponds to the mean level spacing), while the remaining correlation functions depend only on the eigenvalue differences $x_i - x_j$. In particular, posing $s = x_1 - x_2$, for the pair-correlation $R_{2}^{(\beta)}(s)$ we have (see Figure~\ref{2ptcorrelation})
\be
R_{2}^{(\beta)}(s) 
\,=\, \left\{
\begin{array}{lll}
\displaystyle 1 - K^2(s) -\left(\frac{d K}{ds}\right) \, J(s)   
&, & \displaystyle \beta = 1\,\,\,\,\,,\,\,\,({\rm GOE}), \\
& & \\
\displaystyle 1 - K^2(s)   
&,& \displaystyle\beta =2 \,\,\,\,,\,\,\,(\rm{GUE}).
\end{array}
\right.
\label{WD11}
\ee
where 
\be 
K(s) \,=\, \frac{\sin(\pi s)}{\pi s}
\,\,\,\,\,\,\,
,
\,\,\,\,\,\,\,
J(s) \,=\,\int_s^{\infty} K(t) \, dt\,\,\, .
\label{importantkernel}
\ee

Quantities directly related to the correlation functions are the higher-order spacing distributions $P_{k}^{(\beta)}(s)$, where the $k$-step spacing is defined as
\be
s_{n,k} \,=\, e_{n+k} - e_n\,\,\,.
\ee
For those gaps, we obviously have 
\be
s_{n,k} \,=\, (e_{n,k} - e_{n,k-1}) + (e_{n,k-1} - e_{n,k-2}) + \cdots (e_{n+1} - e_n) \,=\, \sum_{i=1}^k s_i \,\,\,,
\label{sumgaps}
\ee 
which, for the average, implies
\be 
\langle s_{n,k}\rangle \,=\, \sum_{i=1}^k s_i \,=\, k \, \langle s_i \rangle \,=\, k \,\,\,,
\ee
since $ \langle s_i \rangle \,=1$. Moreover, we have the identity 
\be
R_{2}^{(\beta)}(s) \,=\, \sum_{k=1}^{\infty} P_{k}^{(\beta)}(s) \,\,\,,
\ee
since $R_{2}^{(\beta)}(s)$ is the probability distribution for having a spacing $s$ between any two eigenvalues of the unfolded spectrum.  

According to Mehta \cite{Mehta1}, if we denote by $E_\beta(j;s)$ the probability that 
an interval of length $s$ contains exactly $j$ eigenvalues for the Dyson ensemble with index $\beta$, 
then the probability density $P_k^{(\beta)}(s)$ of the $k$th neighbour spacing (i.e.,  the distance 
between two eigenvalues with $k-1$ intervening
levels) is given by
\begin{equation}
P^{(\beta)}_k(s) \;=\; \frac{d^2}{ds^2}
\left[
  \sum_{j=0}^{k-1} (k-j)\,E_\beta(j;s)
\right]\,\,.
\label{eq:pk}
\end{equation}
At the origin of this formula, there are the following considerations: 
\begin{enumerate}
\item {\bf Gap probability}. The quantity $E_\beta(j;s)$ is a \emph{gap probability} for the interval $(0,s)$; namely, it enforces that exactly $j$ points fall within the interval.
\begin{itemize}
  \item Differentiating once in $s$ introduces the condition that a point is    located at the right boundary of the interval.
  \item Differentiating a second time enforces that this boundary point is the
  \emph{nearest eigenvalue} at a distance $s$ from the origin, i.e.\ it turns the
  gap probability into a spacing density.
\end{itemize}
Thus, the second derivative in~\eqref{eq:pk} is the probabilistic mechanism that
converts an occupancy probability into a spacing distribution.
\item {\bf Origin of the combinatorial weight $(k-j)$}. Suppose there are $j$ points inside $(0,s)$.  Then the interval $[0,s]$ spans
a block of $k{+}2$ consecutive eigenvalues (the two endpoints plus the $k$
interior ones).  Among the $k{+}1$ gaps formed by these $k{+}2$ levels,
precisely $(k{+}1-j)$ of them can serve as the ``anchoring'' gap coinciding
with the interval $[0,s]$.  This simple combinatorial count yields the weight
$(k+1-j)$ and, shifting $k \rightarrow k-1$, we end up in the factor 
$(k-j)$ in~\eqref{eq:pk}.
\end{enumerate}

\begin{figure}[t]
\begin{center}
\includegraphics[width=0.8\textwidth]{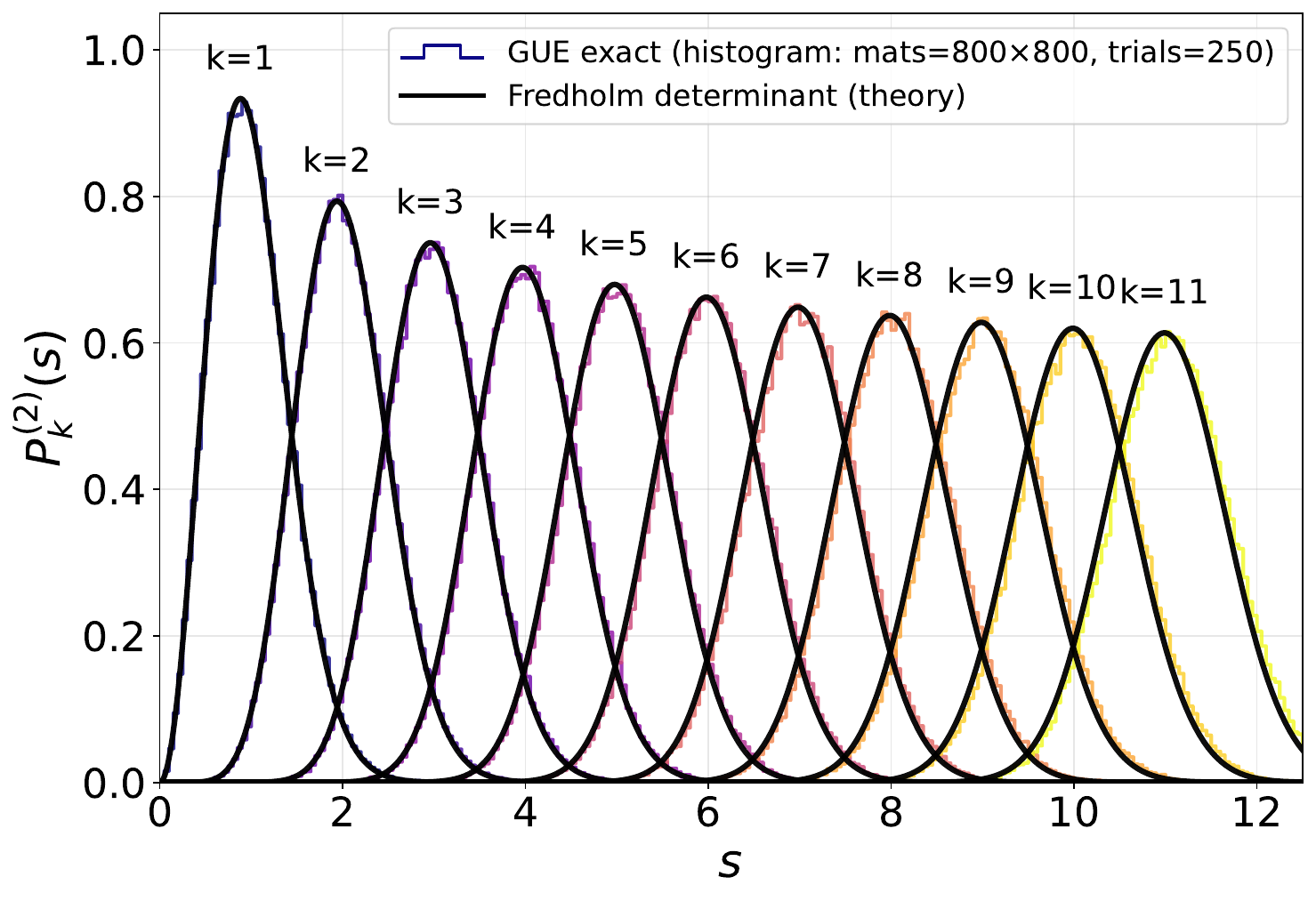} 
\end{center}
\caption{Higher order spacing probability distribution $P_k^{(2)}(s)$ ($k=1,2,\ldots,11)$ for GUE. The histogram is an average over 250 realizations of GUE spectra, each from an $800\times800$ matrix. The $k$ distribution is sharply peaked at $s \simeq k$, reflecting level repulsion and spectral rigidity.}
\label{multigapGUE}
\end{figure}

For the GUE, the quantities defined above admit a Fredholm determinantal representation
\be
D(z,s) \,=\, {\rm Det} (1 - (1-z) {\mathcal K}_{[0,s]}) \,\,\,,
\label{FREDH1}
\ee
where the integral operator ${\mathcal K}_{[0,s]}$ is defined in $L^2([0,s])$ in terms of the kernel $K(x,y)\equiv K(x-y)$ of Eq.\,\eqref{importantkernel} as 
\be
\label{kernel-GUE}
({\mathcal K}_{[0,s]} \, f)(x) \,=\, \int_0^s K(x,y) f(y) dy \,\,\,.
\ee
Indeed, we have 
\be
E_2(k;s) \,=\, \frac{1}{k!} \,\left. \frac{d^k}{d  z^k} D(z,s) \right|_{z=0}\,\,\,.
\ee
As discussed in Appendix \ref{level-space}, the calculation of the Fredholm determinant \eqref{FREDH1} through its eigenvalues can be implemented numerically very efficiently \cite{Bornemann1,Bornemann2}.  
The corresponding probability distribution for the higher spacings is shown in Figure~\ref{multigapGUE}. 

As discussed in Appendix \ref{level-space}, analogous formulas also hold for the GOE, substituting the determinant with the Pfaffian.

\subsection{Poisson distribution for level spacings in integrable systems}\label{Poissonsubsection}
\begin{figure}[t]
\begin{center}
\includegraphics[width=0.5\textwidth]{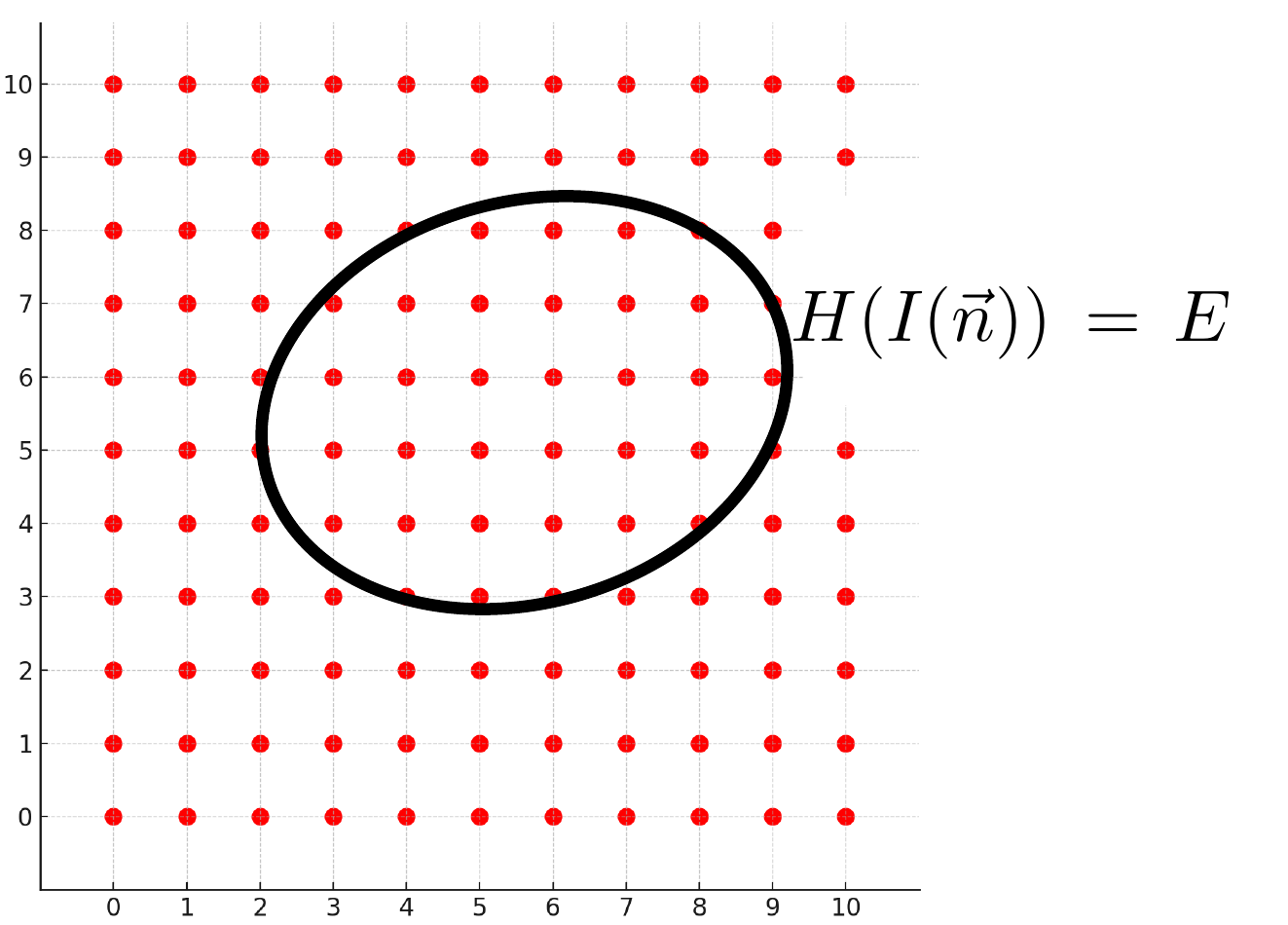} 
\end{center}
\caption{Energy level curve $H(I(\vec{n})) = E$ in the lattice of the quantum numbers ${\vec{n}}$. The curve passes through points with very different quantum numbers. }
\label{lattice}
\end{figure}

\begin{figure}[t]
\begin{center}
\includegraphics[width=0.5\textwidth]{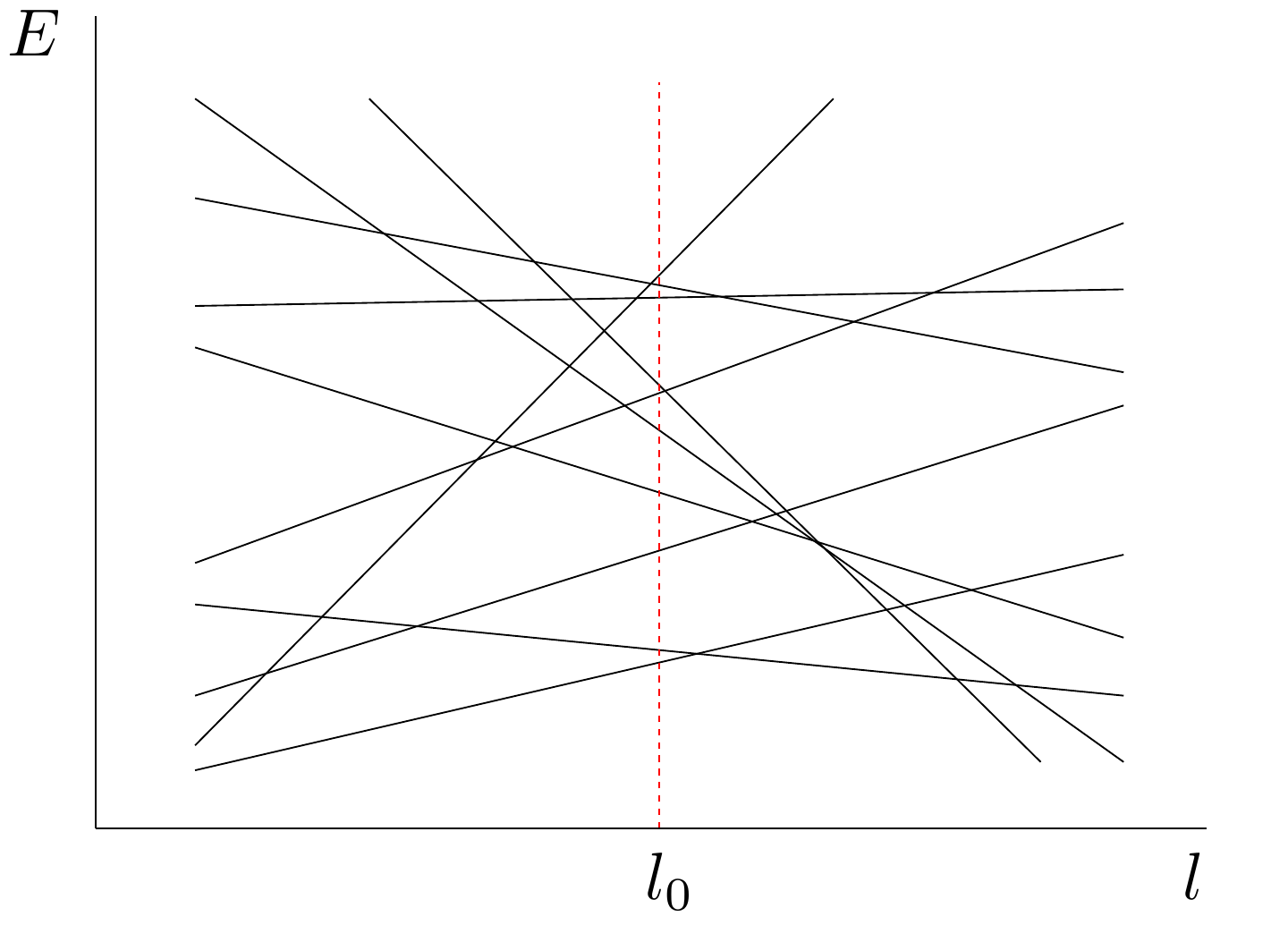} 
\end{center}
\caption{Energy lines as Mikado pick-up sticks in terms of a parameter $l$. The probability distribution at an arbitrary point of the parameter $l$ is a purely geometrical problem.}
\label{randomlines}
\end{figure}

\begin{figure}[t]
\begin{center}
$\begin{array}{ccc}
\includegraphics[width=0.49\textwidth]{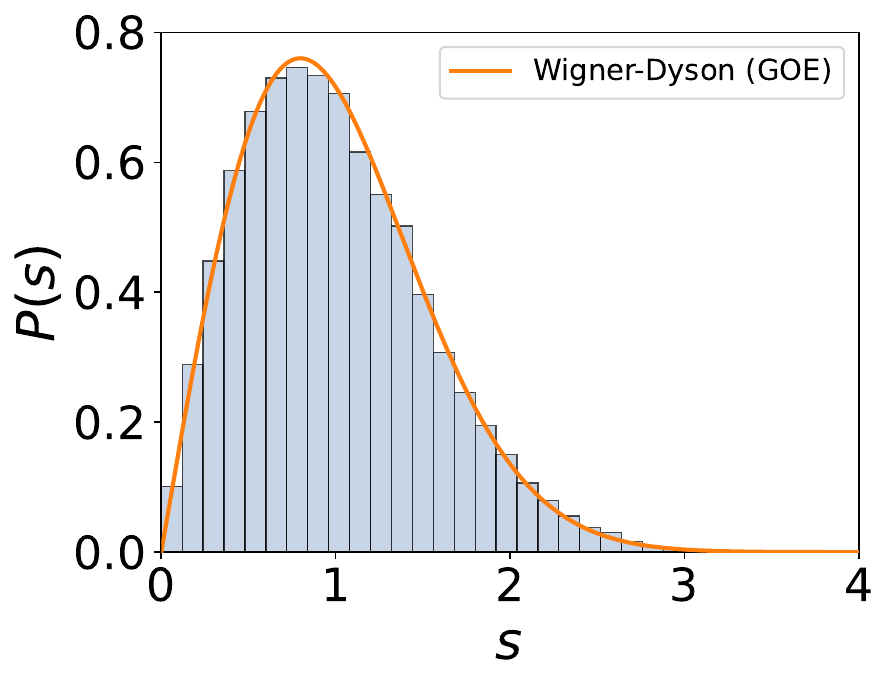} &  
&\includegraphics[width=0.49\textwidth]{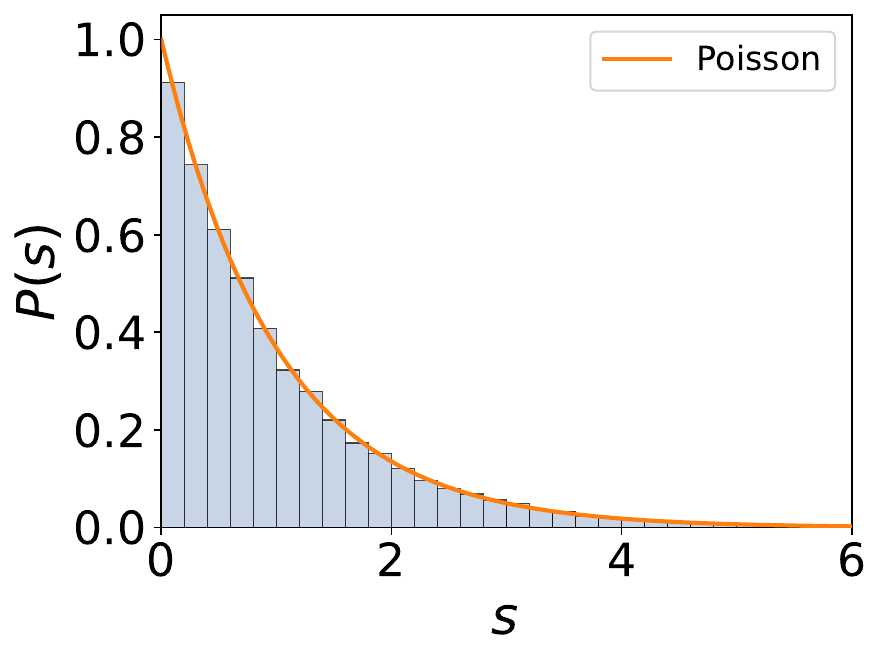} 
 \\
(a) & & (b)  
\end{array}$
\end{center}
\caption{Level spacing distributions: 
(a) GOE Wigner-Dyson; (b) Poisson. In both figures there is also the discretized form of the distributions obtained with an energy resolution $\delta E$.}
\label{distributionfigure}
\end{figure}

In integrable models, the distribution of level spacings is expected to differ markedly from that observed in chaotic systems. 
Berry and Tabor \cite{Berry}, following the previous conjecture of Percival \cite{Percival}, convincingly argued that, in integrable models, 
the extensive (and basically independent) quantum numbers of the conserved charges make energy levels essentially uncorrelated: 
crossings are allowed, there is no level repulsion, and the sequence of spacings behaves like a set of independent and identically distributed (i.i.d.) 
random variables. For an integrable system with a classical analogue possessing $d$ degrees of freedom, semiclassical quantization yields energy eigenvalues of the form
\be
E(\vec{n}) \,=\, H(I(\vec{n})) 
\,\,\,\,\,\,
,
\,\,\,\,\,\,
I_j \,=\, \hbar (n_j + \alpha_j) 
\,\,\,.
\ee
where $\vec{n} \in {\mathbb Z}^d$.  It is worth stressing that basically the same mathematical structure also holds in the Bethe Ansatz formulation of quantum integrable models. 
Since the mapping from integer vectors $\vec{n}$ to energies is essentially arithmetic and uncorrelated, i.e. levels with very different quantum numbers can have very similar energy values (Figure~\ref{lattice}), the local statistics of $E(\vec{n})$ resemble those of a random sequence with Poissonian gaps. The problem is closely connected to geometric probability, specifically to the enumeration of lattice points lying inside closed contours \cite{Kendall-Moran}.
The geometrical origin of the problem and the emergence of the Poisson distribution are clear from the following example inspired by the 
Mikado pick-up sticks: imagine that the $N$ uncorrelated energy levels of an integrable model are parametrised in terms of the     
$N$ random straight lines in the $(\ell,E)$-plane (Figure~ \ref{randomlines}).
\[
E_i(\ell) \,= \, a_i \,\ell + b_i, \qquad i=1,\dots,N,
\]
where $(a_i,b_i)$ are some i.i.d.\ from a continuous joint density $f_{A,B}$.  
 As $\ell$ varies, the set $\{ E_i(\ell)\}$ will exhibit many true crossings (independent slopes imply no hybridization), exactly mirroring the abundance of crossings observed in integrable quantum systems. Let's now fix a vertical slice at $\ell=\ell_0$ and consider the $N$ ordinates
\[
E_i := E_i(\ell_0) = a_i \,\ell_0 + b_i \,\,.
\]
Because the pairs $(a_i,b_i)$ are i.i.d., the $E_i$ are themselves i.i.d.\ with a continuous density
\[
f_Y(E;\ell_0) = \int_{\mathbb{R}} f_{A,B}(a,\,E-a\ell_0)\, da .
\]
Order them as $E_{1} < \cdots < E_{N}$ and define their spacings as
\[
\Delta_i := E_{i+1} - E_{i}. 
\]
To analyse the spacings, we first unfold them by the local mean spacing $1/\bar{\rho}$.
After this rescaling, the distribution of the unfolded gaps reduces to a standard order–statistics problem for i.i.d.\ samples with density 
$f_Y(\:\cdot\: ; l_0)$, a setting treated in full generality by Pyke \cite{Pyke}. Indeed, by Pyke’s spacing theorem, we have 
\begin{itemize}
\item Conditioned on a location $E$, the (unfolded) nearest--neighbour gap is exponential with rate $f_Y(E;\ell_0)$.
\item A ``typical'' spacing is located near $E$ with weight $f_Y(E;\ell_0)\,dE$.
\end{itemize}
Mixing these exponentials over $y$ gives the limiting spacing density
\be
g(s) \;=\; \int_{0}^{1} 
\underbrace{f(x)}_{\text{location density}}\;
\underbrace{\big[ f(x) e^{-s f(x)} \big]}_{\text{conditional spacing density}}
\, dx
\;=\; \int_{0}^{1} f(x)^{2} e^{-s f(x)} \, dx\,\, .
\ee
Two immediate consequences follow:
\begin{itemize}
\item If $f_Y(\:\cdot\:;\ell_0)$ is (locally) flat, then $g(s\mid \ell_0)\approx e^{-s}$ (pure Poisson).
\item In general, $g$ is a mixture of exponentials and still exhibits no level repulsion since the value at $s=0$ is always different from zero 
\[
g(0)=\int f(x)^2 \, dx > 0\,\, .
\]
\end{itemize}
An alternative derivation of the Poisson distribution is provided by the Wigner surmise, given in Eq.\,\eqref{Wignersurmise}. If one assumes that the events 
$A$ and $B$ —namely, the occurrences of energy levels in two disjoint intervals—are statistically independent, then the probability density satisfies 
$\mu(s) = \mu = {\rm const}$, which directly yields the Poisson law for the nearest–neighbour level spacings\footnote{In what follows, we denote the probability density of the Poisson distribution by ${\mathcal P}$, in order to distinguish it from the probability densities associated with random matrix ensembles.} 
\be
{\mathcal P}_1(s) \,=\, e^{-s} \,\,\,.
\label{poissondistribution}
\ee
This is in sharp contrast to the Wigner–Dyson case, where strong correlations between levels preclude such independence and instead give rise to level repulsion. The curves relative to the Poisson and Wigner-Dyson probability distributions are shown in Figure~ \ref{distributionfigure}. For the Poisson distribution, the maximum value occurs at $s=0$, indicating a very strong level of clustering. Therefore, its spectrum generally presents many coincident pairs, triples, quadruplets, etc. of levels. 

\begin{figure}[hb!]
    \centering
    \subfloat[$J_z=1$]{\includegraphics[width=0.3\textwidth]{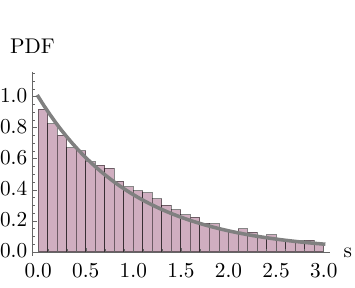}}
    \hfill
    \subfloat[$J_z=0.01$]{\includegraphics[width=0.3\textwidth]{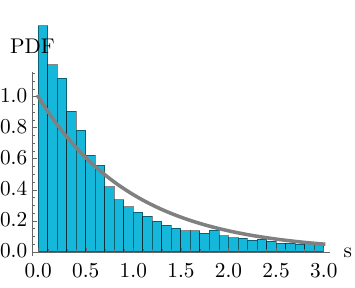}}
    \hfill
    \subfloat[$J_z=0.0001$]{\includegraphics[width=0.3\textwidth]{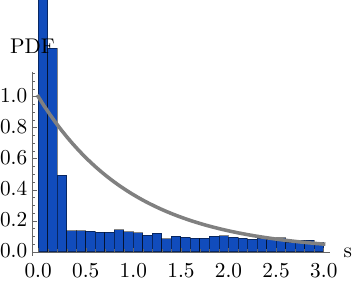}}
    \caption{Level spacing distributions of the integrable XYZ spin chain on $L=14$ sites with $J_x =0.5$, $J_y =0.7$, both couplings kept fixed and varying $J_z$. The graph (a) corresponds to $J_z=1$ while graphs (b) and (c) are close to the value $J_z = 0$ of the free fermion limit, when the original model coincides with the XY model. The Poisson statistic at $J_z =1$ turns into a distribution of free fermion model at $J_z=0$, where the spectrum is given by a set of harmonic oscillators. Contour of the Poisson distribution shown for comparison.}
    \label{oscillators}    
\end{figure}

As emphasized by Berry and Tabor \cite{Berry}, it is worth noting that the spacing distribution of one of the simplest and most familiar integrable systems—the multi-dimensional harmonic oscillator—does not conform, however, to Poisson's law. This exception stems from the arithmetic structure of its spectrum, which is generally given by linear combinations of the oscillator frequencies with integer coefficients
\be
E_{\vec{n}} \,=\, \sum_{j} \hbar \,\omega_j\, \left(n_j + 1/2\right) \\ .
\ee 
If the frequency ratios $\omega_i/\omega_j$ are rational, many different $\vec{n}$ yield the same energy, giving rise to large degeneracies, larger than those predicted by the Poisson distribution.  Indeed, these ``resonances'' destroy the randomness needed for Poisson statistics. If, instead, the frequency ratios are irrational but algebraically related, the spectrum is still highly structured, with stronger correlations and clustering effects. Even when unfolding, the spectrum reflects the lattice arithmetic of linear forms in integers, not the independence assumed in the aforementioned Pyke’s theorem. We refer the reader to the original reference \cite{Berry, BerryLesHouches1981,BerryLesHouches1989} for a discussion of this subtle but important point and to Sections 8 and 9 for some explicit examples. Free fermion systems are of course a typical example of a set of free oscillators and the same considerations apply as well: see Figure~\ref{oscillators} for the evolution of the probability distribution of the gaps in the integrable interacting XYZ model varying $J_z$ 
\begin{equation*}
H=\sum_{j=1}^L\;\left(J_x\sigma^x_j\sigma^x_{j+1}+J_y\sigma^y_j\sigma^y_{j+1}+J_z\sigma^z_j\sigma^z_{j+1}\right)\,\,\,,
\end{equation*}
which reduces to a free fermion system, i.e. the XY model, in the limit $J_z=0$.

\subsubsection{Higher order spacing distributions in integrable models}
\begin{figure}[t]
\begin{center}
\includegraphics[width=0.7\textwidth]{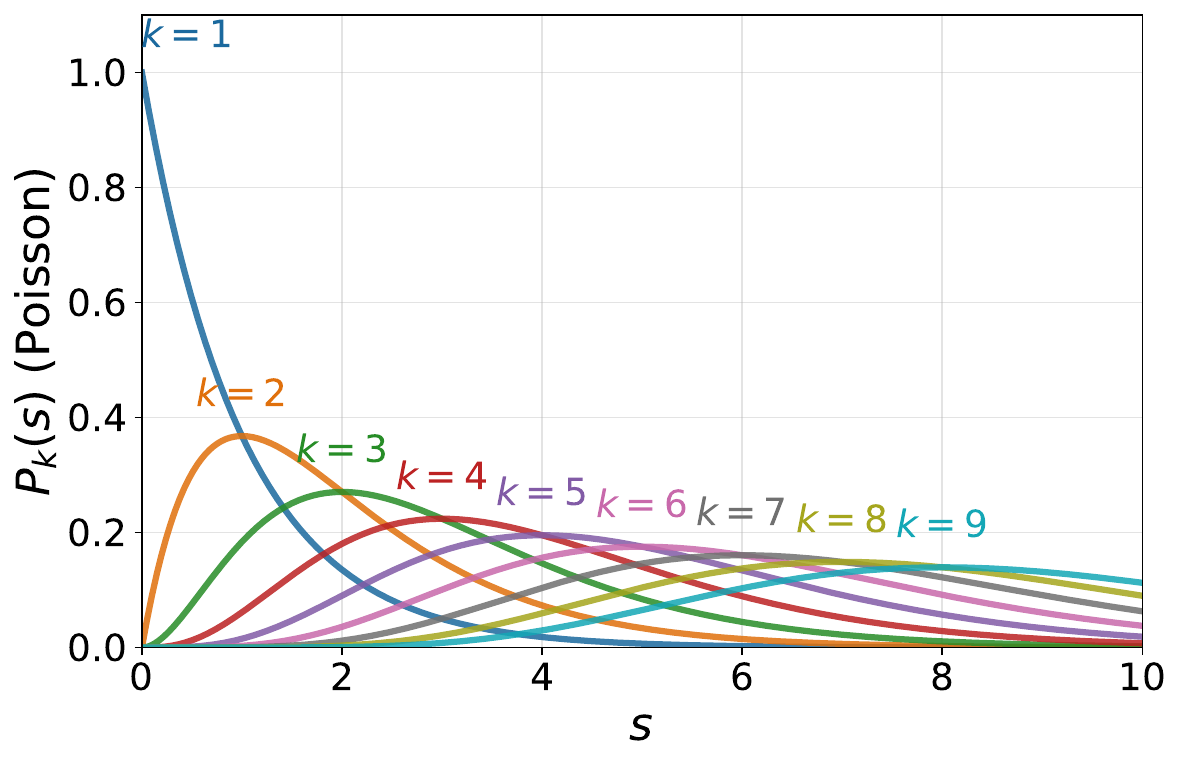} 
\end{center}
\caption{The probability distributions of higher order level spacings for the Poisson case.}
\label{mulgapPoisson}
\end{figure}
It is quite easy to find the probability distributions ${\mathcal P}_{k}(s)$ of higher-order spacings for integrable Hamiltonians. Indeed, in view of Eq.\,(\ref{sumgaps}) and given that each gap $s_i$ is uncorrelated, the ${\mathcal P}_{k}(s)$ are simply the convolution of $k$ Poisson distributions 
\be
{\mathcal P}_k(s) \,=\, 
\left({\mathcal P}_1 \star {\mathcal P}_1 \star {\mathcal P}_1 \cdots \star {\mathcal P}_1\right) (s)\,\,.
\ee
This convolution can be evaluated using the Fourier transform: putting 
\be
\hat {\mathcal P}_1(p) \,=\, \frac{1}{\sqrt{2 \pi}} \,\int_{-\infty}^{\infty} {\mathcal P}_1(x) \, e^{i p x} \, dx \,=\, \frac{1}{\sqrt{2\pi}} \,\frac{1}{1 + i p}\,\,,
\ee
we have indeed 
\be
{\mathcal P}_k(s) \,=\,  \frac{1}{\sqrt{2 \pi}} \,\int_{-\infty}^{\infty} \left(\hat {\mathcal P}_1(p)\right)^k  \, e^{-i p x} \, dp\,=\, \frac{s^{k-1}}{(k-1)!} e^{-s} \,\,\,.
\ee
For the mean and variance of these distributions, we have 
\be
\langle s_k \rangle \,=\, k 
\,\,\,\,\,\,\,\,\,
,
\,\,\,\,\,\,\,\,\,
\langle (s_k - k)^2 \rangle \,=\, k \,\,\,\,.
\ee
As shown in Figure~\ref{mulgapPoisson}, the behaviour of these distributions is markedly different from that observed in chaotic systems. In particular, as 
$k$ increases, the variance of the curves also grows, causing them to broaden progressively. Since for the Poisson distribution the 2-point correlation function $R_2(s)$ is identically equal to 1, we have the identity
\be
R_2(s) \,=\, 1 \,=\, \sum_{k=1}^\infty   {\mathcal P}_k(s) \,\,\,.
\ee

\section{\label{S6-superposition} Superposition of Sequences of Energy Levels}
\begin{figure}[t]
\begin{center}
\includegraphics[width=0.8\textwidth]{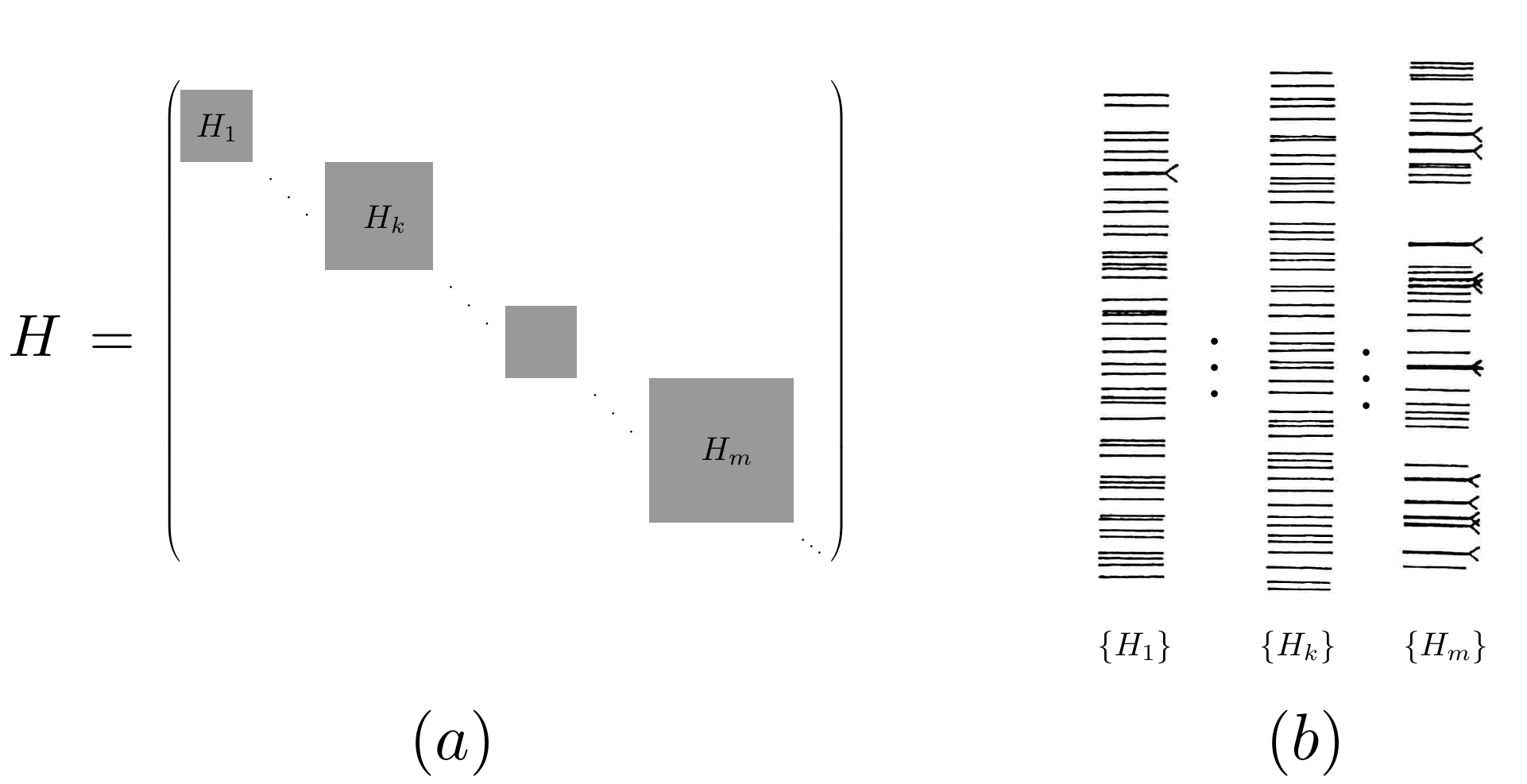} 
\end{center}
\caption{(a) Hidden block form of the Hamiltonian matrix; (b) spectra originating from the various blocks.}
\label{hiddenblockform}
\end{figure}

We then turn to the case of spectra obtained from a set of unrelated levels that has been superimposed, a case that presents a more subtle and potentially deceptive scenario. 
This case occurs when the Hamiltonian has a block structure due to some global symmetry that has not been previously identified. The same situation also occurs in the presence of fragmentation of the Hilbert space. So, imagine that our Hamiltonian has some hidden block form, as shown in Figure~\ref{hiddenblockform}.
Each block of the Hamiltonian produces its own spectrum, which may individually follow either Wigner–Dyson or Poisson statistics. However, if one is unaware of the underlying block structure and simply diagonalizes the full Hamiltonian $H$, the only accessible object is the global spectrum. This immediately raises the central question: once the spectrum of 
$H$ is unfolded, what is the level-spacing distribution obtained in such a case, where the observed sequence of levels is, in fact, a superposition of distinct but hidden spectra? This problem has been addressed by several authors \cite{Mehta1,Mehta2,Porter2,BerryRobnik,Vernier} and in the following, we remind the final formula according to the derivation given by Mehta \cite{Mehta1}. 

Let's first set some notations. Let's define $E(s)$ as the probability that a \emph{randomly chosen interval} of length $s$ of the spectrum is empty. 
Hence, differentiating \eqref{no-level-dI}, we have 
  \be
    -E'(s) = \Pr\big(\,[0,s]\text{ empty AND a level at }s\,\big).
  \ee
  This is the probability density that the interval is empty and a level sits at the boundary. Differentiating again, we have 
  \be
    P(s) = \frac{d^2}{d s^2}E(s)
  \ee
where $P(s)$ is the probability density that the \emph{nearest neighbour} lies at a distance $s$.

Let's now derive the level-spacing distribution for a mixture of spectra. We denote by $\rho_i$ the level density of the $i$-th sequence, such that $p_i(\rho_i s) \rho_i d s$ is the probability that a spacing in the $i$-th sequence has a value between $s$ and $s + ds$. Each function $p_i(x)$ satisfies the two conditions 
\be
\int_0^\infty p_i(x) \, dx \,=\, 1 
\,\,\,\,\,\,\,\,\,\,
, 
\,\,\,\,\,\,\,\,\,\,
\int_0^\infty x\, p_i(x) \, dx \,=\, 1 \,\,\,\,.
\ee
Introducing the three quantities $\Psi_i(x)$, $R_i(x)$  and $E_i(s)$, defined as 
\be
\begin{array}{l}
\displaystyle \Psi_i(x) \,=\, \int_0^x p_i(y) \, dy \,=\, 1 - \int_x^\infty p_i(y) dy = 1 -\int_0^\infty p_i(x+y) \, dy \,\,\,,
\\
\displaystyle R_i(x) \,=\, 1 - \Psi_i(x) \,=\,\int_x^\infty p_i(y) \,dy \,=\, \int_0^\infty p_i(x+y) \, dy \,\,\,,\\
\displaystyle  E_i(x) \,=\, \int_x^\infty R_i(y)\, dy \,=\, \int_x^\infty dy \int_y^\infty p_i(z) \,dz\,=\,
\int_0^\infty \int_0^\infty \,p_i(x + y + z) \, dy\, dz\,\,\,,
\end{array}
\label{W1}
\ee
we have that $\Psi_i(\rho_i s)$ is the probability that a spacing in the $i$-th sequence is less than or equal to $s$, while $E_i(\rho_i s)$ is the probability that a given interval of length $s$ does not contain any level of the $i$-th sequence. 

Superposing together $n$ sequences, the total density is given by 
\be
\rho \,=\, \sum_i^n \rho_i \,\,\,.
\ee
As for the individual sequences, for the superposed sequence, we define the probability $P(x)$, such that $P(\rho s) \rho \, ds$ is the probability that a spacing lies between $s$ and $s + d s$. As above, we also define 
\be
\begin{array}{l}
\displaystyle \Psi(x) \,=\, \int_0^x P(y) \, dy \,=\, 1 - \int_x^\infty P(y) dy = 1 -\int_0^\infty P(x+y) \, dy \,\,\,,\\
\displaystyle R(x) \,=\, 1 - \Psi(x) \,=\,\int_x^\infty P(y) \,dy \,=\, \int_0^\infty P(x+y) \, dy \,\,\,,\\
\displaystyle  E(x) \,=\, \int_x^\infty R(y)\, dy \,=\, 
\int_x^\infty dy \int_y^\infty P(z) \,dz\,=\,
\int_0^\infty \int_0^\infty \,P(x + y + z) \, dy\, dz\,\,\,,
\end{array}
\label{W2}
\ee
$E(\rho s)$ is the probability that a given interval of length $s$ does not contain any level and, for the uncorrelated nature of the superposition, we have 
\begin{equation}
E(\rho s) \,=\, \prod_{i=1}^n E_i(\rho_i s) \,\,\,.
\label{eq-porter-E0-sup}
\end{equation}

Posing $x = \rho s$ and 
\be
f_i \,=\,\frac{\rho_i}{\rho} 
\,\,\,\,\,\,\,\,\,
,
\,\,\,\,\,\,\,\,\,
\sum_{i=1}^n f_i \,=\, 1 \,\,\,,
\ee
we have 
\be
E(x) \,=\, \prod_{i=1}^n E_i(f_i x) \,\,\,,
\ee
and for the final expression of the spacing probability $P(x)$, we have 
\begin{equation}\label{poter-formula}
    \begin{aligned}
        P(x) &= \frac{d^2}{d x^2} E(x) \\
        &= E(x) \,
        \left\{\sum_{i=1}^n f_i^2 \,\frac{p_i(f_i x)}{E_i(f_i x)} 
        +\left[\sum_{i=1}^n f_i \frac{R_i(f_i x)}{E_i(f_i x)}\right]^2 - 
        \sum_{i=1}^n \left(f_i \, \frac{R_i(f_i x)}{E_i(f_i x)}\right)^2
        \right\}.
    \end{aligned}
\end{equation}
For a Poisson distribution, we have 
\be
\label{poter-poisson}
R_P(u) \,=\, E_P(u) \,=\, p(u) \,=\, e^{-u} \,\,\,,
\ee
while for a GOE distribution we have 
\be
\begin{array}{l}
\displaystyle R_{GOE}(u) \,=\, e^{-u^2 \pi/4} \,\,\,, \\
\displaystyle  E_{GOE}(u) \,=\, {\rm erfc}\left(\frac{\sqrt{\pi}}{2} u \right)\,\,\,,
\end{array}
\label{W3}
\ee
where 
\[
{\rm erfc}(t) \,=\, \frac{2}{\sqrt{\pi}} \, \int_t^\infty e^{-y^2} \, dy \,.
\]

As for a GUE distribution,
\be
\begin{array}{l}
\displaystyle R_{GUE}(u) \,=\,{\rm erfc}\left( \frac{2}{\sqrt{\pi}}u\right) +\frac{4}{\pi}u{\rm e}^{-4 u^2/\pi}\,\,, \\
\displaystyle  E_{GUE}(u) \,=\, {\rm e}^{- 4u^2/\pi}-u\, {\rm erfc} \left( \frac{2}{\sqrt{\pi}} u \right)\,\,.
\end{array}
\label{W3}
\ee

\begin{figure}[t]
\begin{center}
\includegraphics[width=0.7\textwidth]{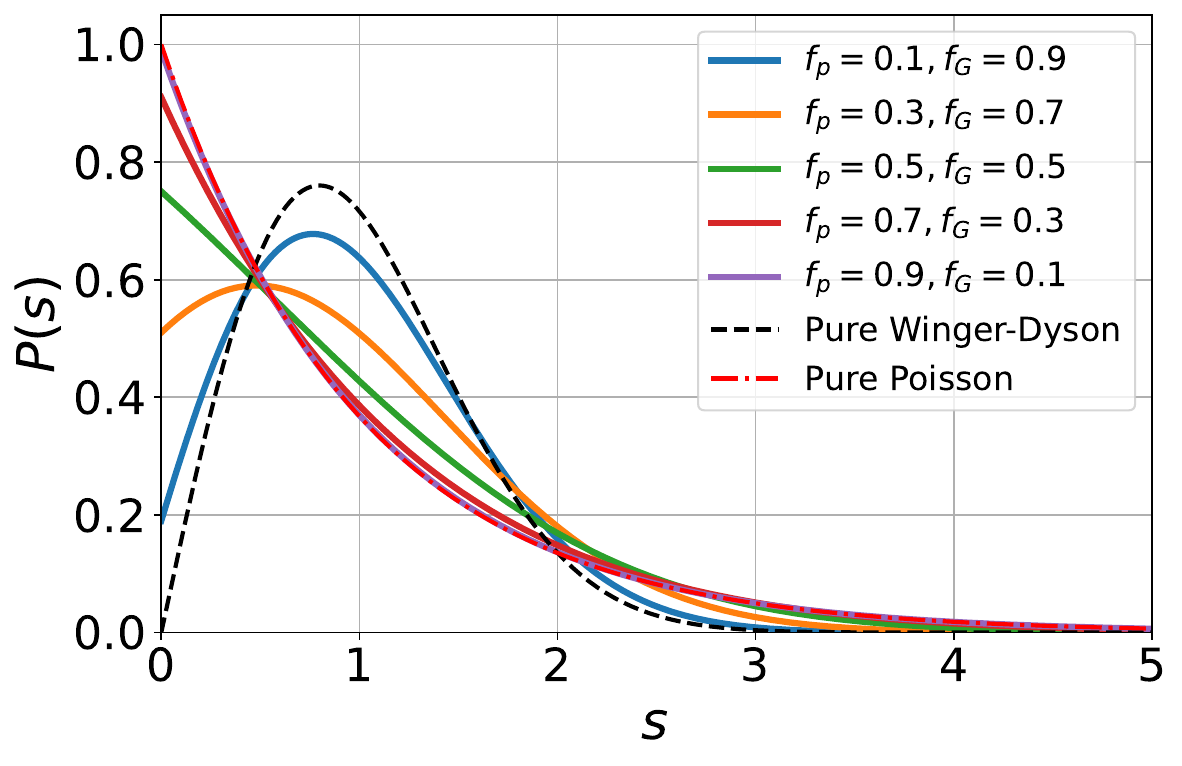} 
\end{center}
\caption{$P(s)$ for a mixture of a Poisson distribution with fraction $f_1$ and a GOE distribution with fraction $f_2 = 1 - f_1$.}
\label{Pois+WD}
\end{figure}
\begin{figure}[h!]
\begin{center}
$\begin{array}{ccc}
\includegraphics[width=0.49\textwidth]{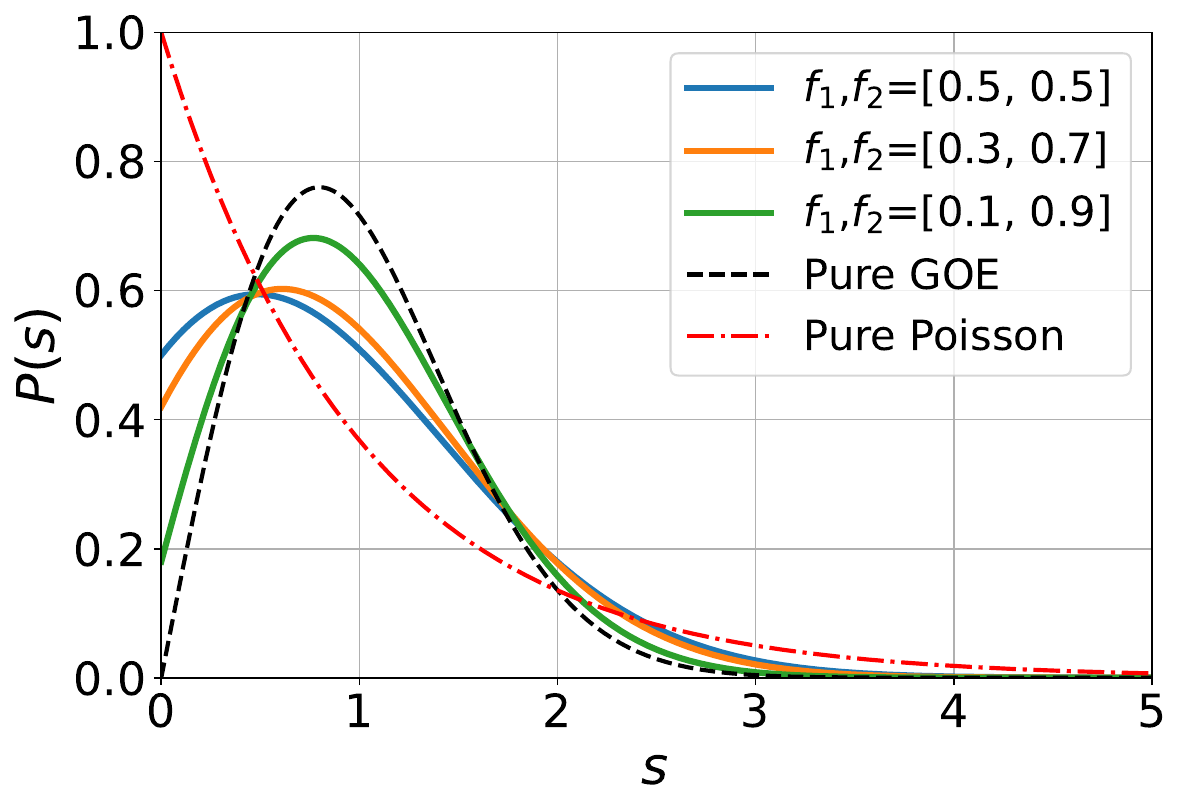} &  
&\includegraphics[width=0.48\textwidth]{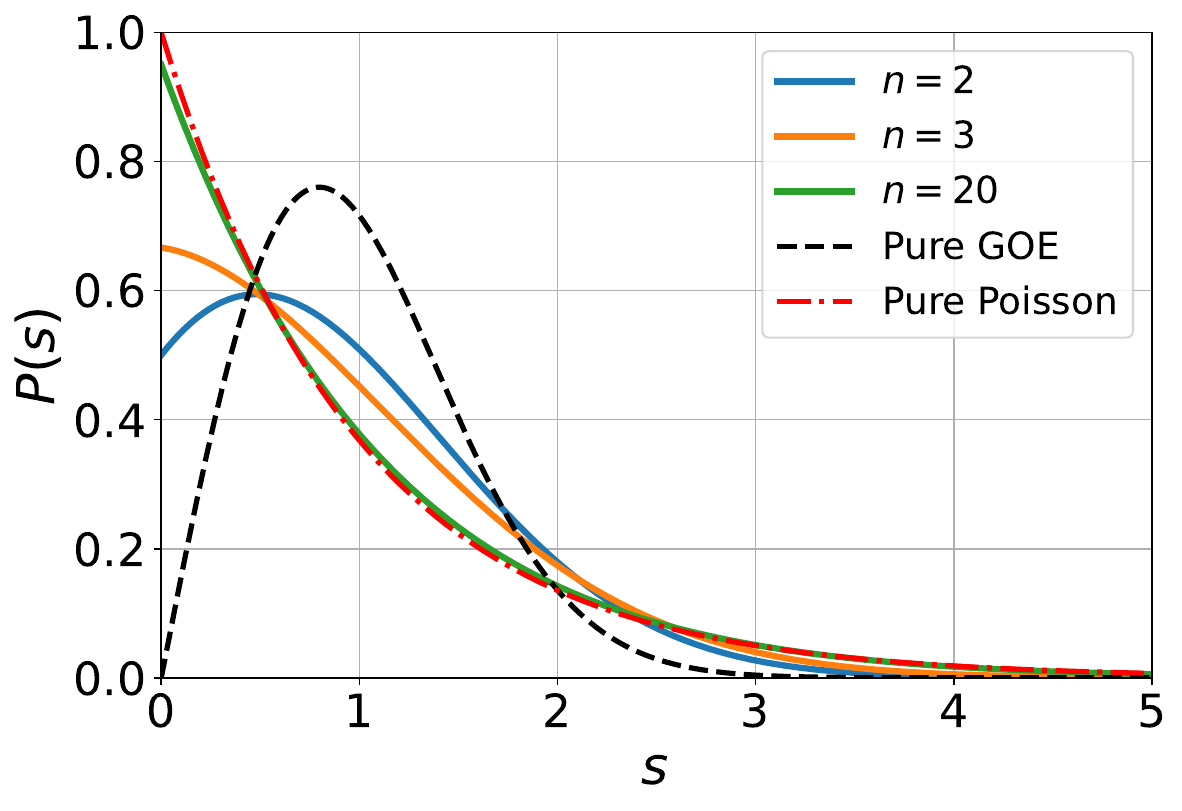} 
 \\
(a) & & (b)  
\end{array}$
\end{center}
\caption{(a)$P(s)$ for a mixture of two GOE distributions with fraction $f_1$ and $f_2$. (b) $P(s)$ for a mixture of $n$ GOE distributions with equal fraction $f_i =1/n$, for $n=2$, $n=3$ and $n=20$.}
\label{poter-GOEs}
\end{figure}

Let's now analyse closely some relevant examples 
\begin{itemize}
\item 
If all $n$ component spectra are Poisson distributed, their superposition likewise yields a Poissonian level–spacing distribution, $P(s) = e^{-s}$, consistent with the 
intuitive expectation that the random superposition of independent random sequences produces a sequence that remains random in character.
\item If the spectrum is made up of two parts, the first with a Poisson distribution with a fraction $f_1$ and the second with a GOE with a fraction $f_2 = 1 - f_1$, varying $f_1$, we have the set of curves shown in Figure~\ref{Pois+WD}. The larger the value of $f_1$, the more pronounced the Poissonian character of the resulting distribution becomes.
\item 
If the spectrum consists of two GOE components with variable relative fractions, the corresponding level–spacing distributions $P(s)$ 
are shown in Figure~\ref{poter-GOEs} (a). Note that, although for a pure GOE one has $P_1(0) = 0$, in the case of a superposition of two GOE spectra, the probability density at the origin instead satisfies $P(0) \neq 0$. Furthermore, as the fraction $f_1$ is increased towards the symmetric partition $f_1 = f_2 = 1/2$,  the value of the curve at the origin changes systematically. As will be discussed hereafter, this represents a generic feature of mixtures of the GOE spectra.
\item let's now consider a spectrum made of a mixture of $n$ GOE.   
Since 
\be
R(0) \,=\, E(0)\,=\, 1\,\,\,,
\ee
we have that at the origin, the probability distribution $P(s)$ is usually different from zero
\be 
P(0) \,=\, 1 - \sum_{i=1}^n f_i^2 \,\,\,,
\label{valuein0}
\ee
with derivative 
\be 
P'(0) \,=\,-1 + 3 \sum_{i=1}^n f_i^2 + \left(\frac{\pi}{2} -2\right) \sum_{i=1}^n f_i^3 
\,\,\,.
\ee
In the case of $n$ sequences with equal fractions $f_i = 1/n$, the resulting distribution approaches the Poissonian form and becomes increasingly indistinguishable 
from it as $n$ grows, see Figure~\ref{poter-GOEs}~(b). Indeed, for $n \rightarrow \infty$, we have $P(0) \rightarrow 1$ and, moreover, for $P(x)$, posing $z=x/n$ we have
\be\label{eq_superposition_GOE}
P(n z) \,=\, E_{WD}^n(z) \left[\frac{1}{n} \frac{P_1(z)}{E_{WD}(z) }+ \left(1-\frac{1}{n}\right) \frac{R^2_{WD}(z)}{E^2_{WD}(z)}\right]\,\,\,. 
\ee
In the limit $n\rightarrow \infty$, i.e. $z \rightarrow 0$, what is within the square brackets becomes equal to $1$, while for the prefactor, expanding it in the series of $z$ we have 
\be 
E_{WD}(z) \simeq 1 - z + \cdots \,=\, 1 - \frac{x}{n} + \cdots 
\ee
and therefore 
\be 
\lim_{n\rightarrow\infty} P(x) \,=\, \lim_{n\rightarrow \infty} \left(1 - \frac{x}{n}\right)^n \,=\, e^{-x}\,\,\,.
\ee
\end{itemize}

In summary, a sufficiently fragmented non-integrable Hamiltonian, composed of many independent Wigner–Dyson subsequences, can mimic a Poissonian spectrum so closely that it becomes practically indistinguishable from that of a genuinely integrable system. This circumstance raises the non-trivial problem of how to unambiguously discriminate between genuine integrability and an apparent Poissonian spectrum arising from the superposition of many chaotic subsequences. We address this problem in the next section, where we see that an important hint may come from the analysis of higher order spacings, discussed hereafter. 

\subsection{Higher order spacing probability distributions}

For the Gaussian ensemble in the RMT, the Fredholm determinant form of a probability generating function of GUE and GOE allows us to calculate the higher-order spacing probability distribution precisely, where the determinant of GUE has already been introduced in section \ref{GUE-determinant}. In this section, we consider the situation of a superposition of several spectra that draw from the same Gaussian ensemble, where the higher-order spacing probability distribution can be obtained exactly via the Fredholm determinant. 

We define a counting function $N([0,s])$, which counts how many energy levels are present in the bulk at $[0,s]$. For the Fredholm  determinant operator $\mathcal{K}$, the probability generating function (PGF) for the number of levels in $[0,s]$ can be written as

\begin{equation}
	D(z;s) \equiv \mathbb{E}\left[z^{N([0,s])}\right]
	= \det\big( I - (1-z){\mathcal K}_{[0,s]} \big)\,\,,
	\label{eq-pgf-single}
\end{equation}
where $\mathbb{E}(X)$ represents the expectation/mean value of the random variable $X$, $\mathcal{K}$ is defined on $L^2([0,s])$. Expanding over $z=0$, the coefficients of the power series give us counting probabilities
\begin{align}
D(z;s) &= \sum_{k=0}^\infty E_{\beta}(k;s)z^k\,\,,\\
E_{\beta}(k;s)&=\Pr\big(N([0,s])=k\big) = \frac{1}{k!} \left. \frac{{\rm d}^{k}}{{\rm d}  z^{k}} D(z;s)  \right|_{z=0}.
\end{align}
The $k$-th level spacing distribution function is 
\begin{equation}
	\label{eq-k-level-spacing-prob}
	P_k^{(\beta)}(s) = \frac{{\rm d}^2}{{\rm d}s^2}\left( \sum_{j=0}^{k-1} (k-j) E_{\beta}(j;s)\right)\,\,.
\end{equation}

We now consider a superposition of $n$ spectra with fractions $f_1,f_2, \cdots,f_n$, \, $\sum_i^n f_i =1$. The counting function that counts the number of levels in $[0,s]$ for a collection of these $n$ spectra is
\begin{equation}
	N_{\mathrm{sup}} ([0,s])= \sum_{i=1}^n N_i([0,f_i s])\,\,,
\end{equation}
i.e., each spectrum $i$ has a contribution at scale $[0,f_is ]$ for the collective spectrum in $[0,s]$. Due to the independence of each individual spectrum, the overall PGF can be factorized into the  product of the PGF corresponding to each component
\begin{equation}
	D_{\mathrm{sup}}(z;s) \equiv \mathbb{E}\left[z^{N_{\mathrm{sup}} ([0,s])}\right]
	= \prod_i^n \det\big( I - (1-z){\mathcal K}_{[0,f_is]} \big)\,\,.
	\label{eq-pgf-sup}
\end{equation}
Therefore, the superposed $k$-level counting probability $E_{\beta}^{\mathrm{sup}}(k;s) $($k=0,1,2,\cdots)$, can be written as
\begin{equation}
	\label{eq-E-sup}
	E_{\beta}^{\mathrm{sup}}(k;s) = \frac{1}{k!} \left. \frac{{\rm d}^{k}}{{\rm d}  z^{k}} D_{\mathrm{sup}}(z;s)  \right|_{z=0}= \sum_{k_1+k_2+\cdots+k_n =k} \prod_{i=1}^m E_{\beta}^{(i)}(k_i;f_is)\,\,.
\end{equation}
One can quickly check that Eq.\eqref{eq-E-sup} reduces to Eq.\eqref{eq-porter-E0-sup} when $k=0$. A similar example for the rules (\ref{eq-pgf-sup}) and (\ref{eq-E-sup}) can be found at \cite{soshnikov2000}.
The $k$-th level spacing probability distribution is
\begin{equation}
	\label{eq-k-level-spacing-prob-sup}
	P_k^{(\beta)}(s) = \frac{{\rm d}^2}{{\rm d}s^2}\left( \sum_{j=0}^{k-1} (k-j) E^{\mathrm{sup}}_{\beta}(j;s)\right)\,\,,
\end{equation}
which is the exact formula we rely on to calculate the higher spacing probability distribution for superposed spectra.

To make the selection rule \eqref{eq-E-sup} more clear, we take an example of a mixture of 3 spectra drawn from the same ensemble-- e.g., GUE or GOE, with fractions $f_1,f_2,f_3$, the counting probability for different $k$ is 
\begin{align}
	k = 0:\quad E_{\beta}^{\mathrm{sup}}(0;s) 
	&= E^{(1)}_{\beta}(0;f_1s)E^{(2)}_{\beta}(0;f_2s)E^{(3)}_{\beta}(0;f_3s)\,\,, \nonumber \\[6pt]
	k = 1:\quad E_{\beta}^{\mathrm{sup}}(1;s) 
	&= E^{(1)}_{\beta}(1;f_1s)E^{(2)}_{\beta}(0;f_2s)E^{(3)}_{\beta}(0;f_3s) \nonumber \\
	&\quad + E^{(1)}_{\beta}(0;f_1s)E^{(2)}_{\beta}(1;f_2s)E^{(3)}_{\beta}(0;f_3s) \nonumber \\
	&\quad + E^{(1)}_{\beta}(0;f_1s)E^{(2)}_{\beta}(0;f_2s)E^{(3)}_{\beta}(1;f_3s)\,\,, \nonumber \\[6pt]
	{k=2}:\quad E_{\beta}^{\mathrm{sup}}(2;s)
	&= \cdots
\end{align}
The $k$-th level spacing probability distribution is 
\begin{align}
	P_1^{(\beta)}(s) &= \frac{{\rm d}^2}{{\rm d}s^2} E_{\beta}^{\mathrm{sup}}(0;s)\,\,, \nonumber \\
	P_2^{(\beta)}(s) &= \frac{{\rm d}^2}{{\rm d}s^2} \left[ 2 E_{\beta}^{\mathrm{sup}}(0;s)+
	E_{\beta}^{\mathrm{sup}}(1;s) \right]\,\,, \nonumber \\
	P_3^{(\beta)}(s) &= \frac{{\rm d}^2}{{\rm d}s^2} \left[ 3 E_{\beta}^{\mathrm{sup}}(0;s)+
	2E_{\beta}^{\mathrm{sup}}(1;s)+E_{\beta}^{\mathrm{sup}}(2;s) \right]\,\,, \nonumber \\
	\cdots ,\nonumber \\
	P_k^{(\beta)}(s) &= P^{(\beta)}_{k-1} + \frac{{\rm d}^2}{{\rm d}s^2} \sum_{j=0}^{k-1} E_{\beta}^{\mathrm{sup}}(j;s) \quad \quad (k=1,2,3,\cdots)\,\,,
\end{align}
where we take $P^{(\beta)}_0(s)=0$.

In the overall PGF Eq.\eqref{eq-pgf-sup} of superposed spectra, we did not specify the determinant kernel $\mathcal{K}_{[0,s]}$, which can actually be the kernel of any Gaussian ensemble ($\beta =1,2$) or Poisson distribution(one should note that the PGF for GOE is a Pfaffian). The $k$-th level counting probability Eq.\eqref{eq-E-sup} and probability distribution Eq.\eqref{eq-k-level-spacing-prob-sup} are, therefore, universal for any superposition of independent spectra drawn from the same Gaussian ensemble. Since the individual spectra composed the mixture are independent of each other, the product rule Eq.(\ref{eq-pgf-sup}) is expected to be universal. Hence, it is valid regardless of whether its constituents are GOE or GUE bulk (with respect to its own determinant kernel) or a Gaussian-Poisson hybrid. 

\begin{figure}[h!]
	\begin{center}
		\includegraphics[width=0.7 \textwidth]{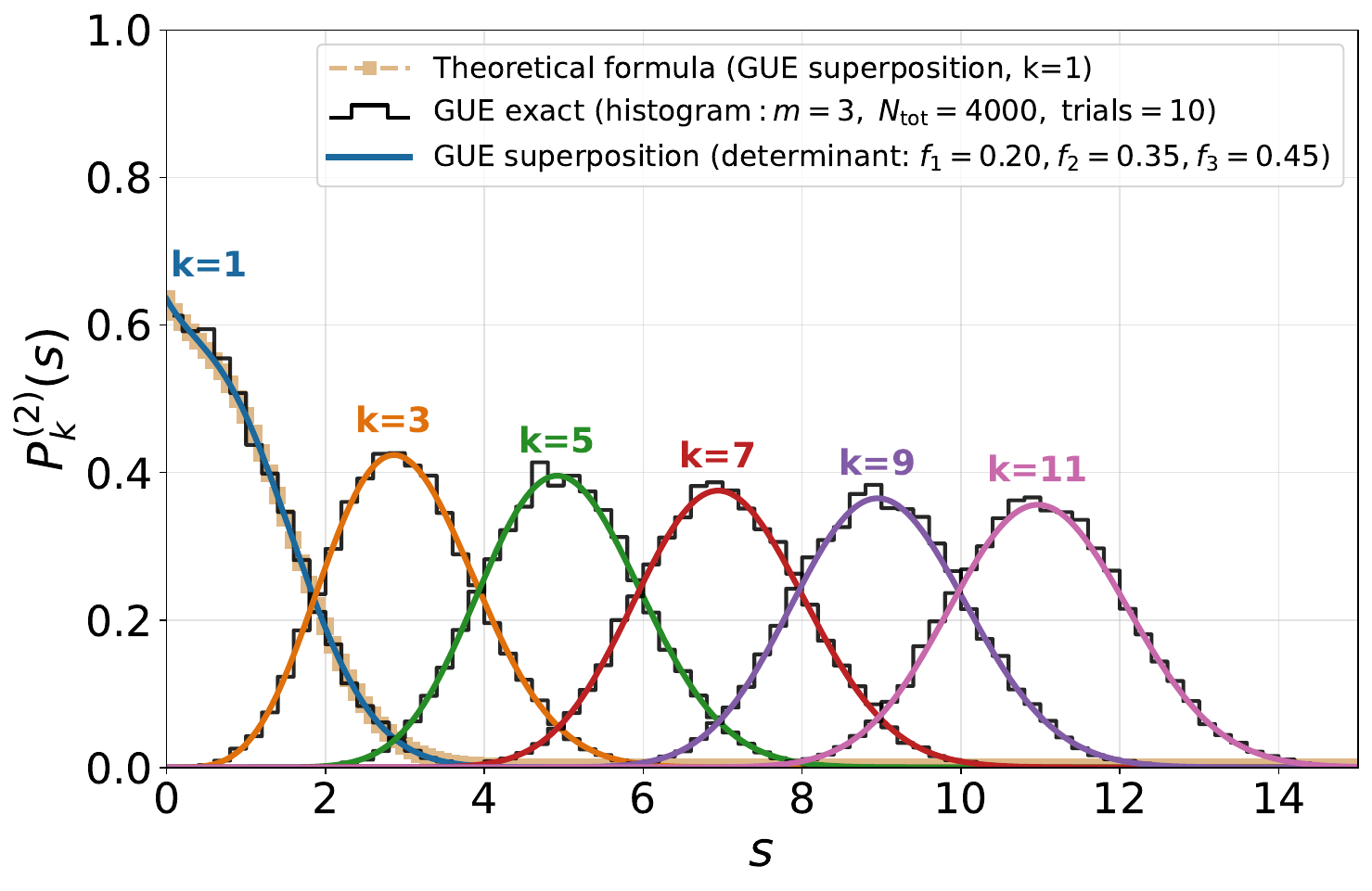}  
	\end{center}
	\caption{$k$-th level spacing probability distribution $P_k^{(1)}(s)$ for superposed 3 GUE spectra with fraction $f_1=0.2,f_2=0.35,f_3=0.45$. The histogram is an average over 10 exact diagonalization realizations for three GOE matrices with total size $4000\times4000$. The theoretical formula is Eq.\eqref{poter-formula} for GUE.  }
	\label{GUE-determinant-ED-nonequal}
\end{figure}

For an exact calculation, in order to know the $k$-th level spacing distribution for a joint spectrum, it is necessary to (i) evaluate the counting probability $E_{\beta}(k;f_is)$ for each individual component on its own fraction; (ii) use  Eq.\eqref{eq-E-sup} to join them together; (iii) obtain the $k$-th level spacing probability distribution in terms of Eq.\eqref{eq-k-level-spacing-prob-sup}. To illustrate this procedure, we show in Figure~\ref{GUE-determinant-ED-nonequal} the higher spacing probability distribution of superposed 3 GUE spectra with non-equal fractions. The exact numerical method that we adopted for computing $k$-th level spacing probability distribution is discussed in Appendix B, where one can also find several examples of superposed GOE spectra.

\section{Protocol\label{S7-protocol}}

The problem we aim to address in this section is the following: given a Hamiltonian matrix $H$ and a chosen resolution $\delta E$ for constructing histograms, can we determine whether the energy levels of $H$ follow a genuinely Poissonian distribution (and are therefore indicative of integrability), or whether we are merely misled by the superposition of a large number of components that produces an apparently Poissonian spectrum? To answer this question, we employ a two–fold protocol: (i) a Monte Carlo decimation procedure designed to filter out levels that may be truly Poissonian; (ii) a comparison between the higher–order statistical distributions of the original spectrum and the corresponding Poissonian predictions.

Let us now discuss these two procedures separately. Concerning the first of them, we will initially describe the Rejection Sampling (RS) subroutine, which will be used extensively in the Spectral Decimation (SD) algorithm. Then we will provide the details of the SD and its heuristic interpretation.

\subsection{Prelude: Rejection Sampling}

\begin{figure}
    \centering
    \subfloat[$\;$]{\includegraphics[width=0.45\linewidth]{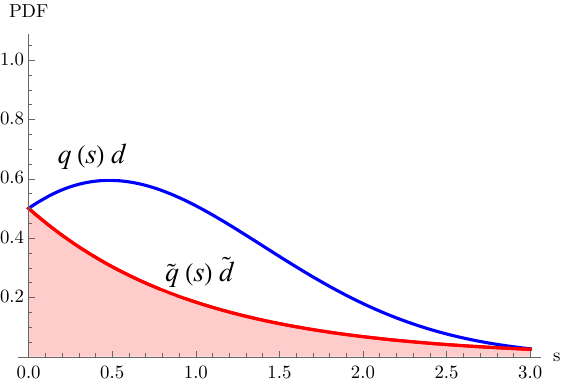}}
    \hfill 
    \subfloat[$\;$]{\includegraphics[width=0.45\linewidth]{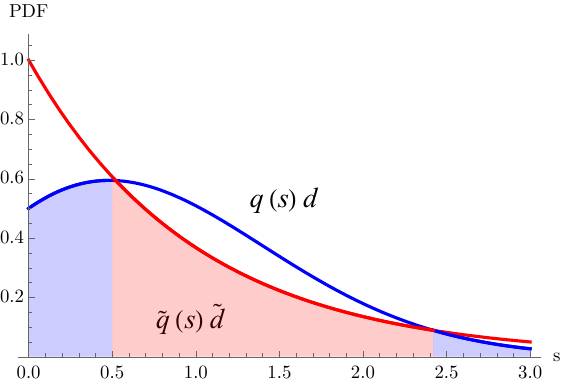}}
    \caption{
  The Rejection Sampling algorithm permits the reconstruction of the target curve $\tilde{q}(s)$, defined on $\tilde{d}$ elements, from the original curve $q(s)$, defined on $d$ elements, only if the rescaled curve $\tilde{q}(s) \tilde{d}$ is entirely contained within $q(s) d$ (as illustrated by the red curve (a)). If this condition is not satisfied, the resulting probability distribution coincides with the portion lying above the coloured region (b). }
    \label{f_rej_sampling}
\end{figure}

The RS algorithm provides a method to construct a target distribution $\tilde{q}(s)$ from samples $\{ s \}$ drawn according to an original distribution $q(s)$. This is achieved by generating an \textit{empirical} distribution $\pi(s)$, which converges to $\tilde{q}(s)$ in the limit of infinitely many samples. 
Such construction is possible only if the target distribution is entirely supported within the original one. In particular, if the target distribution $\tilde{q}(s)$ has $\tilde{d}$ elements and the original distribution $q(s)$ has $d > \tilde{d}$, then $\tilde{q}(s)$ can be exactly reproduced from $q(s)$ without distortion, provided that $\tilde{q}(s) \tilde{d} \leq q(s) d$, for any $s\geq 0$, see Figure~\ref{f_rej_sampling}. 
Thus, the objective is to generate a target distribution $\tilde{q}(s)$ supported on $\tilde{d}$ elements from an original distribution $q(s)$ defined on $d$ elements. Whether the empirical distribution $\pi(s)$ matches the target is not guaranteed by the algorithm, but by the particular choice of $q(s)$, $d$ and $\tilde{d}$.

In what follows, starting from a given list of gaps, we sequentially extract a prescribed number of them and record, for each extraction, the step index $t=1,2,3,… $ at which it occurs. Let $\mathcal S =\{s\}$ denote the initial set of gaps and let ${\mathcal S}_t$ denote the set of gaps available at step t (with ${\mathcal S}_1 = {\mathcal S}$). Let ${\mathcal E}_t$ be the set of gaps extracted up to and including step 
$t$, ${\mathcal A}_t \subseteq {\mathcal E}_t$ the set of accepted gaps and ${\mathcal R}_t \subseteq {\mathcal E}_t$ the subset of rejected gaps. These sets satisfy
\be
\mathcal{E}_t=\mathcal{A}_t \,\dot\cup\, \mathcal{R}_t,\qquad 
\mathcal{A}_t\cap \mathcal{R}_t=\varnothing,
\ee
\be
\mathcal{S}_t=\mathcal{S}\setminus \mathcal{E}_{t-1},\qquad 
\mathcal{S}_{t+1}=\mathcal{S}\setminus \mathcal{E}_t,
\ee
\be
\mathcal{E}_0=\mathcal{A}_0=\mathcal{R}_0=\varnothing,\qquad
|\mathcal{E}_t|=t\quad\text{for } t=1,2,3,\ldots
\ee
By construction, one gap is extracted per step, and this leads to  
$|{\mathcal E}_t| = t$.

At the step $t$, let $\tilde{s}_t$ be a value extracted uniformly 
from ${\mathcal S}_t$. It will be accepted if a uniformly distributed random number $u$ taken from the interval $(0,1]$ satisfies
    \begin{equation}\label{check}
        u \leq r(\tilde{s}_t) = M \, \frac{\tilde{q}(\tilde{s}_t)}{q(\tilde{s}_t)} ,
    \end{equation}
where $M$ denotes a constant ensuring the validity of the inequality $M \tilde q(x) \leq q(x)$ for any $x$. Hence, the parameter $r$ serves as the ``acceptance threshold'', while $M$ has the meaning of the acceptance probability\footnote{Indeed, the total probability of accepting a proposed $x$ is the expectation of $r(x)$ over the initial distribution $q(x)$, $\eta = {\mathbb E}_{x\in q}[r(x)]$. Therefore, $\eta = \int r(x) q(x) dx  = \int M \frac{\tilde{q}(x)}{q(x)} q(x) dx = M \int {\tilde q}(x) dx = M$.} per trial. Hence, if Eq.~\eqref{check} is satisfied, $\tilde{s}_t$ will be added to ${\mathcal A}_t$, forming the new ${\mathcal A }_{t+1}$; otherwise, it will be added to ${\mathcal R}_t$, making the new ${\mathcal  R }_{t+1}$.

Since the initial distribution contains $d$ elements, at most 
$d$ iterations can be executed. If, at some iteration $t_* \leq d$,  
the required $\tilde{d}$ elements are identified, the algorithm terminates and is deemed \textit{successful}. Conversely, if all 
$d$ iterations are exhausted without locating the $\tilde{d}$, the algorithm is said to \textit{fail}.

In our implementation, the constant $M$ is chosen to be $M = q(0)$.  
A crucial observation is that if the original distribution is the Wigner-Dyson probability density function (hence $M = 0$), while the target distribution satisfies $\tilde{q}(0) \neq 0$, then the RS algorithm invariably fails. On the other hand, if the original and target distributions coincide, i.e.\ $q = \tilde{q}$, then choosing $M=1$ the RS algorithm reproduces the same distribution as uniform random sampling. 

A closely related modification of the RS is the well-known 
\textit{Metropolis-Hastings (MH) algorithm}, a general method for sampling from a probability distribution $q(s)$ defined over a finite set $\{s\}$ of $d$ elements. The goal is to obtain a representative 
subset of $\tilde{d} < d$ samples. The algorithm proceeds as follows. Starting from an initial element 
$s_0 \in \{s\}$, one generates at iteration $t$ a candidate element 
$\tilde{s}$ from a proposal distribution $p(\tilde{s}\,|\,s_t)$. 
The candidate is then accepted with probability  
\begin{equation}
    \alpha(s_t \to \tilde{s}) 
    = \min\!\left\{ 1, \; 
    \frac{q(\tilde{s})\,p(s_t \,|\, \tilde{s})}
         {q(s_t)\,p(\tilde{s}\,|\,s_t)} \right\}.
\end{equation}
If the candidate is accepted, the new state is set to 
$s_{t+1} = \tilde{s}$; otherwise, the chain remains at 
$s_{t+1} = s_t$.  
In the simplest symmetric case (the one which we will use later), where the proposal distribution satisfies 
$p(\tilde{s}\,|\,s_t) = p(s_t \,|\,\tilde{s})$, the acceptance probability 
reduces to  
\begin{equation}
    \alpha(s_t \to \tilde{s}) 
    = \min\!\left\{ 1, \; \frac{q(\tilde{s})}{q(s_t)} \right\}.
\end{equation}
It follows that if the candidate $\tilde{s}$ is more probable than 
the current state $s_t$, i.e.\ $q(\tilde{s}) \geq q(s_t)$, 
then the move is always accepted. Conversely, if 
$q(\tilde{s}) < q(s_t)$, the move is accepted with probability 
$q(\tilde{s})/q(s_t)$.  
This conditional acceptance mechanism guaranties that the Markov chain 
constructed by the MH algorithm has $q(s)$ as its stationary distribution, 
thereby allowing one to extract a representative sample of size $\tilde{d}$.

\subsection{Spectral Decimation Algorithm}

The \textit{spectral decimation} algorithm is designed to test whether the energy spectrum of $d$ levels is uncorrelated — i.e., whether the gap statistics follow a genuine Poisson distribution $P(s) = e^{-s}$ 
— or instead represents a mixture of spectra, made, for instance, by several non-integrable spectra characterized by Wigner–Dyson statistics. In the latter case, the algorithm further enables the determination of the number of contributing spectra of the mixture.

\subsubsection*{Input and Halting Conditions}
We assume that energies are already unfolded and therefore gaps are simply $s_i = e_{i+1}-e_{i}$: these are the input data of the algorithm. In principle, the number of levels $d$  
may be exponentially large, making a direct test of whether all gaps follow Poisson statistics both difficult and computationally inefficient. Fortunately, a smaller representative sample of 
${\tilde d}$
 elements can always be extracted from the same underlying distribution using the Metropolis–Hastings algorithm described above. Hence, the analysis may be performed either on the complete set of unfolded gaps or on a suitably chosen subset.
 Hereafter, we shall use the symbol $d$ 
indistinguishably to denote either the complete set of unfolded gaps or a representative subset thereof.

\begin{figure}
    \centering
    \includegraphics[width=0.5\linewidth]{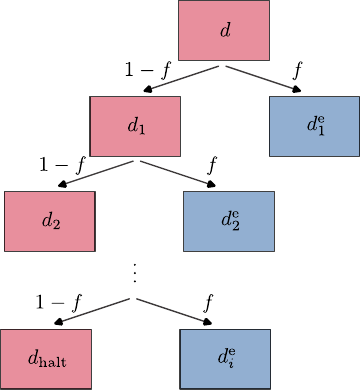}
    \caption{Tree structure of the spectral decimation algorithm. At each iteration, a fraction $f$ is decimated from the spectrum. The algorithm ultimately stops when a size $d_{\text{halt}}$ is reached, unless a halting condition is met.}
    \label{f_illustration_decimation}
\end{figure}

In our implementation of the SD algorithm, $q(s)$ denotes the initial empirical distribution of the $d$ gaps, while the target distribution ${\tilde q}(s)$ is chosen to be the Poisson law ${\tilde q}(s) = e^{-s}$. We can run the SD algorithm using an \textit{extraction fraction} $f$, responsible for the rate of decimation. Namely, if $d_k$ gaps are present at iteration $k$, with $d_1 = d$, then $d_k^{\text{e}} = f d_k$ is the target number of gaps at iteration $k$ required to follow the Poisson distribution. If such a number of gaps is found, the remaining gaps $d_k^{\text{r}} = (1-f) d_k = (1-f)^k d$ will be those used in the next iteration, $d_{k+1} := d_k^{\text{r}}$. See Figure~\ref{f_illustration_decimation} for a schematic visualization of the tree structure of the algorithm.

Hence, the algorithm runs as follows: chosen a final number\footnote{
For the validity of our statistical analysis, it is essential that both $d$ and $d_{\text{halt}}$  be sufficiently large. In practice, typical values are $d \sim 10^6$ and $d_{\text{halt}} \sim 10^3$.} $d_{\text{halt}}$ of gaps and a fraction $f$ of decimation (typically on the order of $f\sim 10^{-2}$), at the $k$-th step of decimation, with $d_k^{\text{r}} > d_{\text{halt}}$, there could be two situations:   
\begin{itemize}
    \item {\textit{Successful finding}}: if, after $t_* < d_k$ iterations, the target number $d_k^{\text{e}} = f d_k$ of gaps consistent with the Poisson law is found, then the remaining gaps $d_k^{\text{r}}$ will be used in the next iteration, $d_{k+1} := d_k^{\text{r}}$, until eventually reaching the number $d_{\text{halt}}$ chosen as a final number of gaps where the algorithm definitely stops. 
    \item {\textit {Unsuccessful finding}}: if all $d_k$ iterations are exhausted without identifying the required $d_k^{\text{e}} = f d_k$ gaps satisfying the Poisson distribution, then the algorithm halts.
    \end{itemize}
Let's discuss the interpretation of these two outputs with the example of an initial distribution of the gaps obtained by joining $N$ non-integrable Wigner-Dyson spectra, each of length $\delta$. The probability density of zero gaps is equal to \cite{Porter2}
\begin{equation}
    q(0) = 1-\frac{1}{N}\,\,,
\end{equation}
and therefore almost indistinguishable from a Poissonian for large enough $N$. The initial sample will then have $d = N \delta$ gaps. Imagine that we now apply the RS algorithm to extract a fraction $f\ll q(0)$ of the original gaps. In infinitesimal interval $[0, \eta]$ nearby the origin, the number of zero-gaps 
\begin{equation}
    d^{(\eta)} = d q(0) \eta
\end{equation}
will then be decimated to
\begin{equation}
    d^{(\eta)}_1 = d q(0)  \eta - d f \eta = d\br{1-\inv{N} - f}  \eta.
\end{equation}
Crucially $d^{(\eta)}_1 < d_{\phantom{1}}^{(\eta)}$ and this justifies the name ``decimation'' for the algorithm. After the $m$-th decimation, the number of zero-gaps will be
\begin{equation}
    d_m^{(\eta)} = d \br{1-\inv{N} -  f \sum_{k=0}^{m-1} (1-f)^k}  \eta = d \br{(1-f)^m - \inv{N}}  \eta.
\end{equation}
The algorithm is said to have decimated all zero gaps if 
\begin{equation}
    m \geq m^\star = - \frac{\log N}{\log (1-f)}.
\end{equation}
The latter equation makes clear the role of the extraction fraction $f$ as a controller of the rate of decimation. Moreover, at iteration $m = m^\star + 1$, the SD will terminate because we have exhausted all possible zero gaps necessary to have a Poisson distribution, a value which could be before the final halting condition is fulfilled, i.e. when $(1-f)^{m^\star} d > d_{\text{halt}}$.

\subsubsection*{More Details on the Algorithm }

At the $k$-th step of the decimation tree — conditional on the success of step $(k-1)$ -- the SD algorithm proceeds as follows:
\begin{enumerate}
    \item \textbf{Empirical Distribution}.
    From the starting sequence of $d_t = (1-f)^t d$ of the new unfolded gaps $\{s_i\}_t$, construct the empirical distribution $q_t(s)$. 
    
The resolution $\delta_t$ of the histogram is fixed using the Freedman-Diaconis rule \cite{FreedmanDiaconis} for the Poisson PDF:
    \begin{equation}
        \delta_t = \left(\frac{d_t}{6}\int_0^\infty \left( \frac{\dd}{\dd s} p(s)\right)^2 ds \right)^{-1/3} 
        = \left(\frac{12}{d_t}\right)^{1/3}.
    \end{equation}
    For the typical case $d = 10^5$, one finds $\delta \approx 8 \times 10^{-3}$.

    \item \textbf{Decimation}.  
    Apply the RS algorithm to extract $d_t^{\text{e}} = f d_t$ gaps following the Poisson distribution. The constant $M$ of the RS is fixed to $M = q_i(0)$, so that zero-gaps are almost surely accepted.

    \item \textbf{First Check}. If the RS algorithm is successful, the algorithm continues at the next $k+1$ step of decimation; otherwise, it stops.

    \item \textbf{Unfolding}. If the RS algorithm is successful, remove the set of extracted gaps, $\{\tilde{s}_i\}_k$, that follows the Poisson distribution, from the original gaps $\{s_i\}_k$. The original ordering of the surviving gaps $\{\sigma_j\}_{k+1} = \{s_i\}_k \setminus \{\tilde{s}_i\}_k$ is preserved.  

    Because the starting gaps are already unfolded, the unfolding of the $\{\sigma_j\}$ is quite simple: it is only sufficient to divide by the mean $\overline{\sigma}$ to obtain the new set 
    \begin{equation*}
    \{s_{i}\}_{k+1} = \left\{\frac{\sigma_i}{\overline{\sigma}}\right\}_{k+1},
    \end{equation*}
    which are $d_t^{\text{r}} = (1-f) d_t$ in number. 

    \item \textbf{Second Check}. If $d_t^{\text{r}} > \tilde{d}_{halt}$, use $\{s_{i}\}_{t+1}$ as the set of gaps for the successive iteration of the SD algorithm. Otherwise, if $d_t^{\text{r}} \leq  {\tilde d}_{\text{h}}$, the algorithm terminates. In this case, the \textit{SD has met the halting threshold ${\tilde d}_{\text{halt}}$.}
\end{enumerate}

\subsubsection*{Physical Interpretation}

By construction, the SD algorithm terminates in two possible ways and it is useful to analyse their implications. 

In the first way, namely if the SD algorithm reaches the halting threshold $d_{\text{halt}}$, it provides a positive, though probabilistic, indication of the uncorrelated nature of the spectrum. In this case, the decimation has reduced the original set of $d$ gaps to a subset of size  ${d}_{\text{halt}}$. For ${d}_{\text{halt}} \ll d$, 
  this implies that, to the best of our numerical resolution, the original spectrum is uncorrelated. More precisely, the number of uncorrelated gaps is at least  $d-{d}_{\text{halt}}$, 
  namely those identified by the SD procedure. One may then define the \textit{fraction of uncorrelated gaps} as
  \be
  f_u \,=\, (1-d_{\text{halt}}/d)\,\,\,.
  \ee
  This fraction can be fixed in advance as a target for the SD algorithm. In this sense, the spectral decimation acts as a \textit{certificate} for the presence of uncorrelated energies.

  Let's now discuss the second way the algorithm can terminate. 
In this case the algorithm may also halt prematurely at step $m$
of the decimation tree. In this event, the RS procedure provides positive—albeit probabilistic—evidence for the presence of multiple independent spectra. Suppose the algorithm halts at iteration $m$, which implies that iteration$(m-1)$ was successful. This occurs because, after the 
$m$-th application of RS, the reservoir of zero gaps is exhausted.  Consequently, the number of independent spectra $N$ can be estimated as 
\begin{equation}
    \left(\frac{1}{1-f}\right)^{m-1} \leq N \leq \left(\frac{1}{1-f}\right)^{m}.  
\end{equation}
This bound can be made arbitrarily small by selecting a sufficiently small extraction fraction $f$. Conversely, when the SD algorithm reaches the halting threshold ${\tilde d}_h$ after $m$ 
 iterations, one can establish a lower bound on the number of copies
 
\begin{equation}
    N\geq \left(\frac{1}{1-f}\right)^{m}.
\end{equation}

\begin{figure}
    \centering
    \includegraphics[width=0.75\linewidth]{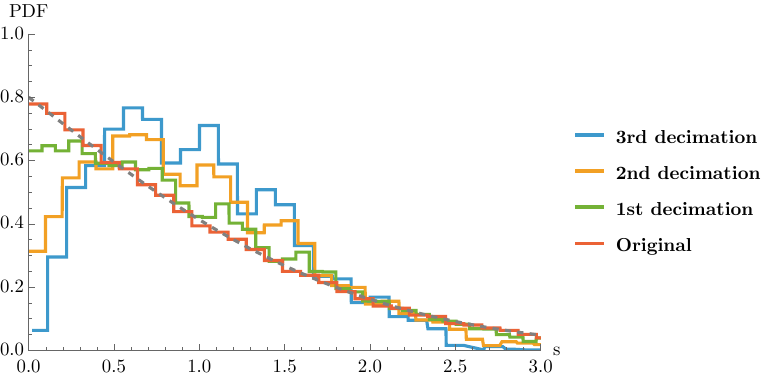}
    \caption{Application of spectral decimation.  
    The empirical PDF is obtained by superposing five Wigner--Dyson distributions generated from GOE spectra of size $d=10^5$.  
    The dashed line represents the analytical PDF from \cite{Porter2, BerryRobnik}, Eq~\eqref{eq_superposition_GOE}.  
    The decimation output after three iterations ($f=1/2$) is also shown: the probability of zero-gaps decreases as they are progressively depleted.}
    \label{f_decimation}
\end{figure}

This second halting way of the algorithm is illustrated in Figure~\ref{f_decimation}, which shows the decimation of five Wigner-Dyson spectra generated from GOE ensembles of size $d=10^5$. With $f = 1/2$, the algorithm halts after three iterations when the RS algorithm fails.

Notice that selecting different values of $f$ 
inevitably leads to different failure iterations 
$m_f$ of the SD algorithm. Hence, the number of iterations itself is not physically meaningful. Reversing the argument, however, the iterations cannot be arbitrarily large if the spectrum consists of $N$ independent Wigner–Dyson components. Indeed, the right-hand side of the previous equation shows that the SD algorithm admits at most
\begin{equation}
    m = O \left(-\frac{\log N}{\log(1-f)}\right)
\end{equation}
iterations, which do not depend on the initial $d$. 

However, the dependence on $d$ is taken into account by the computational cost of one iteration. Assuming a single RS acceptance/refusal to be $O(1)$, the $m$-th iteration of the SD will have polynomial runtime. The best-case scenario is when exactly $d_m^{\text{e}} = f(1-f)^{m-1} d$ calls are made, i.e. $O(d_m^{\text{e}})$, while the worst-case scenario is when all the initial set of gaps is spanned, $O(d_m)$. As we already mentioned, the empirical distribution can be sampled without distortions by a simple RS involving $O(\tilde{d})$ calls, where $\tilde{d}$ is arbitrarily fixed.

In summary, the SD provides answers in both outputs, which have net statistical interpretations that can be made arbitrarily precise by fine-tuning the input parameters $f$, $\tilde{d}_h$. 

\subsection{Higher order spacings}

\begin{figure}[h]
\begin{center}
\includegraphics[width=0.7 \textwidth]{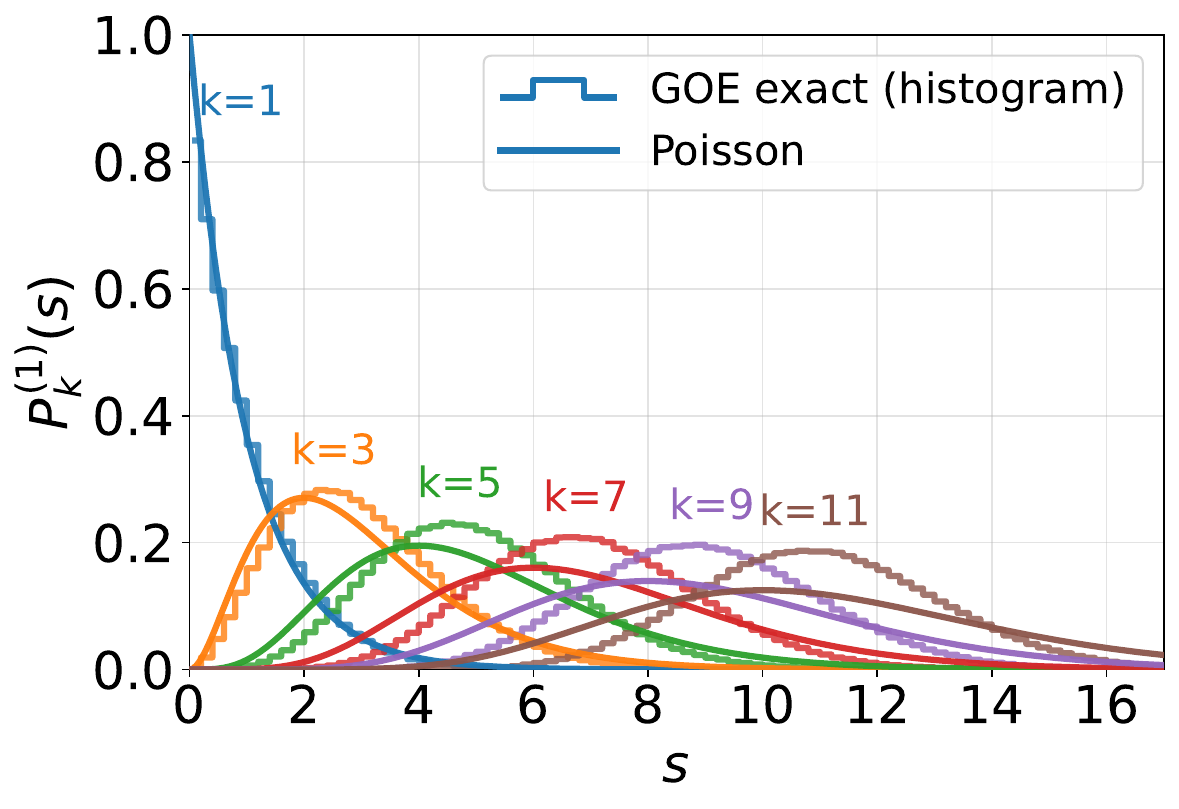} 
\end{center}
\caption{$k$-th level spacing probability distribution $P_k^{(1)}(s)$ for Poisson distribution and a superposition of 10 GOE spectra, where the histogram is an average over 200  realizations of 10 GOE spectra, each from an exact diagonalization of a $400 \times 400$ matrix. }
\label{GOE-ED-Poisson}
\end{figure}

An additional and highly effective diagnostic for distinguishing between a genuine Poisson distribution and a Poisson-like distribution arising from a mixture of spectra is provided by the analysis of higher–order level spacings. In the previous sections, we computed these distributions both for the purely Poisson case and for spectral mixtures, and significant discrepancies emerge that allow one to discriminate between the two situations. For instance, Figure~\ref{GOE-ED-Poisson} displays the higher–order spacing distributions for a mixture of $10$ spectra compared to those of a purely Poisson ensemble. The two sets of curves are clearly distinguishable, at least until a moderate number of spectra in the mixture. Naturally, as the number of superposed spectra tends to infinity, these differences become increasingly subtle and eventually imperceptible, in which case, it is more advantageous to rely on the Monte Carlo decimation method.

\section{Permutation Hamiltonians}\label{s_permutation}
In this section, we describe Hamiltonians based on permutations, i.e. the elementary interaction features the swap of sites. They can be generated on graphs and can feature different boundary conditions.

\subsection{Generalities on the Permutation Group}

In what follows, we focus on the archetype of all finite groups: the permutation group. The purpose of this presentation is not to provide a comprehensive treatment of this group but rather to supply the essential background needed for the numerical analysis presented later in this section. Further details on the structure of the permutation group are collected in Appendix~\ref{a_permutations}, while a more systematic discussion of permutation Hamiltonians will be provided in a forthcoming dedicated publication.

The permutation group of $N$ elements, denoted $\mathcal{S}_N$ is a finite group of order $\lvert \mathcal{S}_N \rvert = N!$ generated by $N-1$ elementary (adjacent) transpositions $P_{i,i+1}$, $i \in \cbr{1, \ldots, N-1}$. A transposition $P_{i,j}$ is an element of the permutation group that exchanges, out of $N$ distinct objects, the ones labelled $i$ and $j$. In terms of adjacent transpositions, $P_{i,j}$ can be obtained by recursion:
\begin{equation}
    P_{i, i+2} = P_{i+1, i+2} P_{i, i+1} P_{i+1, i+2}\,\,.
\end{equation}
All permutations admit a ``word'' in terms of generic transpositions $P_{i,j}$, which may not be unique. Moreover, an elementary word contains only adjacent transpositions $P_{i,i+1}$. This word is said to be minimal if it contains the least amount of $P_{i,i+1}$. For example, for $P_{i, j}$, the minimal word has $2|i-j|-1$ adjacent transpositions.

Each element of the permutation group belongs to a conjugacy class $\mathcal{C}_\alpha$, which is identified by the number of distinct cycles. For example, for $N=9$, the conjugacy class $\sbr{3^2,2,1}$ contains two different cycles where three elements are permuted ($3$-cycle), a $2$-cycle and an element that is not permuted (one-cycle). Transpositions are therefore $2$-cycles. The number of conjugacy classes equals the number $p(N)$ of integer partitions of $N$, whose asymptotic behaviour is given by Eq.~\eqref{eq_integer_partitions_asymptotics}. The conjugacy classes of the permutation group can be represented by Young diagrams, which are also in one-to-one correspondence with the group’s Irreducible Representations. Young diagrams provide a natural and powerful framework for studying irreducible representations, as they encode the combinatorial rules required to compute fundamental quantities. Among the most significant is the dimension of a representation, given by the celebrated hook-length formula, recalled in Appendix~\ref{a_permutations}. For example, the fully symmetric representation $[L]$ and the fully antisymmetric representation $[1^L]$ are both one-dimensional.

A permutation Hamiltonian is a generic linear superposition of permutation elements
\begin{equation}
    H = \sum_{P \in \mathcal{S}_N} \alpha_P \br{P+ P^{-1}}.
\end{equation}
The inclusion of the inverse guaranties the Hermiticity of the operator. Furthermore, it is not necessary to consider higher powers of permutation elements, as the symmetric group $\mathcal{S}_N$ is, by definition, closed under multiplication and already contains all such products. The coupling constants $\alpha_P$ may be treated either as fixed real parameters or as random variables drawn from a prescribed probability distribution.

We consider Hilbert spaces in which each site can be locally occupied by one of $n$ bosonic states, distinguished by their ``colours''. Locally the state will be denoted as 
$|v_a\rangle_i$, where $i$ refers to the site while $a = 1, 2, \ldots n$ refers to the colour index. Consequently, each site transforms according to the fundamental representation of $SU(n)$, and permutation Hamiltonians act by permuting the sites and their superpositions. For a chain of length $L$, the Hilbert space has dimension $d = n^L$, which can be decomposed into symmetry sectors, i.e. ``colour sectors'' $\cbr{L_1, L_2, \ldots L_n}$, where $L_j$ is the number of particles of colour $j\in  \cbr{1, \ldots, n}$ subject to $\sum_{j=1}^n L_j =L$. 
Within these colour sectors, permutation Hamiltonians remain invariant. More precisely, let $U \in SU(n)$ denote an $n \times n$ unitary transformation representing colour mixing. Its tensor product $U^{\otimes L}$ acts on the full Hilbert space as a $d \times d$ matrix and commutes with the Hamiltonian, i.e.
$\br{U^{\otimes L}}^{-1} H U^{\otimes L} = H$. 

In a given colour sector, transpositions $P_{i,j}$ are represented in the computational basis—namely, the product basis where each site is assigned a specific colour—-as sparse matrices with a single nonzero entry per row. Remarkably, within these colour sectors the Hamiltonian admits a further decomposition into irreducible representations of $\mathcal{S}_L$. Since permutations among sites carrying the same colour leave the state invariant, bosonic many-body wave functions can be labelled by irreducible representations of the subgroup
$\mathcal{S}_{L_1}\otimes \mathcal{S}_{L_2} \otimes \ldots \otimes \mathcal{S}_{L_n} \subset \mathcal{S}_L$
where $L_i$ denotes the number of sites of colour $i$. These representations are preserved by the Hamiltonian, which thus acquires a fragmented block structure as 
in the figure 
\begin{equation}
  H \,=\, \vcenter{\hbox{\includegraphics[width=0.25\linewidth]{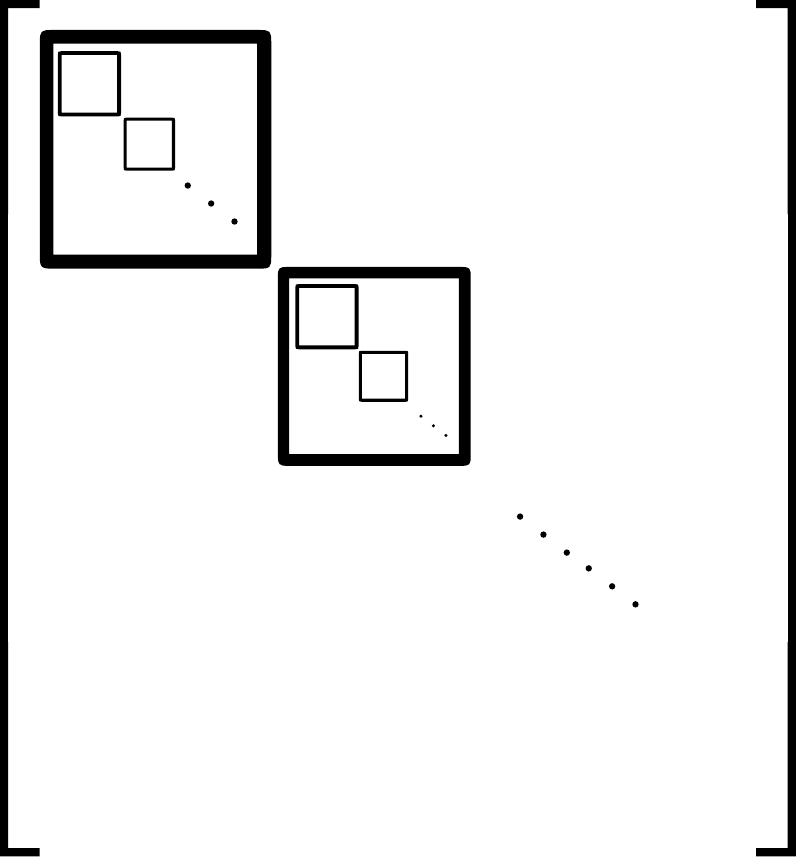}}}\; .
\end{equation}

Within each irreducible representation, the adjacent transpositions $P_{i,i+1}$ can be represented as real, unitary, and Hermitian matrices \cite{natafmila}. Further details regarding their numerical implementation are provided in Appendix~\ref{a_permutations}.

\subsection{Hamiltonians on Graphs}

In permutation Hamiltonians, the interactions act solely on the sites, irrespective of the specific structure of the local Hilbert space. This property makes it possible to investigate a wide variety of models while preserving both the dimensionality and the intrinsic structure of the symmetry blocks.

As a first example, consider Hamiltonians defined on graphs $G$. Here, the nodes represent sites, and the links correspond to permutations. 
The degree of a node is defined as the number of links (or edges) incident upon that node. A simple undirected edge connecting sites $i$ and $j$ is represented by the transposition $P_{i,j}$, and the Hamiltonian can be directly constructed from the adjacency matrix $A_{ij}$ of the graph, defined as follows: for a graph $G = (V,E)$ with $|V| = N$ vertices, the adjacency matrix is the $N \times N$ matrix $A = (a_{ij})$ defined by
\[
a_{ij} \;=\;
\begin{cases}
1 & \text{if there is an edge between vertices $i$ and $j$\,\,,} \\[6pt]
0 & \text{otherwise\,\,.}
\end{cases}
\]
For undirected graphs without self-loops (the case we are interested in), $A$ is real and symmetric, with vanishing diagonal entries. The Hamiltonian $H_G$ on a graph $G$ is then given by 
\be
H_G \,=\,\sum_{i < j} J_{i j} \, A_{i j} \, P_{i j}\,\,,
\label{genericGraphHamiltonian}
\ee
where the coupling constants $J_{i j}$ are either fixed real numbers or random variables drawn from a given probability distribution.

\subsubsection*{Sutherland Permutation Hamiltonians}
The minimum number of edges that generates a graph with $L$ nodes is $L-1$ and the Hamiltonian
\begin{equation}
    H_2^{o} \,= \, \sum_{i=1}^{L-1} J_{i}\, P_{i, i+1}
\end{equation}
corresponds to the Sutherland permutation model with open boundary conditions. 
The same Hamiltonian with periodic boundary conditions
\begin{equation}
    H_2^{p} = \sum_{i=1}^{L} J_i\,P_{i, i+1 \operatorname{mod} L}\,\,,
\end{equation}
is the graph where all nodes have incidence 2. In both Hamiltonians, the coupling 
constant $J_i$ may be either fixed real parameters or random variables drawn from a prescribed probability distribution. Consider the homogeneous case in which $J_i = J, \,
\forall i$. If $|a\rangle_i$ and $| b \rangle_i$ denote two generic colours at the site $i$, the individual operator $P_{i,i+1}$ has the following set of eigenvalues and eigenvectors
\be
P_{i,i+1}\,\left(\frac{|v^a_i \, v^b_{i+1}\rangle \pm | v^b_i \,v^a_{i+1}\rangle}{\sqrt{2}}
\right) \,=\, \pm \, \left(\frac{|v^a_i \,v^b_{i+1}\rangle \pm |v^b_i \,v^a_{i+1}\rangle}{\sqrt{2}}
\right)\,\,.
\ee
Therefore, the symmetric combination of colours of the neighbouring sites has the highest eigenvalues, while the antisymmetric one (if $a \neq b$) has the minimal eigenvalue. Hence, if $J < 0$, the ground state eigenstate of the Sutherland Permutation Hamiltonians is given by the fully symmetric representation,
\be
| SG \rangle \,=\, \sum_{P\in S_N} |v^{a_1}_{P(1)} v^{a_2}_{P(2)} \ldots v^{a_N}_{P(N)}\rangle \,\,\,.
\ee
It is a different story if $J$ is instead positive $J > 0$. Indeed, the minimum energy $E^*= - JL $ is obtained when each permutation operator $P_{i,i+1}$ can simultaneously take the value $-1$; this is, however, only possible if the number of different colours $n$ is equal to the number of sites $L$. If this is the case, the relative state corresponds to the totally anti-symmetric IR (corresponding to the vertically longest Young Tableau) and can be written as a Slater determinant built in terms of any set of $L$ different colours $|v_a\rangle_i$ 
\EQ
| v^{(1)},\ldots,v^{L} \rangle_- =\frac{1}{\sqrt{L!}} 
\left| 
\begin{array}{ccc}
|v^{(1)}\rangle_1 & \cdots &|v^{({L})}\rangle_1\\
|v^{(1)}\rangle_2 & \cdots &  |v^{({L})}\rangle_2\\
\cdots & \cdots &  \cdots  \\
|v^{(1)} \rangle_{L} & \cdots & |v^{({L})}\rangle_{L}
\end{array}
\right|\,\,\,.
\label{Slaterdeterminant}
\EN
However, for all other colour sectors, it is impossible to satisfy the minimum of all local $P_{i,i+1}$ and to find the ground state energy and the wave function of the ground state, one must employ a nested Bethe Ansatz approach \cite{Sutherland}.

\subsubsection*{Class Operator Hamiltonians}
On the other hand, the maximum degree that a node can have is $L-1$. It gives rise to the fully connected Hamiltonian
\begin{equation}
    H_2^{\text{class}} \,= \, \sum_{i< j} P_{i,j}\,\,\,.
\end{equation}
Such a Hamiltonian is also called a ``class operator'' \cite{elliottdawber}. Indeed, it is invariant under the left and right multiplication of any group element $g$
\begin{equation}
H_2^{\text{class}} \,=\, g^{-1} \, H_2^{\text{class}}\, g\,\,\,.
\end{equation}
It is easy to construct other class operator Hamiltonians. Indeed, if a conjugacy class $C_i$ consists of the group elements $\{g_1, g_2, \ldots g_{d_i}\}$, we can define the Hamiltonian ${\mathcal C}_i$ as 
\begin{equation}
    {\mathcal C}_i \,=\, \frac{1}{d_i} (g_1 + g_2 + \cdots g_{d_i}) \,\,\,.
    \label{CCC}
\end{equation}
For all Hamiltonians constructed in this way, we have $g^{-1} {\mathcal C}_i g$ for all group elements $g$. Hence, the ${\mathcal C}_i$ commutes with all elements of the Permutation Group. On the other hand, any element ${\mathcal C}$ of ${\mathcal S}_L$ , which commutes with all elements of the group, must be expressed as a linear combination of these class operators, ${\mathcal C} =\,c_1 {\mathcal C}_1 + c_2 {\mathcal C}_2 + \cdots$. The subspace, composed of sums of the class operators, belongs to the center ${\mathbb Z}$ of the group. There is a close connection between the algebra of the class operators and the Verlinde algebra \cite{Verlinde:1988sn}. Indeed, since the product ${\mathcal C}_i {\mathcal C}_j$ commutes with every element of the group, it must belong to the center of the group; therefore, there must exist constants $c_{ij}^k$ such that 
\begin{equation}
{\mathcal C}_i \,{\mathcal C}_j \,=\, \sum_k c_{i j}^k \, {\mathcal C}_k \,\, .
\label{algebraC}
\end{equation}
In view of this equation, ${\mathcal C}_i$ can be regarded as a linear mapping ${\mathbb Z} \rightarrow {\mathbb Z}$ with associated matrices, whose entries are $({\mathcal C}_i)_j^k = c_{i j}^k$. From the associativity condition of the algebra \eqref{algebraC}, all these matrices commute and, therefore, can be simultaneously diagonalized. Denoting with 
${\mathcal P}^I$ the projector operator on the Irreducible Representation $I$ of the permutation group, we have 
\be
{\mathcal C}_i \,{\mathcal P}^I \,=\, \left(\frac{\chi_i^I}{\chi_0^I}\right) \, {\mathcal P}^I
\,\,\,,
\label{eigensystemC}
\ee
where $\chi_i^I$ is the character of the class $i$ in the Irreducible Representation $I$ while $\chi_0^I \,=\,\chi_{\{e\}}^I \,=\, {\rm dim}\, I$. In summary, the common eigenvectors of the ${\mathcal C}_i$ are the projector operators ${\mathcal P}^I$ and the eigenvalues $\lambda_i^I = \chi_i^I/\chi_0^I$ are proportional to the characters.

\subsubsection*{Random Graphs} 
\begin{figure}
    \centering
    \subfloat[$r=0$]{\includegraphics[width=0.3\textwidth]{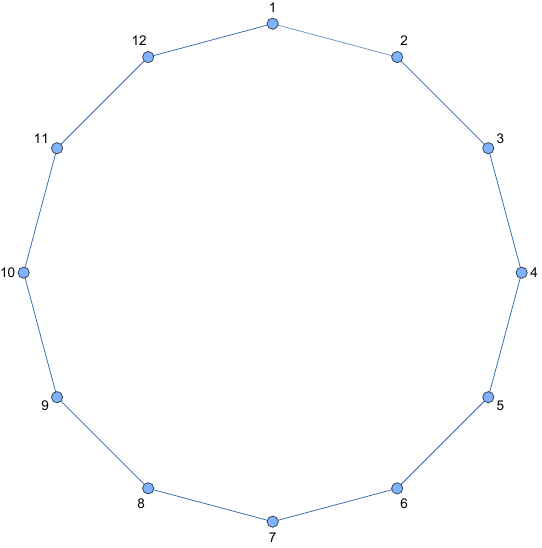}} \hspace{0.01\textwidth}
    \subfloat[$r=1/3$]{\includegraphics[width=0.3\textwidth]{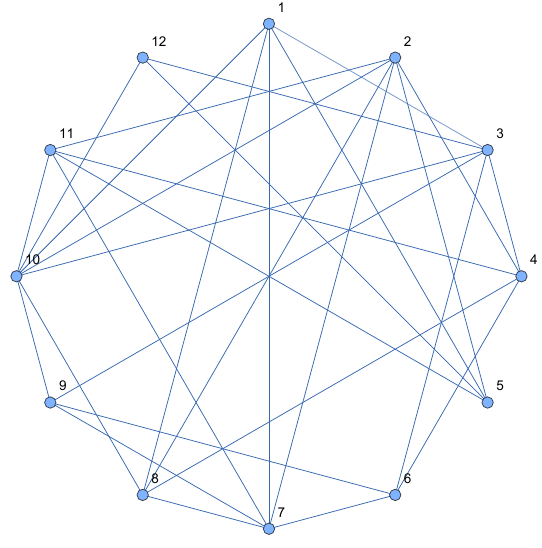}} \hspace{0.01\textwidth}
    \subfloat[$r=1$]{\includegraphics[width=0.3\textwidth]{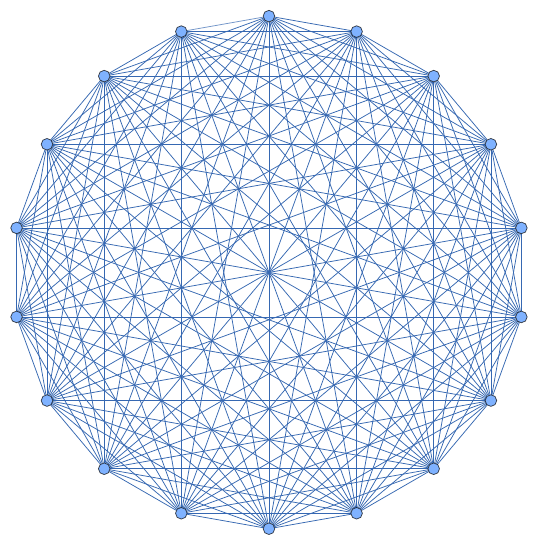}} 

    \caption{Different permutation Hamiltonian on graphs. While the cases $r=0$ and $r=1$ are integrable, the intermediate cases are not.}
    \label{f_graphs}
\end{figure}

\begin{figure}
    \centering
    \subfloat[]{\includegraphics[width=0.3\textwidth]{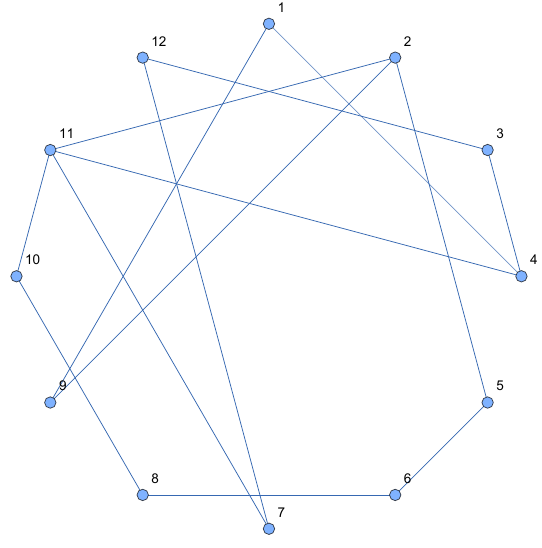}} \hspace{0.01\textwidth}
    \subfloat[]{\includegraphics[width=0.3\textwidth]{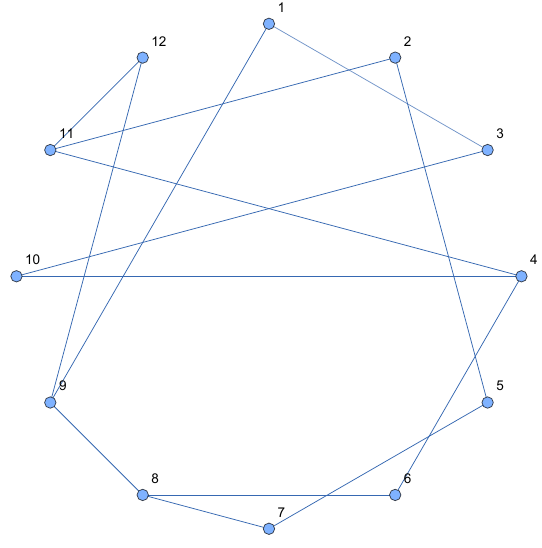}} \hspace{0.01\textwidth}
    \subfloat[]{\includegraphics[width=0.3\textwidth]{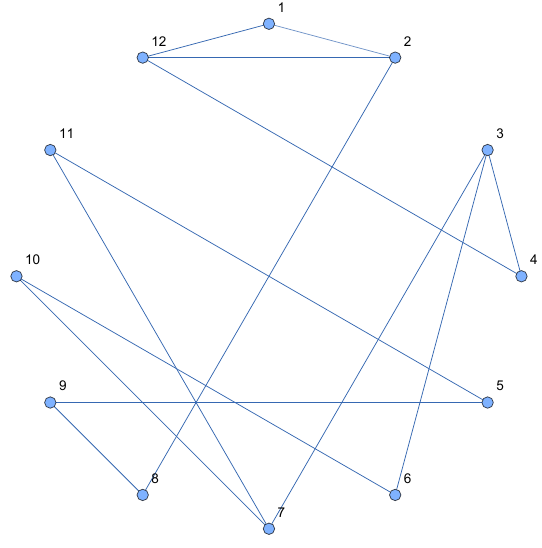}} 

    \caption{Different realizations of permutation Hamiltonian on graphs fixed the parameter $r=0.05$.}
    \label{f_realizations}
\end{figure}

The Hamiltonians discussed so far are ``rigid'' in the sense that their structure is fully determined by the underlying graph. However, one may also consider \textit{random} graphs, characterized by an average node degree parameterized by $r$:
\begin{equation}
    \mathcal{I}(r) \,=\,  \br{1-r} 2 + r (L-1).
\end{equation}
The parameter runs in $[0,1]$ and both 0 and 1 have only one possible Hamiltonian. For generic $r$, if $[x]$ denotes the integer part of the real number $x$, $\lfloor{rL/2}\rfloor = E$ is the number of edges, the number of Hamiltonians $H_G$ that one can generate on such random graphs is $E!/L!(E-L)!$. Examples of these graphs are provided in Figure~\ref{f_graphs} varying $r$, while in Figure~\ref{f_realizations} some realizations for $r=0.05$ are given. 

Except for the limiting cases $r=0$ and $r=1$, where the Hamiltonian is integrable, intermediate values of $r$ generally yield non-integrable dynamics. The Hamiltonian 
$\hat H$ is obtained as the average Hamiltonian of all the Hamiltonians $H_G$ of the random graphs with a given parameter $r$:
\begin{equation}
\hat H \,=\, \sum_{G} H_{G} \,\,\,,
\end{equation}
has an unfolded spectrum of the averaged energies that is equally spaced with a Gaussian distribution around the value of the gap $s = 1$.

\subsection{The Role of Boundary Conditions}
Let us now consider a one-dimensional Hamiltonian. For a fixed matrix size, it is computationally easier to diagonalize a Hamiltonian with open boundary conditions than with periodic boundary conditions, as the corresponding matrix is generally sparser. Nevertheless, diagonalizing Hamiltonians with periodic boundary conditions remains computationally feasible, as we are going to discuss. In a homogeneous system, periodic boundary conditions imply translational invariance. To obtain accurate level statistics, we know that it is essential to factor out global symmetries; translational invariance, when present, is one such symmetry that must be accounted for. For permutation Hamiltonians, the translation operator is given by
\begin{equation}
    \mathsf{T} = P_{1,\;2}P_{2,\;3}\ldots P_{L-2,\;L-1} P_{L-1,\;L} =\prod_{j=1}^{L-1} P_{j, \; j+1}\,\,.
\end{equation}
As the generator of the cyclic group \(\mathbb{Z}_L\), the translation operator has eigenvalues that are $e^{i \delta}$, where $\delta = 2\pi \kappa / L$ and $\kappa \in \{0, \ldots, L - 1\}$. Within a given irreducible representation \(\alpha\), the dimension \(m_\kappa^\alpha\) of the sector corresponding to momentum \(\kappa\) cannot be easily predicted \emph{a priori}, although its average value is \(d_\alpha / L\), where \(d_\alpha\) is the dimension of the representation.

Using the tools presented in Sec.~\ref{s_symmetries}, it is possible to compute the eigenvalues associated with each momentum sector in a memory-efficient manner for any given irreducible representation. This algorithm avoids explicitly constructing the full translation matrix \(\mathsf{T}\); instead, it requires only the application of individual transpositions---highly sparse operators---on a vector. The iterative algorithm proceeds as follows:

\begin{enumerate}
    \item \textbf{Initialize with a sample vector} \(\ket{v_i}\). This vector may be chosen as a standard basis vector \(\ket{e_i}^\ell = \delta_i^\ell\), though doing so often results in linear dependence among the generated vectors after only a few iterations---significantly fewer than \(m_\kappa\). To mitigate this, a random vector \(\ket{v_i}\), with components randomly chosen in the standard basis, is preferable. For \(\kappa = 0\) and \(\kappa = L/2\) (when \(L\) is even), the eigenvectors are real, so \(\ket{v_i}\) can have real coefficients. For all other values of \(\kappa\), the eigenvectors are complex, and the coefficients of \(\ket{v_i}\) should be complex as well.
    
    \item \textbf{Project onto the momentum sector} by extracting the \(\kappa\)-component \(\ket{\bm{\kappa}'_i}\) from \(\ket{v_i}\) using Eq.~\eqref{fourier1}.
    
    \item \textbf{Check the linear independence} of \(\ket{\bm{\kappa}'_i}\) with respect to the previously obtained vectors \(\ket{\bm{\kappa}_j}\) for \(j < i\). If \(\ket{\bm{\kappa}'_i}\) is linearly independent, retain its orthogonal component and denote it \(\ket{\bm{\kappa}_i}\). Otherwise, terminate the algorithm. The resulting set of vectors \(M_\kappa = \{\ket{\bm{\kappa}_1}, \ldots, \ket{\bm{\kappa}_{m_\kappa}}\}\) then spans the eigenspace associated with momentum \(\kappa\).
\end{enumerate}

Once the matrices \(M_\kappa\) have been constructed for a given irreducible representation \(\alpha\), the Hamiltonian can be projected onto the \(\kappa\)-momentum sector. While the full Hamiltonian has dimension \(d_\alpha \times d_\alpha\), the projected Hamiltonian \(H^{(\kappa)}\) has a reduced dimension \(m_\kappa^\alpha \times m_\kappa^\alpha\), given by:
\begin{equation}
    H^{(\kappa)} = M_{\kappa} H M_\kappa^\dagger\,\,.
\end{equation}
This projected matrix is typically denser and complex-valued (yet remains Hermitian), but the number of entries is cut off by a factor $O(L^2)$.

With moderate computational resources, it is possible to carry out exact diagonalization within sufficiently large irreducible representations and still obtain meaningful statistical results—see Table~\ref{tab_dimensions} for illustrative examples.
\begin{table}
    \centering
    \newcolumntype{L}[1]{>{\raggedright\arraybackslash}p{#1}}
    
    \begin{tabular}{|L{2.8cm}|L{2.8cm}|L{2.8cm}|}
    \hline
    Irreducible Representation & Dimension of Representation & Dimension of 0-Momentum Sector \\
    \hline
    $\sbr{11^2},\; L=22$ & $58786$ & $2652$
    \\
    \hline
    $\sbr{6^3} ,\; L=18$ & $87516$ & $4862$
    \\
    \hline
    $\sbr{4^4}, \; L=16$ & $24024$ & $1522$\\
    \hline
    \end{tabular}
    \caption{Dimensions of the largest irreducible representations studied for $n=2,3,4$. The dimension of the zero-momentum sector has been found through the algorithmic method of Section~\ref{s_permutation}.}
    \label{tab_dimensions}
\end{table}

Moreover, a generic translation invariant Hamiltonian of support $k$ can be written as
\begin{equation}
    H_k = \frac{1}{L} \sum_{j=0}^{L-1} \mathsf{T}^j h_{1,\ldots, k} \mathsf{T}^{-j}\,\,,
\end{equation}
where $h_{1,\ldots, k}$ is between the first $k$ sites. On a given momentum sector, $H_k$ has matrix elements
\begin{equation}
  \braket{v_\kappa|  H_k^{(\kappa)} |u_\kappa} =  \braket{v_\kappa| h_{1,\ldots, k} |u_\kappa}\,\,, \quad H_k^{(\kappa)} =  M_{\kappa} h_{1,\ldots, k} M_\kappa^\dagger\,\,.
\end{equation}
This last equation is quite remarkable because it implies that only one term of the Hamiltonian, $h_{1,\ldots, k}$, has to be stored in the computational memory. Moreover, the matrix $M_{\kappa}$ does not depend on the particular form of $H_k$, but only on the irreducible representation chosen; therefore, it can be stored and used for different permutation Hamiltonians.

\section{Examples of Permutation Hamiltonians}\label{examples}

We focus on permutation Hamiltonians with periodic boundary conditions and analyse them within a given momentum sector, as described previously. The two simplest Hamiltonians with two-site interactions are exemplified by the Sutherland permutation model with periodic boundary conditions, denoted by $H_2$, which takes the form
\begin{equation}
    h_2 = P_{1,2}\,\,,
\end{equation}
and the fully-connected class operator
\begin{equation}
    h_2^{\text{class}} = \frac{1}{L-1} \sum_{j>1} P_{1, j}\,\,.
\end{equation}

The Sutherland permutation model is known to be integrable for all $SU(n)$ \cite{Sutherland}, and its reduction to $n=2$ coincides with the antiferromagnetic Heisenberg chain, see Section~\ref{s_symmetries}. On the other hand, the fully-connected class operator is ``trivially integrable'', because in each irreducible representation, the Hamiltonian is proportional to the identity by Schur's lemma. 

We can generalize both models to arbitrary support by considering the operator $\mathscr{P}_{i_1, \ldots, i_k}$ which sums together all the $k$-cycles of the sites $i_1$, $\ldots$, $i_k$. $\mathscr{P}_{i_1, i_2, i_3}$ is given by
\begin{equation}
    \mathscr{P}_{i_1, i_2, i_3} = P_{i_1, i_2, i_3} + P_{i_1, i_3, i_2}\,,
\end{equation}
which visually can be represented as
\begin{equation}
    \mathscr{P}_{i_1, i_2, i_3} = \vcenter{\hbox{\includegraphics[width = 0.15\textwidth]{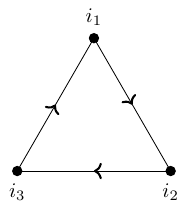}}} + \vcenter{\hbox{\includegraphics[width = 0.15\textwidth]{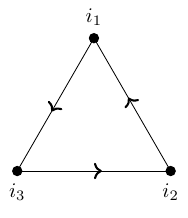}}}.
\end{equation}
Analogously for $\mathscr{P}_{i_1, i_2, i_3, i_4}$ we have the graphical representation
\begin{equation}
    \begin{aligned}
        \mathscr{P}_{i_1, i_2, i_3, i_4} &= &\vcenter{\hbox{\includegraphics[width = 0.15\textwidth]{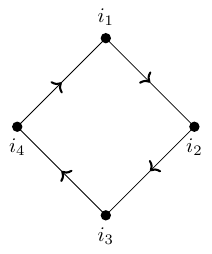}}} + \vcenter{\hbox{\includegraphics[width = 0.15\textwidth]{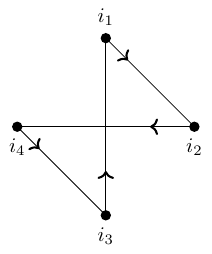}}} + \vcenter{\hbox{\includegraphics[width = 0.15\textwidth]{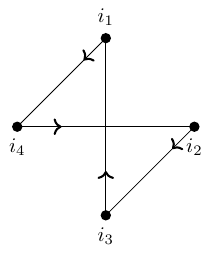}}} &+ \\
        & &\vcenter{\hbox{\includegraphics[width = 0.15\textwidth]{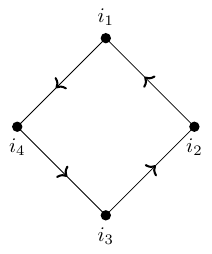}}} + \vcenter{\hbox{\includegraphics[width = 0.15\textwidth]{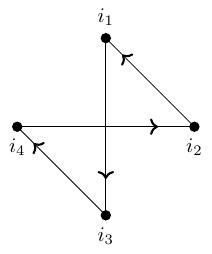}}}  + \vcenter{\hbox{\includegraphics[width = 0.15\textwidth]{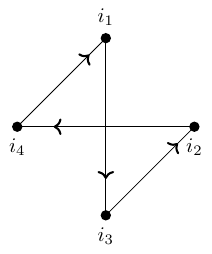}}} &.
    \end{aligned}
\end{equation}
Each $\mathscr{P}_{i_1, \ldots, i_k}$ has $(k-1)!$ terms, and for each element in the sum, its inverse is present to preserve Hermiticity. We can then define $k$-site Hamiltonians $H_{k}^{\text{site}}$ with
\begin{equation}
    h_{k}^{\text{site}} = \mathscr{P}_{1,2,\ldots, k}
\end{equation}
and $k$-class Hamiltonians $H_k^{\text{class}}$ with
\begin{equation}
    h_{k}^{\text{class}} = \frac{1}{(k-1)!} \sum_{i_2\neq i_3\neq\ldots\neq i_k > 1} \mathscr{P}_{1, i_2, \ldots, i_k}\,\,.
\end{equation}

As before, the $H_k^{\rm class}$ are trivially integrable, whereas numerical evidence indicates that the $H_k^{\text{site}}$  are generically non-integrable for 
$k= 3, \ldots, L-1$ in any $SU(n)$, since their level statistics follow the Wigner–Dyson distribution—see the plots at the end of this section.

The last Hamiltonians we consider are long-range, all-to-all ones. We start from the Inozemtsev hyperbolic chain model \cite{SerbanStaudacherIno, klabbersIno, BarbaIno, BarbaIno2}
\begin{equation}
    h_W = \sum_{j=2}^L \wp_{L, \pi/\tau}\br{j-1} P_{1, j}\,\,,
\end{equation}
where $\wp$ is the Weierstrass function defined by its two periods $L$ and $1/\tau$
\begin{equation}
    \wp_{L, \pi/\tau}\br{z} = \frac{1}{z^2} + \sum_{m, n \in \mathbb{Z}-\cbr{0}} \sbr{\br{z-mL-i\frac{n\pi}{\tau}}^{-2} - \br{mL+i\frac{n\pi}{\tau}}^{-2} }.
\end{equation}
We are interested in two limits. The first one is when $\tau \to 0$: in this case the effective Hamiltonian reduces to the trigonometric Haldane-Shastry model  \cite{Haldane1,Shastry1} with
\begin{equation}
    h_{HS} = \br{\frac{\pi}{L}}^2 \sum_{j=2}^L \br{\sin \frac{\pi (j-1)}{L}}^{-2} P_{1, j}\,\,.
\end{equation}
On the other hand, if we take the limit $L\to \infty$, we obtain the hyperbolic Inozemtsev chain
\begin{equation}
    h_{hI} = \tau^2 \sum_{j=2}^L \br{\sinh \tau(j-1)}^{-2} P_{1, j}\,\,.
\end{equation}
One also retrieves the Heisenberg model as $\tau\to \infty$ \cite{SerbanStaudacherIno}. 

An important feature of this family of Inozemtsev models is that \cite{BarbaIno,BarbaIno3,BarbaIno4}, while they admit exact solutions, their level statistics are not necessarily Poissonian. In particular, the Haldane–Shastry model exhibits almost equispaced degenerate energy levels, reminiscent of those of the multidimensional harmonic oscillator. By contrast, the hyperbolic Inozemtsev chain displays Poissonian level statistics. One may also consider the Haldane–Shastry chain with an arbitrary, non-extensive period relative to the system size; in this case, the level statistics follow the Wigner–Dyson distribution. To the best of our knowledge, no other local models exhibit the same degree of pathological behaviour as the Inozemtsev chain for what concerns the level statistics.

\subsection{Unfolding}
In our numerical analysis of many-body spectra, we have developed a procedure to obtain unfolded energies by fitting the energy density with Chebyshev polynomials. Since any Hamiltonian can be rescaled by an overall coupling constant $J$,  it is necessary to fine-tune $J$ so that the spectrum lies entirely within the domain of validity of the polynomials, namely $\sbr{-1,1}$.

In general, if the Hamiltonian has $T$ terms, i.e. $H = \sum_{i} h_i$, one chooses 
\begin{equation}
J = \br{\sum_{i} \max_j\lvert{\lambda_{j}^i}\rvert}^{-1},
\end{equation}
where $\lambda^i_j$ is the $j$-th eigenvalue of $h_i$. In most cases, the spectrum of the individual Hamiltonian terms is easily accessible. For instance, in $SU(2)$ 
spin chains with $T$ terms, the local terms are Pauli matrices, which are both Hermitian and unitary, and thus have eigenvalues $\pm 1$,  implying $J = 1/T$. 
Similarly, in $SU(n)$  permutation Hamiltonians generated by operators permuting (even cyclically) subsets of sites, the largest eigenvalue is $+ 1$
in the fully symmetric combination of basis states, and again one obtains $J=1/T$. 

Once the energies are rescaled to lie within $\sbr{-1,1}$, the fitting of the energy density $\rho_0(E)$ can be performed. Because this quantity is subject to statistical fluctuations, it is advantageous to fit its integral, namely the empirical cumulative density of states (Eq.~\eqref{eq_cumulative_doe}), with a linear combination of Chebyshev polynomials. For this method to be effective, it suffices that the level spacing is chosen in such a way that the cumulative density is strictly increasing and does not exhibit extended plateaus. The number of terms retained in the fit is not fixed a priori; instead, we employ an estimator, the reduced $\chi^2$, which is required to fall below a prescribed threshold. If the fit using the first 
$n$ Chebyshev polynomials yields a reduced $\chi^2$ above the acceptance value, the procedure is repeated with $n+1$ terms. Since the cumulative density is monotonic in 
$\sbr{-1,1}$, its derivative—the energy density—provides a smooth interpolation of the empirical distribution. In practice, it is not necessary to evaluate the derivative explicitly because the level spacings can be extracted directly from the unfolded energies as $s_i = e_{i+1}-e_i$, which are already normalized to a unit mean.

\begin{figure}
    \centering
     \subfloat[$\;$]{\includegraphics[width=0.45\linewidth]{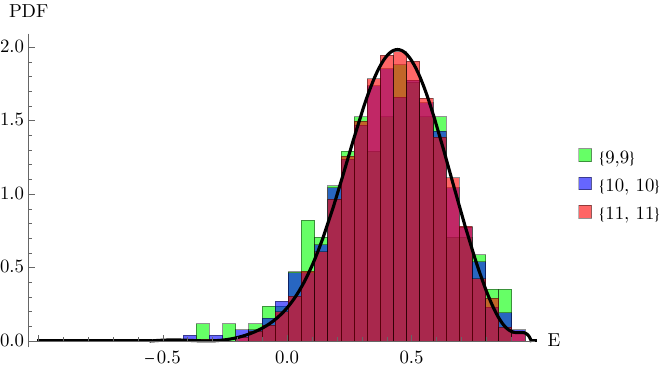}}
     \hspace{0.1\linewidth}
     \subfloat[$\;$]{\includegraphics[width=0.35\linewidth]{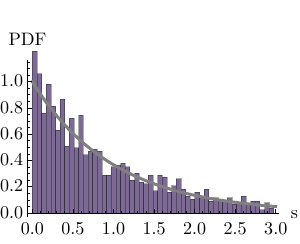}}
    \caption{On the left, the fit of the density of energy using Chebyshev polynomials of a zero-magnetization symmetry sector of the antiferromagnetic Heisenberg chain of various lengths: 18 (green), 20 (blue), 22 (red). The empirical density collapses to a continuous function as the length increases. On the right, the gap statistics obtained after unfolding is Poissonian because the model is integrable.}
    \label{f_example_fit}
\end{figure}

In Figure~\ref{f_example_fit} we provide an example of the method, considering the $SU(2)$ antiferromagnetic Heisenberg chain in some symmetry sectors, which are chosen such that in increasing the length of the chain, the density collapses to a continuous function -- see Appendix~\ref{a_permutations} for details.

\subsection{Numerical Results}

We now present the level statistics obtained through the unfolding procedure. The largest irreducible representations for $SU(n)$, $n=2$, 3 and 4, are those in Table~\ref{tab_dimensions}. 

\begin{figure}
    \centering
    \subfloat[Haldane Shastry, $\tau = 0$]{\includegraphics[width=0.3\textwidth]{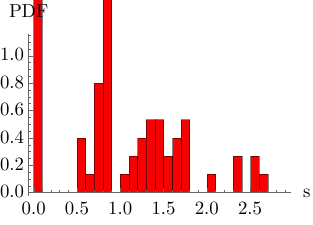}}
    \hfill
    \subfloat[Inozemtsev, $\tau = \pi/2$]{\includegraphics[width=0.3\textwidth]{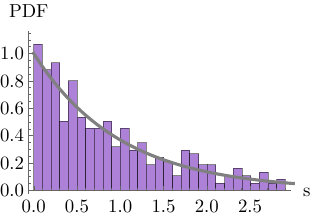}}
    \hfill
    \subfloat[Hyperbolic, $\tau = 1$]{\includegraphics[width=0.3\textwidth]{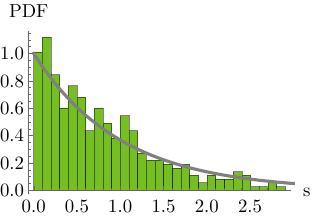}}
    \caption{Level statistics of the Inozemtsev chain and its limits, for $L=15$ in the irreducible representation $[5^3]$.}
    \label{f_inozemtsev}
\end{figure}

We start from the $SU(2)$ antiferromagnetic Heisenberg chain, as already reported in Figure~\ref{f_example_fit}. All momentum sectors exhibit the same Poissonian behaviour. On the other hand, in Figure~\ref{f_inozemtsev} we present level statistics for the Inozemtsev chain and its limits. For the Haldane-Shastry case, we observe two peaks at $s=0$, indicating a degenerate spectrum, and at $s=1$, the hallmark of equally spaced levels\footnote{More precisely, they are close to being  equally spaced in the thermodynamic limit; see details in \cite{BarbaIno3,BarbaIno4}}.

\begin{figure}
    \centering
    \subfloat[Even Sector]{\includegraphics[width=0.3\textwidth]{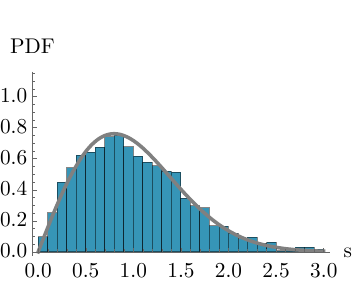}}
    \hfill
    \subfloat[Zero-momentum Sector]{\includegraphics[width=0.3\textwidth]{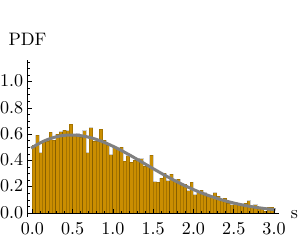}}
    \hfill
    \subfloat[Odd Sector]{\includegraphics[width=0.3\textwidth]{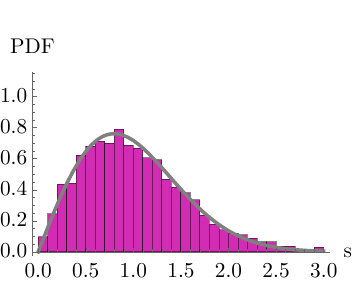}}
    \caption{Analysis of the zero-momentum sector of the 3-site Hamiltonian $H_3$ in the irreducible representation $[6^3]$ of $SU(3)$ ($L=18$). According to the reflection operator the sector splits into an even and odd one, which if put together generate the intermediate statistics for the superposition of 2 GOEs, see Eq.~\eqref{eq_superposition_GOE}. }
    \label{f_k_site}
\end{figure}

For $k$-site Hamiltonians, we diagonalize momentum sectors separately for $k = 3,4$ in the representations of $SU(3)$ and $SU(4)$, respectively. While for a generic momentum sector the Hamiltonians are not integrable and their statistics resemble Wigner-Dyson, the zero-momentum sector is intermediate, with a zero-gap probability around $1/2$. Following \cite{Porter}, we expect at least two degenerate, non-integrable spectra. This is indeed the case, since there exists this additional operator
\begin{equation}
    \label{eq_reflection}
    \mathsf{S} = \prod_{j = 1}^{L/2-1} P_{j, L-j+1}\,\,,
\end{equation}
which operates a reflection around the center of the chain. Since $\mathsf{S}^2 = \operatorname{id}$, reflections generate a $Z_2$ subgroup of the zero-momentum sector (also the $\pi$-momentum one if the chain is even). We label the two symmetry sectors ``even'' and ``odd'', according to the usual convention for parity (see Section~\ref{s_symmetries}). In these sectors, the Hamiltonian is non-integrable, see Figure~\ref{f_k_site}. This numerical analysis excludes the possibility of integrable $k$-site Hamiltonians apart from $k=2$ and $k=L$. Moreover, presently we do not have indications of integrable Hamiltonians made of three-site permutations.

\section{Conclusions}\label{conclusions}
 
Determining whether an infinite Hermitian matrix defines an integrable or chaotic quantum system is a central problem in many-body physics. In this work, we have approached this question from a statistical perspective, focusing on the spectral properties of the Hamiltonian. While the hallmark of integrability is traditionally identified with Poissonian statistics of the unfolded level spacings, we emphasize that similar behaviour may also emerge from mixtures of distinct spectral components. This ambiguity motivates the search for sharper criteria capable of distinguishing the spectrum of a genuinely integrable model from that of a system composed of non-integrable subsectors.

To this end, we have proposed a two-pronged protocol. First, a Monte Carlo decimation algorithm selectively filters the energy levels; by tracking the evolution of the spacing distribution under successive sweeps, one can assess whether an apparent Poissonian behaviour is robust or merely accidental. Second, we analyse higher-order spacing distributions, which provide discriminating signatures between Poisson, GOE, GUE, and mixed spectra.

We have demonstrated the effectiveness of this protocol through a range of examples, including quantum Hamiltonians constructed from permutation groups and realized on diverse graphs with different boundary conditions. Hence, our results establish a practical statistical framework for disentangling true integrability from spectral mimicry.

\section*{Acknowledgments} 
We acknowledge financial support from PNRR M4C2I1.3 PE\_00000023\_NQSTI Grant, Spoke A2, funded by NextGenerationEU. AS would like to thank the BIRS center for their kind hospitality during the workshop ``Exact Solutions in Quantum Information: Entanglement, Topology, and Quantum Circuits''. GM thanks Vladimir Kravtsov and Ivan Kostov for discussions on random matrices. GM and AS would also like to thank Matthias Staudacher for discussions on the Inozemtsev chain and all other participants of the Workshop on Integrability, in Favignana (June 2025). AH appreciated a discussion with Pavlo Gavrylenko and is thankful to Mark Arildsen for his help with programming. 

After we published the preprint of this work, the manuscript \cite{Bhosale2} appeared, which has partial overlap with the present one.

 \appendix
 \section{Conserved Charges}\label{AppendixA}
In this appendix, we aim to discuss the structure that the matrices associated with the conserved charges assume in the local basis \eqref{Hilbertdis}. It is convenient to consider a particular quantum integrable model (in our case, the Sinh-Gordon model, see \eqref{sh-G}) in which the conserved charges can be explicitly derived and then extract from this example the general features of the matrices.

The proper setting to approach the problem is the Hamiltonian formulation of a quantum field theory, in which the dynamical variables are the field $\varphi(x,t)$ and the conjugate field $\Pi(x,t)$: they enter the Hamiltonian 
\be
H\,=\,\int dx \left[\frac{1}{2} \Pi^2 + \frac{1}{2} (\nabla \varphi)^2 + V(\varphi) \right] \,\,\,,
\label{hamm}
\ee
and the equal-time commutation relation 
\be
\left[ \varphi(x,t),\Pi(y,t)\right] \,=\,i\,\delta(x-y) \,\,\,.
\label{commrel}
\ee
These operators satisfy the equation of motion 
\begin{eqnarray}
&& \frac{\partial \varphi}{\partial t} \,=\,-i \,\left[\varphi,H\right] \,=\,\Pi \,\,\,,\\
&& \frac{\partial \Pi}{\partial t} \,=\,-i \,\left[\Pi,H\right] \,=\,\nabla \varphi - \frac{d V}{d \varphi} \,\,\,.
\label{eqsmotion}
\end{eqnarray}
Eliminating $\Pi$, we obtain the familiar equation of motion for the field $\varphi(x,t)$
\be
\square \varphi + \frac{d V}{d\varphi} \,=\,0 
\,\,\,\,\,
,
\,\,\,\,\,
\square = \frac{\partial^2}{\partial t^2} - \frac{\partial^2}{\partial x^2} \,\,\,.
\label{kgeqs}
\ee
Let us see now how to derive the (classical) expression of the conserved ${\mathcal Q}_s$ of a particular quantum integrable model, the Sinh-Gordon model. In the light-cone coordinates $\sigma$ and $\tau$ 
\[
\sigma\,=\,\frac{1}{2} (x-t)\,\,;
\hspace{3mm}
\tau\,=\,\frac{1}{2} (x+t)\,\,\,.
\]
and rescaling the field $\phi$, the equation of motion \eqref{kgeqs} for the field $\phi$ assumes the form   
\be
\partial_{\sigma} \,\partial_{\tau} \,\phi(\sigma,\tau) \,=\,\,\sinh(\phi) \,\,\,. 
\label{MOTOSHSIGMATAU}
\ee
There is a conserved charge ${\mathcal Q}_s$ if there exists a current with components $(J^0_s,J^1_s)$ satisfying the equation $\partial_{\mu} J^{\mu}_s =0$. This can be written in light-cone coordinates, defining $J^0_{s} = T_{s+1} + \Theta_{s-1}$ and $J^1_{s} = T_{s+1} 
- \Theta_{s-1}$. For the densities $T_{s+1}[\phi]$ and $\Theta_{s-1}[\phi]$, we have 
\be
\frac{\partial}{\partial \sigma} T_{s+1}[\phi] \,=\, \frac{\partial}{\partial\tau} \Theta_{s-1}[\phi] \,\,\,.
\label{dsdtau}
\ee
The index $s$ refers to the spin of this current, related to the difference of the partial derivatives
$\partial_{\tau}^n$ and $\partial_{\sigma}^k$ present in the expression of the densities, $s = n - k $. The charge $Q_s$  
\be
{\mathcal Q}_s \,=\,\int_{-\infty}^{\infty} J^0_s \,dx\,=\,
\int \left[T_{s+1}  + \Theta_{s-1}  \right]\,dx \,\,\,, 
\ee 
is a conserved quantity since, for Eq. \eqref{dsdtau}, it satisfies 
\be
\frac{d {\mathcal Q}_s}{d t} \,=\,0 \,\,\,.
\ee 
To explicitly find the densities $T_{s+1}[\phi]$ and $\Theta_{s-1}[\phi]$, let's define the field 
$\hat \phi(\sigma,\tau)$, solution of the so-called {\em B\"acklund transformations}
\begin{eqnarray}
\partial_{\sigma}(\hat\phi-\phi)\,=\,
2\, \epsilon \,\sinh\left(\frac{1}{2}(\hat\phi+\phi)\right)\,\, ,
\label{Backlund}\\
\partial_{\tau} (\hat\phi+\phi)\,=\,
\frac{2}{\epsilon} \,\sinh\left(\frac{1}{2}(\hat\phi-\phi)\right)
\,\, . \nonumber
\end{eqnarray}
Assuming that $\phi(\sigma,\tau)$ is a solution of the equation of motion, Eqs. \eqref{Backlund} provide another solution. In fact, acting with $\partial_{\tau}$ on the first of them and using the second equation, we have  
\[
\partial_{\tau} \,\partial_{\sigma} (\hat\phi -\phi) \,=\,2\,\,\sinh\frac{1}{2}
(\hat\phi - \phi)\,\cosh\frac{1}{2} (\hat\phi +\phi) \,=\,
\left[\sinh ( \hat\phi) - \sinh( \phi) \right]\,\,\,.
\] 
The field $\hat\phi(z,\bar z,\epsilon)$ can be expressed as a power series of the parameter $\epsilon$
\be
\hat\phi(\sigma,\tau,\epsilon)\,=\,\sum_{n=0}^{\infty} 
\phi^{(n)}(\sigma,\tau)\,\epsilon^n\,\,,
\label{series}
\ee
where $\phi^{(n)}(\sigma,\tau)$ can be computed by plugging it into \eqref{Backlund}
and comparing term to term in $\epsilon$. For the first term, we 
have\footnote{$\phi_{\tau} \equiv \partial_{\tau} \phi$ and 
$\phi_{\sigma} \equiv \partial_{\sigma} \phi$. In the following, we will also use $\phi_t \equiv \partial_t \phi$ and $\phi_x \equiv \partial_x \phi$.}
\be
\begin{array}{lll}
\hat\phi^{(0)} \,=\,\phi & \,\,,\,\, & \phi^{(1)} \,=\,2 \phi_{\tau} \,\,\,,\\
\hat\phi^{(2)} \,=\,2 \phi_{\tau \tau} & \,\,,\,\, & \phi^{(3)} \,=\,2 \phi_{\tau \tau \tau } 
- \phi_{\tau}^3/3 \,\,\,, \\
\hat\phi^{(4)} \,=\,2 \phi_{\tau \tau \tau \tau} - 2 \phi_{\tau}^2\,\phi_{\tau \tau} & \,\,,\,\,& \cdots 
\end{array}
\label{espansionephihat}
\ee 
The existence of this series expression gives us the possibility to obtain an infinite number of conservation laws starting from a finite number of them. To this aim we can use, for instance--
\be
\left(\frac{1}{2} \psi_{\tau}^2\right)_{\sigma} + (1 - \cosh\psi )_{\tau} \,=\,0 \,\,\,,
\label{sigmataufundation}
\ee 
or a similar equation 
\be
\left(\frac{1}{2} \psi_{\sigma}^2\right)_{\tau} + (1 - \cosh\psi )_{\sigma} \,=\,0 \,\,\,,
\label{tausigmafundation}
\ee 
whose validity can be easily checked by employing the equation of motion \eqref{MOTOSHSIGMATAU} 
satisfied by the field $\psi$. Using, for instance, Eq.  \eqref{sigmataufundation} and substituting Eq. \eqref{espansionephihat}, we obtain an infinite number of conserved densities. The first non trivial expressions (i.e. those that cannot be expressed as total derivatives) are 
\begin{eqnarray}
&& T_2 \,=\,\frac{1}{2} \phi_{\tau}^2 \nonumber \\
&& T_4 \,=\, 2 \phi_{\tau\tau}^2 + 2 \phi_{\tau}\,\phi_{\tau\tau\tau} \\
&& T_6 \,=\, 2 \phi_{\tau\tau\tau}^2 + 4 \phi_{\tau\tau} \,\phi_{\tau\tau\tau\tau} 
- 6 \phi_{\tau\tau}^2 \phi_{\tau}^2 - 2 \phi_{\tau}^3 \phi_{\tau\tau\tau} + 
2 \phi_{\tau} \phi_{\tau\tau\tau\tau\tau} \nonumber
\end{eqnarray}  
and 
\begin{eqnarray}
&& \Theta_1 \,=\,(\cosh\phi -1 )  \nonumber \\
&& \Theta_3 \,=\, 2  \phi_{\tau}^2\,\cosh\phi + 2 \phi_{\tau\tau}\,\sinh\phi \\
&& \Theta_5 \,=\, 4 \phi_{\tau}\,\phi_{\tau\tau} \cosh\phi + \frac{4}{3} \phi_{\tau}^3\,\sinh\phi + (2 \phi_{\tau\tau\tau} - \frac{1}{3} \phi_{\tau}^3) \sinh\phi \,\,\, . 
 \nonumber
\end{eqnarray}
Conserved densities of negative values of $s$ are obtained by simply substituting the index $\tau$ with $\sigma$. In general, it can be proved that non trivial conservation laws are obtained for all odd values of $s$
\be
s \,=\,1, 3,5,\ldots 
\label{conservedspin}
\ee 
The set of these values of $s$ constitutes the spectrum of the conserved charges. It is also possible to show that the classical expressions of the conserved currents, appropriately modified, keep their meaning also at the quantum level and that the corresponding charges are in involution, i.e. they commute with each other  
\be
[\,{\mathcal Q}_s,{\mathcal Q}_{s'}\,] \,=\,0 \,\,\,.
\ee  
Conserved charges that are invariant under parity transformation $x \rightarrow -x$ and time reversal 
$t \rightarrow - t$ are given by the combination $\hat{\mathcal Q}_s = 1/2 ({\mathcal Q}_s + {\mathcal Q}_{-s})$. The Hamiltonian, for instance, is given by 
\begin{eqnarray}
H & \,=\, & \frac{1}{2}({\mathcal Q}_1 + {\mathcal Q}_{-1}) \,=\,\int \left[\frac{1}{4} (\phi^2_{\tau} + \phi^2_{\sigma}) + (\cosh\phi -1) \right] dx  = \nonumber \\
& \,=\, & \int \left[\frac{1}{2} (\phi^2_t + \phi^2_x) + V(\phi) \right] dx = 
\int \left[\frac{1}{2} \Pi^2 + \frac{1}{2} (\nabla\phi)^2 + V(\phi) \right] dx  \,\,\,.
\end{eqnarray}
It is easy to see that the conserved charges ${\mathcal Q}_{\pm s}$ can be expressed as polynomials in the variables $\Pi(x)$, $\phi(x)$ and higher space derivatives of these fields, alias 
\be
{\mathcal Q}_{\pm s} = {\mathcal P}_{\pm s} \left[\Pi(x), \partial_x^l \Pi(x),\phi(x),\partial_x^m \phi(x)
\right]\,\,\,.
\ee
To find the explicit expressions of these polynomials, one has to do the following steps: 
\begin{enumerate}
\item firstly, to express the derivatives $\partial_\sigma$ and $\partial_{\tau}$ as 
\be
\partial_{\sigma} \,=\,\partial_x - \partial_t 
\,\,\,\,\,\, 
, 
\,\,\,\,\,\,
\partial_{\tau} \,=\,\partial_x + \partial_t 
\ee
\item 
secondly, to use the equation of motion \eqref{eqsmotion} of the Hamiltonian formalism and   substitute in all expressions in which they appear $\partial_t \phi \rightarrow \Pi$ and $\partial_t \Pi = \left(\nabla^2 \phi - \frac{dV}{d\varphi}\right)$. 
\end{enumerate}
Following these rules, one arrives at the final expression of the conserved charges in terms of 
$\Pi(x)$, $\phi(x)$ and their higher space derivatives. In the local basis \eqref{Hilbertdis}, the matrix elements of the terms that involve the field $\phi$ and its derivatives are diagonal. Off-diagonal terms come from the hopping action induced by the conjugate field $\Pi$ and its higher powers $\Pi^k$ (see Eq. \eqref{actionPi} in the main text). Following the rules given above, these powers in $\Pi$ are originated by those terms in ${\mathcal Q}_{\pm s}$ that contain odd-derivatives in $\tau$ of the field $\phi$. Consider, for instance, the term $\phi_{\tau} \phi_{\tau\tau\tau}$ in 
$T_4$: applying the above rules and neglecting, for simplicity, issues related to the non-commutativity of the fields $\phi$ and $\Pi$, the higher derivative in $\Pi$ is obtained by keeping track of the higher derivative in $t$ of the field $\phi$ 
\begin{eqnarray}
&& \phi_{\tau} \phi_{\tau\tau\tau} \rightarrow \left[(\partial_x + \partial_t ) \phi \right]\,
\left[ (\partial_x + \partial_t )^3 \phi \right] 
\rightarrow \left[\partial_t \phi \right] \left[\partial^3_t \phi \right]\\
&=& \Pi \,\partial^2_t \Pi = \Pi \,\partial_t (\nabla^2 \phi - \sinh\phi) =
\Pi \,(\nabla^2 \Pi - \Pi \cosh\phi) \nonumber \,\,\,.
\end{eqnarray} 
It is easy to see that there is a bound on the higher power of $\Pi$ that can enter ${\mathcal Q}_s$: this bound comes directly from the spin $s$ of this conserved charge. In fact, for a given spin $s$, the higher power of $\Pi^k$ can only come from a term $(\phi_{\tau})^s \rightarrow \Pi^s$, 
if present in the expression of the conserved current. Therefore, for all conserved charges ${\mathcal Q}_s$ we have 
\be
k \leq s \,\,\,.
\label{bound}
\ee
This means that the $N \times N$ matrix representation of the conserved charges ${\mathcal Q}_s$ in the local basis is a sparse matrix, with the maximum number of non-zero entries given by 
\be
\rho \,\leq \, \frac{s}{N \log q} \,\log N \,\,\,.
\ee

\section{Higher order spacing in random matrix\label{level-space}}

As we discussed in the main text, evaluating the higher spacing probability distribution for superposed spectra requires the numerical computation of the determinant of the Gaussian ensemble. In what follows, we will present two numerical strategies for calculating the determinants of GUE and GOE: the eigenvalue expansion method and the contour integral method. Both rely on Nystr\"om-type discretization.  Detailed implementation guidelines and technical aspects can be found in \cite{Bornemann1,Bornemann2}.

\subsection*{GUE $\beta = 2$}

The bulk eigenvalue correlation for GUE is governed by a determinantal point process \cite{johansson2006}.  The PGF for the number of energy levels in $[0,s]$ is (see \cite{deift2000})
\begin{equation}
	D(z;s)
	= \det\big( I - (1-z){\mathcal K}_{[0,s]} \big)\,\,,
	\label{PGF-GUE}
\end{equation}
where the ${\mathcal K}_{[0,s]}$ is the integral operator on ${L^2([0,s])}$ with kernel 
\begin{equation}
	\label{kernel-GUE}
	K(x,y)= \frac{\sin(\pi (x - y))}{\pi (x - y)} , \quad x,y \in[0,s],
\end{equation}
i.e.,
\begin{equation}
	({\mathcal K}f)(x) = \int_0^s K(x,y) f(y) dy .
\end{equation}

In order to determine the higher spacing probability distribution $P_k^{(2)}(s)$, we can make use of the following two numerical methods: 

\subsubsection*{(a) Eigenvalue product expansion}

\begin{itemize}
\item We first discretize the sine-kernel \eqref{kernel-GUE} by using the Nystr\"om-type method, which approximates the integral operator on $L^2([0,s])$ as a finite $m \times m$ symmetric matrix $K_m$, i.e.,
\begin{equation}
D(z;s)
= \det\big( I-(1-z) K_m \big) \approx  \det\big( I-(1-z)  \big(\sqrt{\omega_i} K(x_i,y_j) \sqrt{\omega_j} \big)_{i,j=1}^m \big)\,\,\,.
\end{equation} 
The $\{\omega_i\}$ ($i=1,2,\cdots,m$) are Gauss–Legendre \cite{Bornemann1} quadrature weights.

\item Secondly, we compute the eigenvalues $\{\lambda_j(s)\}$ ($0 \leq \lambda_j(s) \leq 1$) of $K_m$. The counting probability $E_k$ can now be expressed as
	\begin{equation}
		D(z;s)
		= \sum_{k=0}^{m} E_2(k;s)  z^k = \prod_{j=1}^{m} (1 - (1-z) \lambda_j(s)) = E_2(0;s)\; \sum_{k=1}^{m} e_k^{(m)}\big(a_1,a_2,\cdots,a_m \big)z^k\,\,,
	\end{equation}
where
\begin{equation}
	E_2(0;s) = \prod_{j=0}^{m} (1-\lambda_j)\,\,,\qquad\quad a_j = \frac{\lambda_j}{1-\lambda_j}\,\,\,.
\end{equation}
$e_k$ are elementary symmetric polynomials, defined as
\begin{equation}
	e_k^{(m)}(x_1,x_2,\ldots, x_m) =\sum_{1 \leq i_1 < i_2 < \cdots < i_k \leq m} x_{i_1} x_{i_2} \cdots x_{i_k}\,\,\,.
\end{equation}
with convention $e_0^{m}=0$ and $e_k^{(m)}=0$ for $k<0$ and $k>m$.
Hence, we obtain the counting probability
\begin{equation}
	\label{EK-egenval-approach}
	E_2(k;s) =E_2(0;s)\; e_{k}^{(m)}\big(a_1,a_2,\cdots,a_m \big) \qquad k=1,2,\cdots,m\,\,\,
\end{equation}
In addition, we can use the recursive equations of the elementary symmetric polynomials 
  \begin{equation}
	e_k^{(n)} (x_1,\cdots,x_n)= e_{k}^{(n-1)}(x_1,\cdots,x_{n-1}) + x_n\, e_{k-1}^{(n-1)}(x_1,\cdots,x_{n-1}).
\end{equation}
for computing them efficiently.

\item Finally, the higher spacing distribution function $P^{(2)}_k(s)$ can be obtained by taking the second derivative of $\{E_2(k;s)\}$ with respect to $s$, as given in Eq.\eqref{eq-k-level-spacing-prob} in the main text. 

The eigenvalue product expansion method is effective for small $k$, but it suffers from instability for large $k$. The main reason is that it requires full diagonalization of the kernel matrix $K_m$, but all the eigenvalues are small and $0 \leq \lambda_j(s) \leq 1$. Therefore, the counting probabilities $\{E_2(k;s)\}_{k\geq 0}$ are very sensitive to  precision of eigenvalues, while, for large $k$, they involve many products of eigenvalues, and hence demand a high computed eigenvalue accuracy.

\end{itemize}

\subsubsection*{(b) Contour integral method}

In this method, no diagonalization of kernel matrix is needed. The numerical steps are as follows:

\begin{itemize}
	\item We first discretize the sine-kernel \eqref{kernel-GUE} by the Nystr\"om-type method, similar as what we used in eigenvalue product expansion method, which lead to
	\begin{equation}
		D (z;s)
		= \det \big( I-(1-z) K_m \big) \approx  \det \big( I-(1-z)  \big(\sqrt{\omega_i} K(x_i,y_j) \sqrt{\omega_j} \big)_{i,j=1}^m \big)\,\,.
	\end{equation} 
\item Secondly, we calculate $E_{2}(k;s)$ by Cauchy integrals at $z=0$,
\begin{equation}
	E_{2}(k;s)= \frac{1}{k!} \left. \frac{{\rm d}^{k}}{{\rm d}  z^{k}} D(z;s)  \right|_{z=0} = \frac{1}{k!} D^{(k)}(0;s)\,\,,
\end{equation}

\begin{equation}
	\label{eq:cauchy}
	D^{(k)}(0;s)
	= \frac{k!}{2\pi i}\oint_{|z|=\rho} \frac{D(z;s)}{z^{k+1}}{\rm d}z
	= \frac{k!}{2\pi\rho^k}\int_0^{2\pi} e^{-ik\theta}
	D \left(\rho e^{i\theta};s\right){\rm d}\theta\,\,,
\end{equation}
Using the trapezoidal rule with $N$ equispaced angles $\theta_l =2\pi l/N\,(l=0,1,\cdots,N-1)$, the counting probability is
\begin{align}
	E_{2}(k;s) \approx \frac{1}{N \rho^k} \sum_{l=0}^{N-1} D(\rho e^{i\theta_l};s )e^{-ik\theta_l}\,\,,
\end{align} 
where the determinant $D \left(\rho e^{i\theta_l};s\right)$ can be computed via the LU decomposition. The $E_2(k;s)$ is exactly the discrete/fast Fourier transform of the samples $D(\rho e^{i\theta_l};s )$. For numerical stability, the contour radius $\rho \in [0,1]$ here is chosen properly to ensure that the $\rho^{-k}$ is well-behaved and the contour $|z|=\rho$ does not pass through any zeros of $D(\rho e^{i\theta};s)$ .

\item Finally, the $P^{(2)}_k(s)$ can be numerically calculated through Eq.\eqref{eq-k-level-spacing-prob}.

The contour integral method is not as intuitive as the eigenvalues product method, but it is stable for large $k$ and easy to control the precision. The only requirement is the stable evaluation of the determinant over the complex plane. Therefore, it is better to use the contour integral method to evaluate higher $P^{(2)}_k(s)$.

\end{itemize}

\subsection*{GOE $\beta = 1$}

For the GOE ensemble, the bulk eigenvalue correlations form a Pfaffian point process \cite{kim2021,ortmann2017,anderson2010}. The corresponding PGF is
\begin{equation}
	D(z;s)
	= \text{Pf}  \big( J - (1-z)J\mathcal{K}_{[0,s]} \big)\,\,.
	\label{PGF-GOE}
\end{equation}
where $J$ is a $2 \times 2$ canonical skew-symmetric matrix and $\mathcal{K}_{[0,s]}$ is a matrix-valued operator on $L^2([0,s])$ with kernel
\begin{equation}
	J = \begin{pmatrix} 0 & 1 \\ -1 & 0 \end{pmatrix}\,\,, \quad 
    (\mathcal K)_{x,y} = \begin{pmatrix}
		S(x,y) & (DS)(x,y) \\[6pt]
		(IS)(x,y)-\epsilon(x,y) & S(y,x)
	\end{pmatrix}, \quad x,y \in [0,s]\,\,,
\end{equation}
in the  in the $N\rightarrow \infty$ limit. Here, the kernel matrix has entries
\begin{align}
    &S(x,y) = K(x,y)\,\,, 
\quad (DS)(x,y) = \frac{\partial}{\partial x} K(x,y)\,\,,\\
&(IS)(x,y) = \int_y^x K(t,y) {\rm d}t\,\,, \quad \epsilon(x,y) = \frac{1}{2}\text{sign}(x-y),
\end{align}
where the $K(x,y)$ is the sine-kernel \eqref{kernel-GUE}.

In the bulk scaling limit, as proved by Mehta in \cite{Mehta1}, one can introduce two generating functions
\begin{align}
	D_{\pm}(z;s) &= \det  \big(1 - (1-z)\mathcal{K}^{\pm}_{[-s/2,s/2]}\big), 
	\label{PGF-GOE-even-odd}
\end{align}
where $\mathcal{K}^{\pm}$ denotes a scalar Fredholm operator acting on $L^2([-s/2,s/2])$ with kernels 
\begin{equation}
	K^{\pm}(x,y)=\tfrac12\Big(K(x,y)\pm K(x,-y)\Big)\,\,,\qquad x,y\in[-s/2,s/2]\,\,.
\end{equation}
These kernels are just the orthogonal decomposition of the ordinary sine kernel operator in the even/odd subspace after a shift from $L^2([0,s])$ to $L^2([-s/2,s/2])$ (sine kernel is shift invariant).  We then define two reference counting probabilities in the even/odd subspaces
\begin{align}
		E_\pm(k;s) = \frac{1}{k!}\frac{{\rm d}^k}{{\rm d}z^k} D_\pm(z;s)\Big|_{z=0}\,\,, \qquad k=0,1,2,\dots
\end{align}
The count probabilities $\{E_1(k;s)\}_{k\ge0}$ can therefore be computed through
$\{E_+(k;s)\}$ and $\{E_-(k;s)\}$ via the even/odd recursion \cite{Mehta1}
\begin{align}
	E_1(0;s)      &= E_+(0;s)\,\,,\\
	E_1(2k-1;s) &= E_-(k-1;s) - E_1(2k-2;s),\qquad k\ge 1,\\
	E_1(2k;s)   &= E_+(k;s)    - E_1(2k-1;s),\qquad k\ge 1.
\end{align}
Once the counting probabilities $\{E_1(k;s)\}_{k\ge0}$ that we need are computed, one can immediately get the $k$-level spacing probability distribution $P^{(1)}_k(s)$ through \eqref{eq-k-level-spacing-prob}. The two numerical strategies--- eigenvalue product expansion and contour integral method, we used for GUE, are carried over unchanged here for the calculation of $E_{\pm}(k;s)$ according to $D_\pm(z;s)$.  

\subsection*{Poisson}
The counting probability for a Poisson process is 
\begin{equation}
    E_{\textbf{Poisson}}(k;s) = \frac{s^k {\rm e}^{-s}}{k!}.
\end{equation}
One can see that the $E_P(u)$ in (\ref{poter-poisson}) is exactly $E_{\textbf{Poisson}}(0;s)$ here.

\begin{figure}[t]
\begin{center}
$\begin{array}{ccc}
\includegraphics[width=0.48\textwidth]{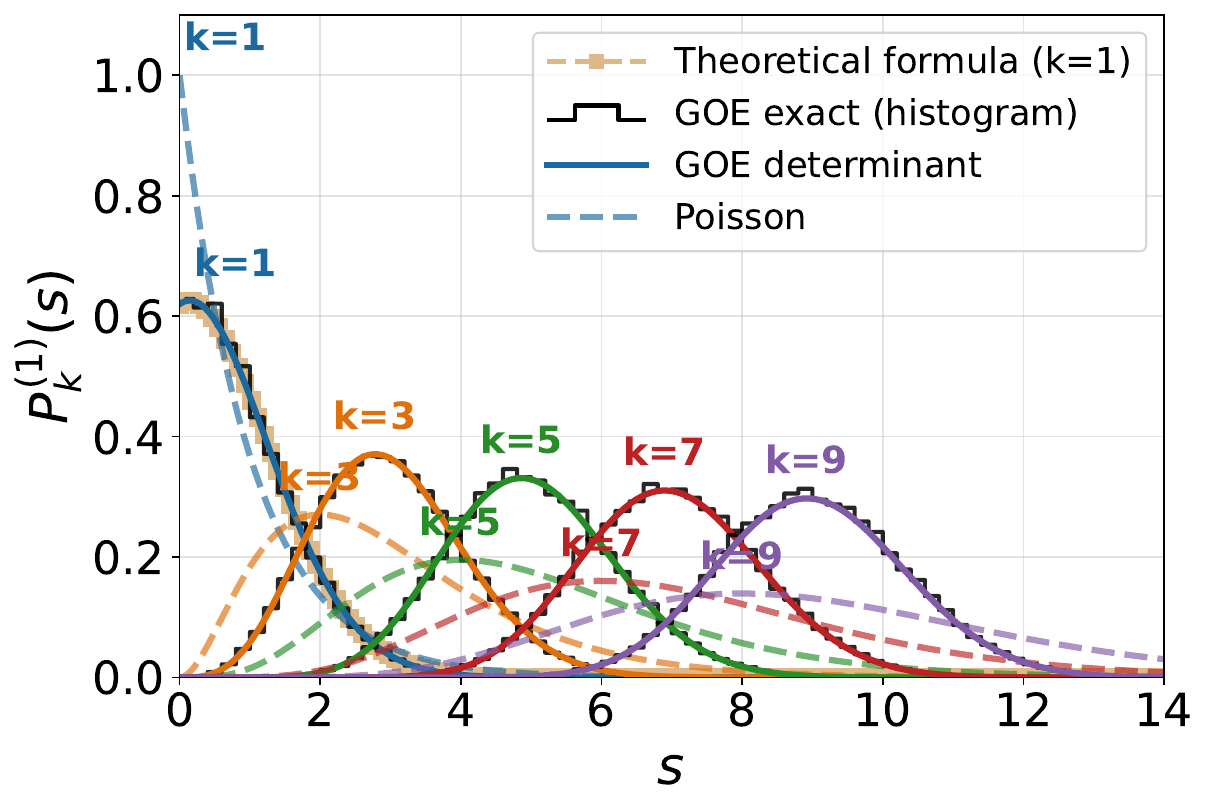} &  
&\includegraphics[width=0.48\textwidth]{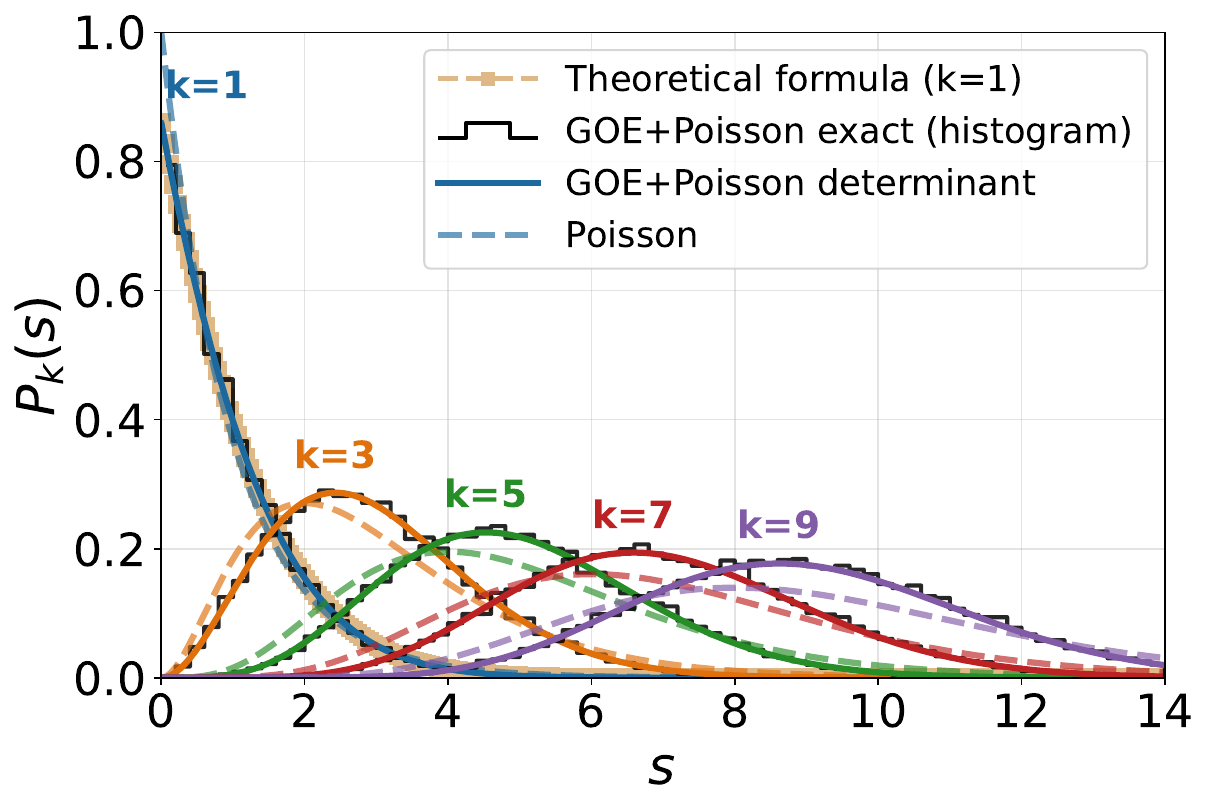} 
 \\
(a) & & (b)  
\end{array}$
\end{center}
\caption{ $k$-th level spacing probability distribution $P_k(s)$ for a mixture of (a) 3 GOE spctra with fractions $f_1=0.2,\;f_2=0.3,\;f_3=0.5$; (b) 1 Poisson spectrum and 2 GOE spctra  with fractions $f_{\text{Poisson}}=0.5$, GOE:$\;f_1=f_2=0.25$. The theoretical formula is Eq.\eqref{poter-formula} for the mixture. The GOE exact and GOE$+$Poisson exact is an average over 10 realization of (a) 3 GOE spectra with total size $N_{tot}=4000$; (b)  1 Poisson and 2 GOE spectra with total size $N_{tot}=4000$.}
\label{GOE-GOE-Poisson-mixture}
\end{figure}

The counting probability distribution functions $E_{\beta}(k;s)$ for GUE, GOE and Poisson processes can be computed through the numerical methods we discussed in this appendix. Hence, we can get any $k$-th spacing probability distribution $P_k(s)$ for superposed spectra with any fractions by using selection rules (\ref{eq-E-sup}) and (\ref{eq-k-level-spacing-prob-sup}). In Figure~\ref{GOE-GOE-Poisson-mixture}, we show examples of the higher spacing probability distribution for (i) the superposition of 3 GOE spectra; (ii) Poisson-GOE mixture, both with unequal fractions.

\subsection*{Convolution method}

\begin{figure}[t]
\begin{center}
$\begin{array}{ccc}
\includegraphics[width=0.48\textwidth]{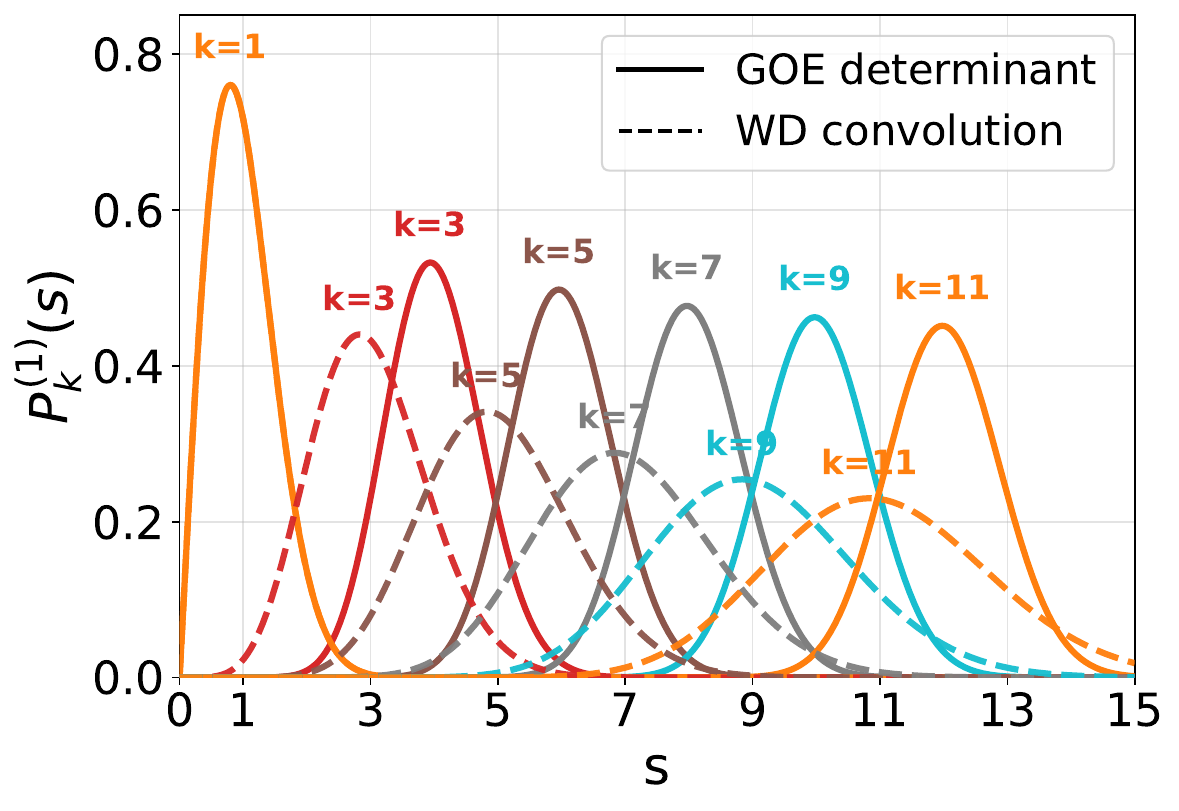} &  
&\includegraphics[width=0.48\textwidth]{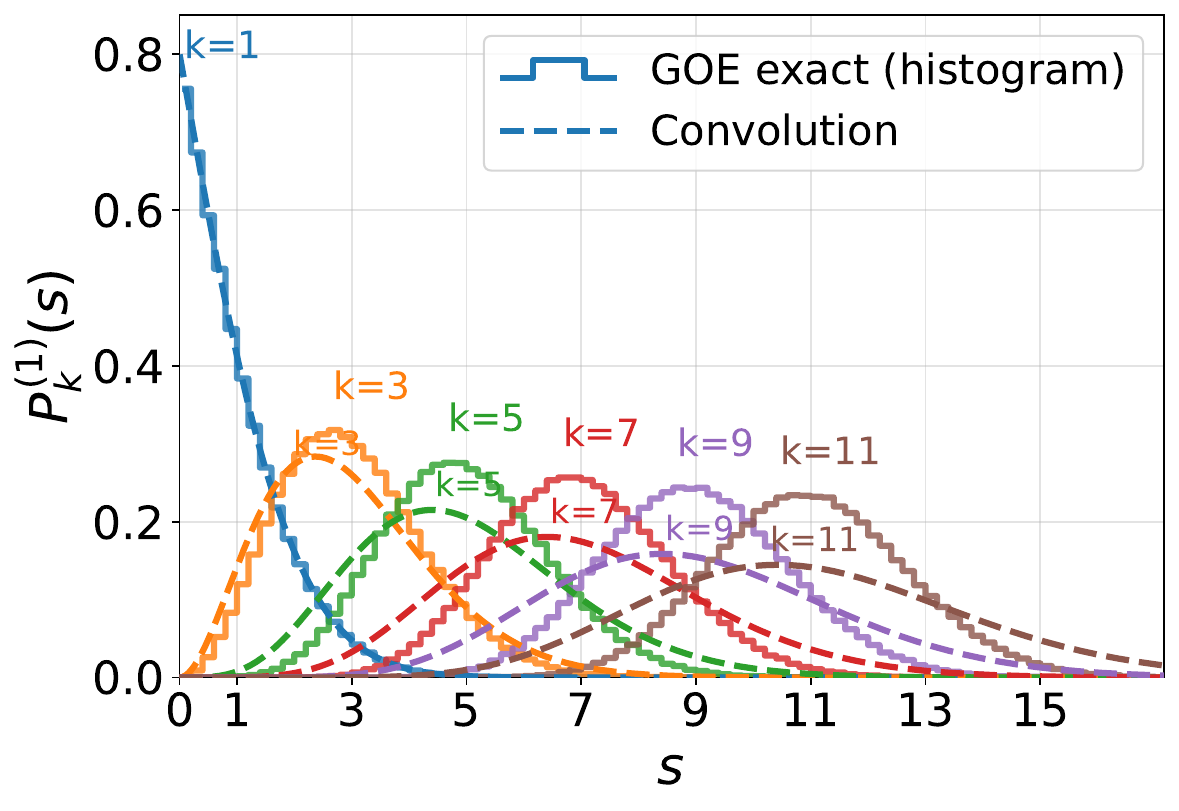} 
 \\
(a) & & (b)  
\end{array}$
\end{center}
\caption{(a) $k$-th level spacing probability distribution $P_k^{(1)}(s)$ for a single GOE determinant and Winger-Dyson convolution. (b) $k$-th spacing probability distribution $P_k^{(1)}(s)$ for a superposition of 5 GOE spectra and probability convolution according to Eq.\eqref{poter-formula} with equal fractions $f=1/5=0.2$, where the histogram is an average over $200$ realization of superposed 5 GOE spectra, each from an exact diagonalization of a $400 \times 400$ matrix.}
\label{GOE-determinant-convolution}
\end{figure}

The exact computation of $k$-th level spacing distributions differs in a significant way from a natural approximation  and it is worth explaining such a discrepancy. To start with, we obviously have
\be
s_{n,k} \,=\, (e_{n,k} - e_{n,k-1}) + (e_{n,k-1} - e_{n,k-2}) + \cdots (e_{n+1} - e_n) \,=\, \sum_{i=1}^k s_i \,\,\,,
\ee 
which, for the average, implies
\be 
\langle s_{n,k}\rangle \,=\, \sum_{i=1}^k s_i \,=\, k \, \langle s_i \rangle \,=\, k \,\,\,,
\ee
since $ \langle s_i \rangle \,=1$. The approximation for the probability distributions of these higher order spacings  relies on treating each individual level spacing as statistically independent from the others. Under this assumption, the probability distribution $P_k(s)$ of the $k$-step spacing is obtained as the convolution of $k$ Wigner–Dyson nearest-neighbour spacing distributions
\be
P_k(s) \,=\, 
\left(P^{WD}_1 \star P^{WD}_1\star P^{WD}_1 \cdots \star P^{WD}_1\right)(s) 
\ee
namely 
\be
P_k(s) \,=\, \int_0^s \cdots \int_0^s \left(\prod_{i=1}^k P^{WD}_1(x_i)\right) \, \delta\left(s - \sum_{i=1}^k x_i\right) \, dx_1 dx_2 \ldots dx_n  \,\,\,.
\ee
Its analytic expression for $k=2$ and the GOE ($\beta =1$) is given by 
\be
P_{k=2}(s) \,=\, \frac{\pi}{16} e^{-\frac{\pi s^2}{4}} \, \left[4 s + \sqrt{2} e^{\frac{\pi s^2}{8}} \,(\pi s^2 - 4) \,{\rm erf}\left(\frac{1}{2} \sqrt{\frac{\pi}{2}} s\right)\right]
\ee
where 
\[
{\rm erf}(x) \,=\, \frac{2}{\sqrt{\pi}} \, \int_0^x e^{-t^2} \, dt \,\,\,.
\]
For higher $k$ one can evaluate the convolution using the Fourier transform: putting 
\be
\hat P^{WD}_1(p) \,=\, \frac{1}{\sqrt{2 \pi}} \,\int_{-\infty}^{\infty} P^{WD}_1(x) \, e^{i p x} \, dx
\ee
we have indeed 
\be
P_k(s) \,=\,  \frac{1}{\sqrt{2 \pi}} \,\int_{-\infty}^{\infty} \left(\hat P^{WD}_1(p)\right)^k  \, e^{-i p x} \, dp\,\,\,.
\ee
For large $k$, by the Central Limit Theorem the higher-spacing probability distribution is given by the normal distribution
\be
P_k(s) \simeq \frac{1}{2 \pi k \sigma_1(\beta)} \, \exp\left[-\frac{(s - k)^2}{2 k \sigma_1^2(\beta)}\right]
\ee
where 
\be
\sigma_1^2(\beta) \,=\, 
\left\{
\begin{array}{lll}
\displaystyle \frac{4}{\pi} -1 \,=\, 0.27234..&, & \displaystyle \beta = 1\,\,\,, \\
& & \\
\displaystyle \frac{3 \pi}{8} -1 \, =\, 0.17810... &,& \displaystyle\beta =2 ,\,\,,,
\end{array}
\right.
\label{sigma11}
\ee
This "independent–gaps" approximation, although simple, essentially neglects residual correlations among adjacent spacings. The variance of the 
corresponding curves in the "independent-gaps" approximation is predicted to grow as $\sqrt{k}$ while in reality. 

In Figure~\ref{GOE-determinant-convolution}, we compare the $k$-th level spacing probability distribution of a single and superposed 5 GOE spectra with convolution-based predictions obtained by probability distribution of Wigner-Dyson and Eq.\eqref{poter-formula}. As $k$ increase, the convolution curves  broaden rapidly, by contrast, the level spacing distribution for the single and superposed GOE remain noticeably rigid.

\section{Permutation Group and Its IR's}\label{a_permutations}

We devout this appendix to present more information about the permutation group of $L$ elements $\mathcal{S}_L$ and its Irreducible Representations (IR) on the Hilbert space of permutation Hamiltonians, i.e. those in which the local Hilbert space is made of one particle with a given colour out of $n$ distinct colours. More details can be found in classic textbooks \cite{elliottdawber,hamermesh,fultonharris,stoneGolbart, diaconis} and classical references \cite{young,rutherford}.

Each element $P$ of the $\mathcal{S}_L$ belongs to a conjugacy class $\mathcal{C}_\alpha$, which is uniquely fixed by the number of cycles. A cycle of length $n$ permutes $n$ elements and it is assigned a ``parity'' $(-1)^{n-1}$. The permutation group splits then into two parity sectors and therefore two subgroups -- the alternating one and the symmetric one. This parity can be checked promptly by computing the action of $P$ on the fully antisymmetric combination generated by the vector $\ket{1, \ldots, L}$, corresponding to the only vector of the fully antisymmetric representation $[1^L]$. 

Let us now describe how the generators $P_{i, i+1}$ are realized as orthogonal matrices in a given irreducible representation $\beta$, following the standard procedure of \cite{elliottdawber}. 

A particular ``standard'' basis for an irreducible representation has vectors $\ket{s_i}$ that are characterized by $L-1$ ``quantum numbers'': $\ket{L_i} = \ket{\beta, \beta^i_1, \ldots, \beta^i_{L-2}}$. Here $\beta^i_j$ is the irreducible representation of $\mathcal{S}_{L-j}$ under which the standard vector $\ket{s_i}$ transforms when $\mathcal{S}_{L}$ is restricted to the subgroup $\mathcal{S}_{L-j}$ generated by the first $(L-j)$ elementary transpositions. 

Visually, a standard vector $\ket{s_i}$ is represented by a standard Young tableau, obtained by filling an empty diagram with the integers $\cbr{1,\ldots,L}$ according to the following rule: if the last $j$ boxes are removed from the tableau, the corresponding diagram corresponds to one of $\beta_j^i$. Their number is exactly $d_\beta$. Another possible enumeration of the standard vectors, equivalent to the Young tableaux, is given by Yamanouchi symbols. The $j$th entry of the symbol indicates a row of the original Young diagram. For that row, the rightmost box will be dropped in the reduction $\mathcal{S}_{L-j}\to\mathcal{S}_{L-j-1}$.

The number of tableaux obtained through this rule coincides with the dimension $d_\beta$ of the representation, which can also be promptly computed through the hook's formula, as noted in the main text. The hooks can be directly computed from the Young diagram in this way. For each box $i$ of the diagram, its hook is defined as 
\begin{equation}
    H_i=1+R_i+B_i,
\end{equation}
where $R_i$ is the number of boxes to the right of $i$, and $B_i$ is the one below. Then the formula reads
\begin{equation}
    d_\beta = \frac{L!}{\prod_{i\in \text{boxes}} H_i}.
\end{equation}
For example, both the symmetric, $[L]$, and the antisymmetric, $[1^L]$, have dimension one.

Young provided a theorem to construct the elementary nearest neighbouring transpositions $P_{i\; i+1}$ in a given irreducible representation $\beta$. The matrices $IR_{\beta}\br{P_{i\; i+1}}\equiv P_{i\; i+1}^\beta$ are chosen to be real and unitary (orthogonal). Crucially, the construction refers to the standard basis -- also known as ``orthogonal'' basis in the literature. It is known that for the regular representation the rotation matrix from the physical and standard bases is given by symmetrizers and antisymmetrizers read from the Young tableaux -- see the Supplementary Material of \cite{natafmila}.

Fixed a standard basis vector $\ket{s_j}$, the $j$-th row of $P_{i\; i+1}^\beta$ will contain at most two entries. Therefore, transposition matrices are sparse, and the maximum number of nonzero entries in total is $2/d_\beta$. For the $j$-th row, we find among the basis the vector $\ket{s_k}$ whose tableau coincides with that of $\ket{s_j}$ upon the exchange of the integer entries $i$ and $i+1$. Between the two tableaux, one computes the axial distance \cite{elliottdawber} $\ell^i_{j, \; k}$, and the entries of the $j$th and $k$th rows form an $SU(2)$ matrix
\begin{equation}
    \begin{bmatrix}
        \braket{s_j|P_{i\; i+1}^\beta |s_j} & \braket{s_k|P_{i\; i+1}^\beta |s_j}\\
        \braket{s_j|P_{i\; i+1}^\beta |s_k} & \braket{s_k|P_{i\; i+1}^\beta |s_k} 
    \end{bmatrix} =
    \begin{bmatrix}
        -\br{\ell^i_{j, \; k}}^{-1} & \sbr{1- \br{\ell^i_{j, \; k}}^{-2}}^\half \\
        \sbr{1- \br{\ell^i_{j, \; k}}^{-2}}^\half & \br{\ell^i_{j, \; k}}^{-1}
    \end{bmatrix}.
\end{equation}

We conclude this appendix by commenting on the structure of the Hilbert space of dimension $d=n^L$. The decomposition presented in Section~\ref{s_permutation} is special because it is left invariant by the action of a permutation Hamiltonian. More generally, an operator on the Hilbert space can be decomposed in the irreducible representations of $SU(n)^{\otimes L}$, i.e. by ``spins'' or equivalently by a Young diagram of $SU(n)$. Confusion may arise, since Young diagrams are also used for $\mathcal{S}_{L}$. Generic $SU(n)$ Young diagrams are labelled by the partitions of $L$, with at most $n$ addends $p(L)|_{n}$. To differentiate these irreducible representations from the ones of $\mathcal{S}_{L}$, we fill a Young diagram of $SU(n)$ with a light gray colour and denote its associated partition by double square brackets, e.g.
\begin{equation*}
\sbr{\sbr{3,2,1}}_n = {\ydiagram[*(lightgray)]{3,2,1} }^{\; n}
\end{equation*}
denotes an irreducible representation with $L=6$ and generic $n\leq L$. 

An irreducible representation $\mathfrak{IR}_\alpha$ of $SU(n)$ has dimension $\mathfrak{f}_\alpha$ and multiplicity $\mathfrak{m}_\alpha$. These coefficients can be read from the Young diagram \cite{natafmila}. To compute $\mathfrak{m}_\alpha$, it is sufficient to divide $L!$ by the product of the hooks read from the Young diagram. For $\mathfrak{f}_\alpha$, fill all the diagonal boxes with $n$, all the boxes to the right of the diagonal with $n+\gamma$, where $\gamma$ is the number of boxes to the diagonal counted horizontally, and all the boxes to the left of the diagonal with $n-\delta$, where $\delta$ is the number of boxes to the diagonal counted vertically. To compute $\mathfrak{f}_\alpha$, multiply the entries obtained in this way and divide by the product of the hooks. Summing over all irreducible representations yields the dimension of the Hilbert space: $d = n^L = \sum_\alpha \mathfrak{m}_\alpha \mathfrak{f}_\alpha$.

As mentioned above, for the permutation Hamiltonian, one can find other invariant subspaces, each labelled by the colour sector $\cbr{L_1, L_2, \ldots, L_n}$,\,\, $\sum_{i}L_i = L$. These subspaces are spanned by the vectors of the ``physical basis'' $\ket{i_1, \ldots, i_L}$, where $L_1$ sites (not necessarily adjacent) are occupied by colour 1, $L_2$ sites by colour 2, and so on. Each subspace has dimension $f_{L_1, L_2, \ldots, L_n} = L!/(L_1!\ldots L_n!)$. Naturally, the number of colour sectors and their dimensions must match $n^L$.

The number of distinguishable colour sectors equals to $\mathscr{P}(L)|_n$, and each of them appears with a nontrivial multiplicity, since $H$ puts all colours on the same footing. To find the multiplicity $m_{L_1,\ldots, L_n}$, with $L_1 \geq L_2\geq \ldots \geq L_n$, denote by $\ell_s$ the number of $N_i$'s such that $L_i = s$. Then, the multiplicity is the number of ways $n$ ``boxes'' can be filled with the indistinguishable numbers $s = \cbr{0, 1, \ldots, L}$, each appearing $\ell_s$ times:
\begin{equation}
    m_{L_1,\ldots, L_n} = \frac{n!}{\ell_0! \ell_1! \ldots \ell_N!}.
\end{equation}
Summing over partitions, we retrieve the dimension of the Hilbert space
\begin{equation}
    d = n^L = \sum_{\cbr{L_1, \ldots, L_n} \in \mathscr{P}(L)|_n} m_{L_1,\ldots, L_n} f_{L_1,\ldots, L_n}.
\end{equation}
In Table~\ref{t_decomposition} we provide an example of the two aforementioned decompositions for $n=3$ and $L=4$.

\begin{table}[t]
\centering
\ytableausetup{smalltableaux}

\begin{subtable}[c]{0.46\textwidth} 
\centering
\begin{tabular}{|c|c|c|}
\hline
Irreducible Representation & $\mathfrak{f}_\alpha$ & $\mathfrak{m}_\alpha$ \\
\hline
${\ydiagram[*(lightgray)]{4}}^{\;3}$ & 1 & 15 \\
${\ydiagram[*(lightgray)]{3,1}}^{\;3}$ & 3 & 15 \\
${\ydiagram[*(lightgray)]{2,2}}^{\;3}$ & 2 & 6 \\
${\ydiagram[*(lightgray)]{2,1,1}}^{\;3}$ & 3 & 3 \\
\hline
\end{tabular}
\caption{$SU(n)$}
\end{subtable}
\begin{subtable}[c]{0.46\textwidth}
\centering
\begin{tabular}{|c|c|c|}
\hline
Colour Sector & $f_{n_1,n_2,n_3}$ & $m_{n_1,n_2,n_3}$ \\
\hline
$\cbr{4,0,0}$ & 1 & 3 \\
$\cbr{3,1,0}$ & 4 & 6 \\
$\cbr{2,2,0}$ & 6 & 3 \\
$\cbr{2,1,1}$ & 12 & 3 \\
\hline
\end{tabular}
\caption{Colour Sector}
\end{subtable}

\caption{Decomposition of the Hilbert space of $L=4$, $n=3$ according to $SU(n)$ (a) and colour sectors (b).}
\label{t_decomposition}
\ytableausetup{nosmalltableaux}
\end{table}

\begin{figure}
    \centering
    \includegraphics[width=\linewidth]{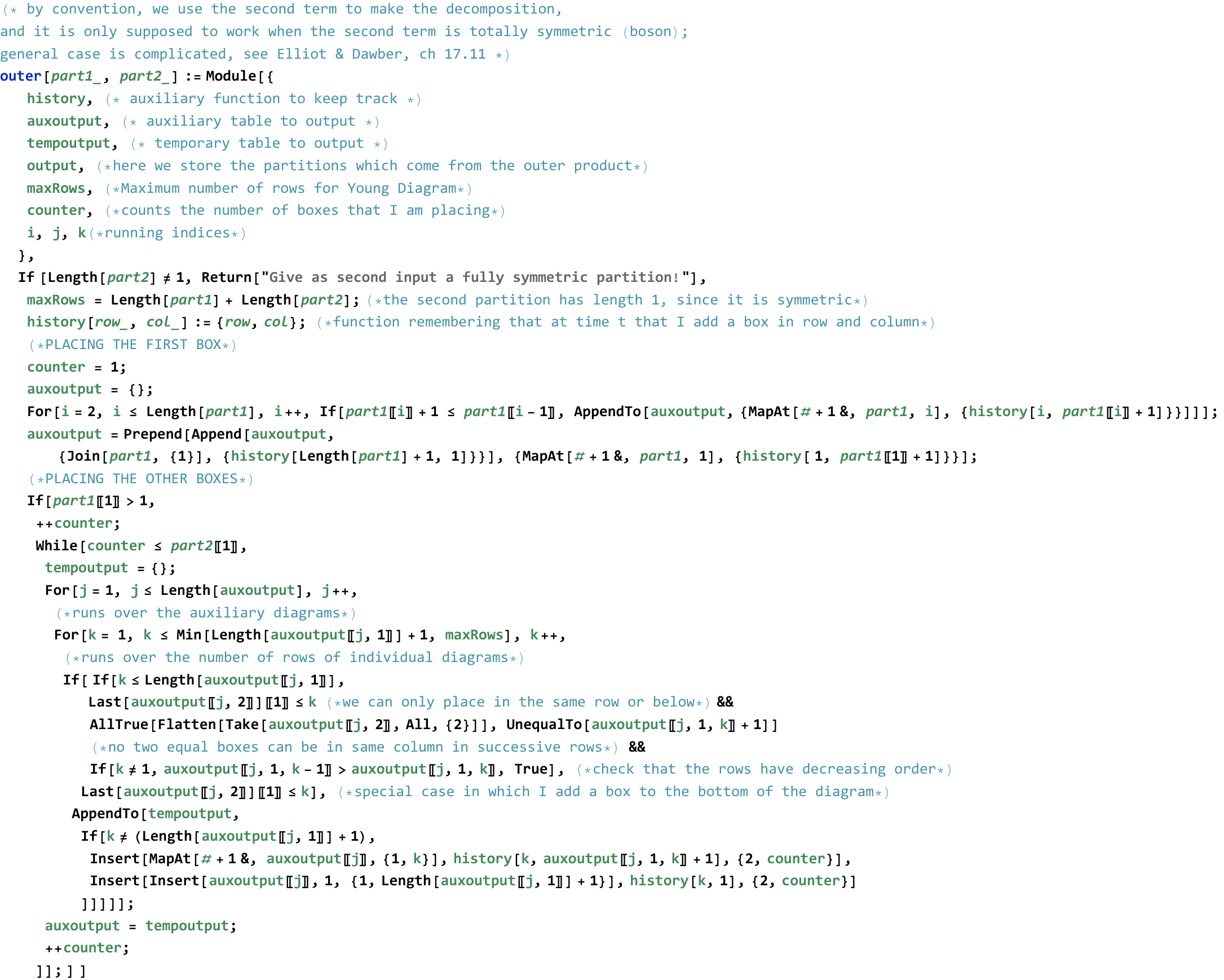}
    \caption{Outer product algorithm, using Wolfram Mathematica\textsuperscript{\textregistered} 14.3.}
    \label{f_code}
\end{figure}

As we mentioned in Section~\ref{s_permutation}, colour sectors can be further split into irreducible representations of $\mathcal{S}_L$. The reason is that many-body wave functions of $L$ bosons in the colour sector $\cbr{L_1, \ldots, L_n}$ transform according to irreducible representations of $\mathcal{S}_{\cbr{L_1, \ldots, L_n}} = \mathcal{S}_{L_1}\otimes \ldots \otimes \mathcal{S}_{L_n}$, which is a subgroup of $\mathcal{S}_L$. It is well known that the irreducible representations of $\mathcal{S}_{\cbr{L_1, \ldots, L_n}}$ are those of $\mathcal{S}_L$. They can be found through the outer product formula \cite{elliottdawber}. For the case at hand, particles are bosons and do not have mixed statistics with respect to $\mathcal{S}_L$, the outer product formula admits a nice algorithm, which we provide in Figure~\ref{f_code}. 

The outer product formula is provides in essence the symmetry transformation of the an $L$ boson wave function subjected to the constraints given by the colour sector $\cbr{L_1, \ldots, L_n}$. Therefore, in the decomposition not all irreducible representations appear, only those with a number of rows $r \leq n$. While the irreducible representations $\alpha$ appearing in a certain outer product formula can be predicted on the basis of symmetry, it is not straightforward to obtain the multiplicities $m_\alpha$ -- this is exactly the purpose of the algorithm in Figure~\ref{f_code}. An obvious check is that the dimension of the colour sector should coincide with $\sum_\alpha m_\alpha d_\alpha$. 

For $SU(L)$, the outer product decomposition of the colour sector $\cbr{1, 1, \ldots, 1}$ where all particles are of different colour coincides with the regular representation of $\mathcal{S}_L$, where all irreducible representations $\alpha$ appear $m_\alpha = d_\alpha$ times, such that $L! = |\mathcal{S}_L| = \sum_\alpha d_\alpha^2$. For small $n$, on the other hand, very interesting representations are those for which $L = mn$. There is one irreducible representation in the colour sector $\cbr{m, m, \ldots, m}$ (i.e. where all particle colours are equally distributed) which is $\sbr{m^n}$. This is the representation where the ground state of the antiferromagnetic Heisenberg and Sutherland permutation Hamiltonians lies, and one may regard it as a Hilbert space with subleading corrections. Using the hook formula, one finds asymptotically:
\begin{equation}
    d_{\sbr{m^n}} = (nm)! \prod_{j=0}^{n-1} \frac{j!}{(n+j)!} \underset{n\gg 1}{\sim} m^{nm} \frac{G(m+1)}{n^{m(m-1)/2}},
\end{equation}
where $G(z) = \prod_{k=0}^{z} \Gamma(k+1)$ is the Barnes $G$ function.

\newpage
\printbibliography
\end{document}